\documentclass[twoside]{pltptentat}
\pdfoutput=1
\usepackage{natbib}
\usepackage{graphicx}
\usepackage{amssymb}
\usepackage{bm}
\usepackage{amsmath}

\bibpunct[, ]{(}{)}{,}{a}{}{,}
\begin{document}

\begin{frontmatter}



\title{TURBULENT DYNAMICS IN ROTATING HELIUM SUPERFLUIDS}


\author{V.B. Eltsov\ead{ve@boojum.hut.fi}}, \author{R. de Graaf, R. H\"anninen, M. Krusius, R.E. Solntsev}
\address{Low Temperature Laboratory, Helsinki University of Technology, P.O.Box 5100, FI-02015-TKK, Finland}
\author{V.S. L'vov}
\address{Department of Chemical Physics, The Weizmann Institute of Science,\\ Rehovot 76100, Israel}
\author{A.I. Golov}, \author{P.M. Walmsley}
\address{School of Physics and Astronomy, The University of Manchester,\\ Manchester M13 9PL, UK}


\begin{abstract}

New techniques, both for generating and detecting turbulence in
the helium superfluids $^3$He-B and $^4$He, have recently given
insight in how turbulence is started, what the dissipation
mechanisms are, and how turbulence decays when it appears as a
transient state or when externally applied turbulent pumping is
switched off. Important simplifications are obtained by using
$^3$He-B as working fluid, where the highly viscous normal
component is practically always in a state of laminar flow, or by
cooling $^4$He to low temperatures where the normal fraction
becomes vanishingly small. We describe recent studies from the low
temperature regime, where mutual friction becomes small or
practically vanishes. This allows us to elucidate the mechanisms
at work in quantum turbulence on approaching the zero temperature
limit.

\end{abstract}

\end{frontmatter}

\newpage
\thispagestyle{empty}
\tableofcontents
\newpage
\setlength{\parindent}{5mm}

\section{Introduction}
\label{Intro}

The transition to turbulence is the most well known example of all
hydrodynamic transitions. It has been marveled for centuries,
since dramatic demonstrations can be seen everywhere where a
sudden change in the flow occurs, owing to a constriction in the
flow geometry, for instance. For fifty years it has been known
that turbulence also exists in superfluids \citep{Vinen:2007},
although by its very nature a superfluid should be a
dissipation-free system. In many situations it is found that on
the macroscopic level superfluid vortex dynamics mimics the
responses of viscous hydrodynamics. This is one of the reasons why
it has been thought that superfluid turbulence might provide a
short cut to better understanding of turbulence in general. From
the developments over the past fifty years we see that this has
not become the case, superfluid turbulence is a complex phenomenon
where experiments have often been clouded by other issues,
especially by vortex formation and vortex pinning. Nevertheless,
the topic is fascinating in its own right: When the flow velocity
is increased, the inherently dissipation-free superfluid is
observed to become dissipative and eventually turbulent. This is
particularly intriguing in the zero temperature limit where the
density of thermal excitations approaches zero and vortex motion
becomes undamped down to very short wave lengths (of the order of
the vortex core diameter).

There are two isotropic helium superfluids in which turbulence has
been studied, namely the B phase of superfluid $^3$He ($^3$He-B)
and superfluid $^4$He ($^4$He II). In the anisotropic A phase of
superfluid $^3$He ($^3$He-A) dissipation is so large that
conventional superfluid turbulence is not expected at the now
accessible temperatures above $0.1\,T_{\rm c}$ \citep{Finne:2003}.
Instead rapid dynamics and large flow velocities promote in
$^3$He-A a transition in the topology and structure of the axially
anisotropic superfluid order parameter field, a transition from
linear line-like vortices to planar sheet-like vortices
\citep{Eltsov:2002}. Turbulence has also been studied in
laser-cooled Bose-Einstein condensed cold atom clouds, although so
far only theoretically \citep{Parker:2005,Kobayashi:2008}, but it
is expected that experiments will soon follow. Here we are
reviewing recent work on turbulence in rotating flow in both
$^3$He-B and $^4$He II, emphasizing similarities in their
macroscopic dynamics.

A number of developments have shed new light on superfluid
turbulence. Much of this progress has been techniques driven in
the sense that novel methods have been required, to make further
advances in a field as complex as turbulence, where the available
techniques both for generating and detecting the phenomenon are
not ideal. Three developments will be discussed in this review,
namely (i) the use of superfluid $^3$He for studies in turbulence,
which has made it possible to examine the influence of a different
set of superfluid properties in addition to those of superfluid
$^4$He, (ii) the study of superfluid $^4$He in the zero
temperature limit where the often present turbulence of the normal
component does not complicate the analysis, and (iii) the use of
better numerical calculations for illustration and analysis.

From the physics point of view, three major advances can be listed
to emerge: In superfluid $^3$He one can study the transition to
turbulence as a function of the dissipation in vortex motion
\citep{Eltsov:2006a}, known as mutual friction. The dissipation
arises from the interaction of thermal excitations with the
superfluid vortex, when the vortex moves with respect to the
normal component.  In classical viscous flow such a transition to
turbulence would conceptually correspond to one as a function of
viscosity. This is a new aspect, for which we have to thank the
$^3$He-B Fermi superfluid where the easily accessible range of
variation in mutual friction dissipation is much wider than in the
more conventional $^4$He II Bose superfluid. We are going to make
use of this feature in Sec.~\ref{VortexInstability}, where we
examine the onset of superfluid turbulence as a function of mutual
friction dissipation \citep{Finne:2006a}.

Secondly, in Sec.~\ref{s:FrontMotion} we characterize the total
turbulent dissipation in superfluid $^3$He as a function of
temperature, extracted from measurements of the propagation
velocity of a turbulent vortex front \citep{Eltsov:2007}. A
particular simplification in this context is the high value of
viscosity of the $^3$He normal component, which means that in
practice the normal fraction always remains in a state of laminar
flow.

Finally, our third main topic in Sec.~\ref{TurbulentDecay} are the
results from recent ion transmission measurements in superfluid
$^4$He \citep{Walmsley:2007,Walmsley2008}, where the decay of
turbulence is recorded from 1.6 to 0.08\,K. Here turbulent
dissipation can be examined in the true zero temperature limit
with no normal component. As a result we now know that turbulence
and dissipation continue to exist at the very lowest temperatures.
Although the dissipation mechanisms of $^4$He or $^3$He-B in the
$T \rightarrow 0$ limit are not yet firmly established
\citep{Vinen:2000,Vinen:2001,Svistunov:2005b}, it is anticipated
that these questions will be resolved in the near future
\citep{Vinen:2006,Svistunov:2008b}.

Phrased differently, our three studies address the questions (i)
how turbulence starts from a seed vortex which is placed in
applied vortex-free flow in the turbulent temperature regime
(Sec.~\ref{VortexInstability}), (ii) how vortices expand into a
region of vortex-free flow (Sec.~\ref{s:FrontMotion}), and (iii)
how the vorticity decays when the external pumping is switched off
(Sec.~\ref{TurbulentDecay}). The common feature of these three
studies is the use of uniformly rotating flow for creating
turbulence and for calibrating the detection of vorticity.
Turbulence can be created in a superfluid in many different ways,
but a steady state of constant rotation does generally not support
turbulence. Nevertheless, at present rotation is the most
practical means of applying flow in a controlled manner in the
millikelvin temperature range. In this review we describe a few
ways to study turbulence in a rotating refrigerator. Superfluid
hydrodynamics supports different kinds of flow even in the zero
temperature limit, so that turbulent losses can vary greatly both
in form and in magnitude, but generally speaking, the relative
importance of turbulent losses tends to increase with decreasing
temperature. Two opposite extremes will be examined: highly
polarized flow of superfluid $^3$He-B, when a vortex front
propagates along a rotating cylinder of circular cross section
(Sec.~\ref{s:FrontMotion}), and the decay of a nearly homogeneous
isotropic vortex tangle in superfluid $^4$He
(Sec.~\ref{TurbulentDecay}), created by suddenly stopping the
rotation of a container with square cross section.

\begin{figure}[t]
\begin{center}
\centerline{\includegraphics[width=.49\linewidth]{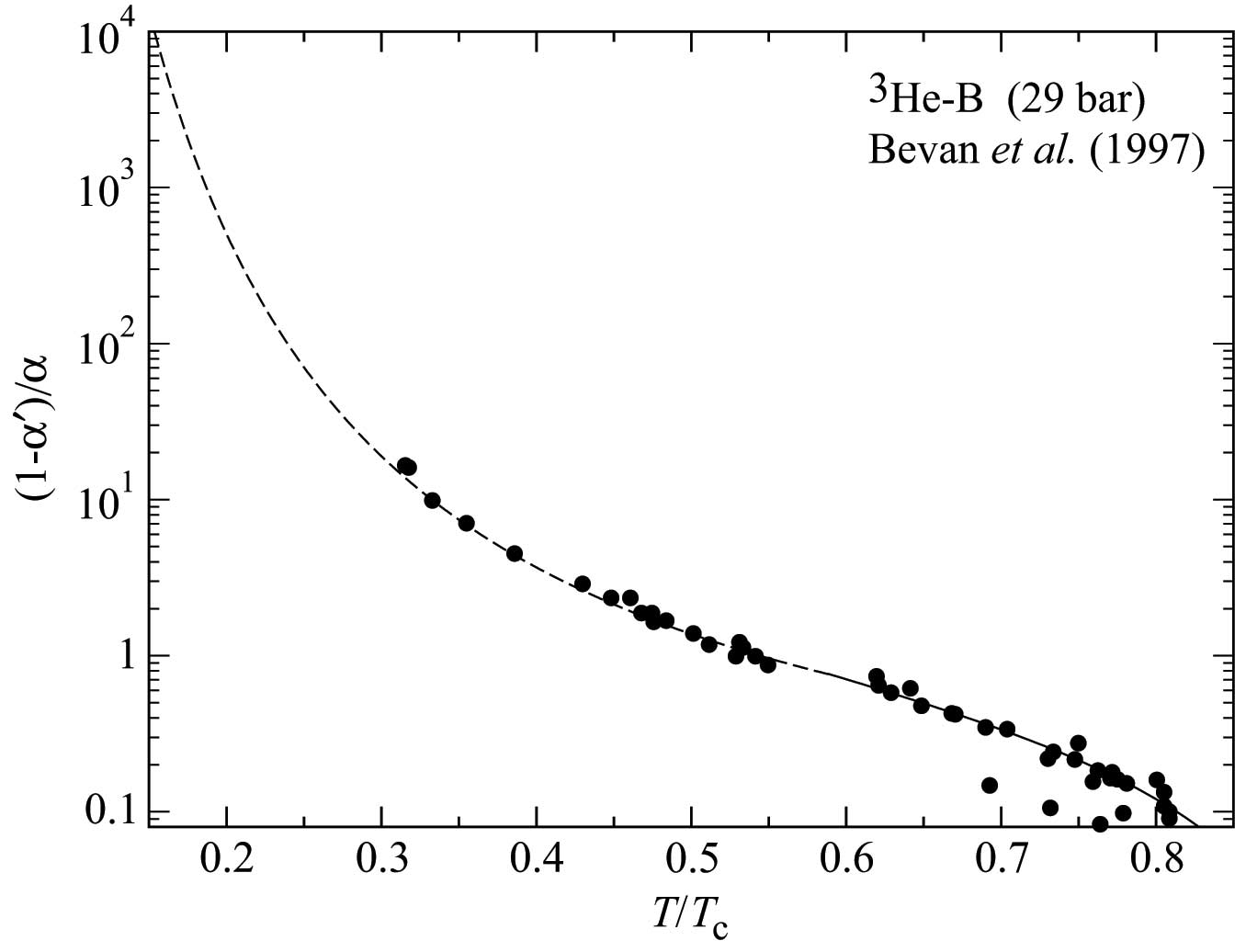}
\includegraphics[width=.5\linewidth]{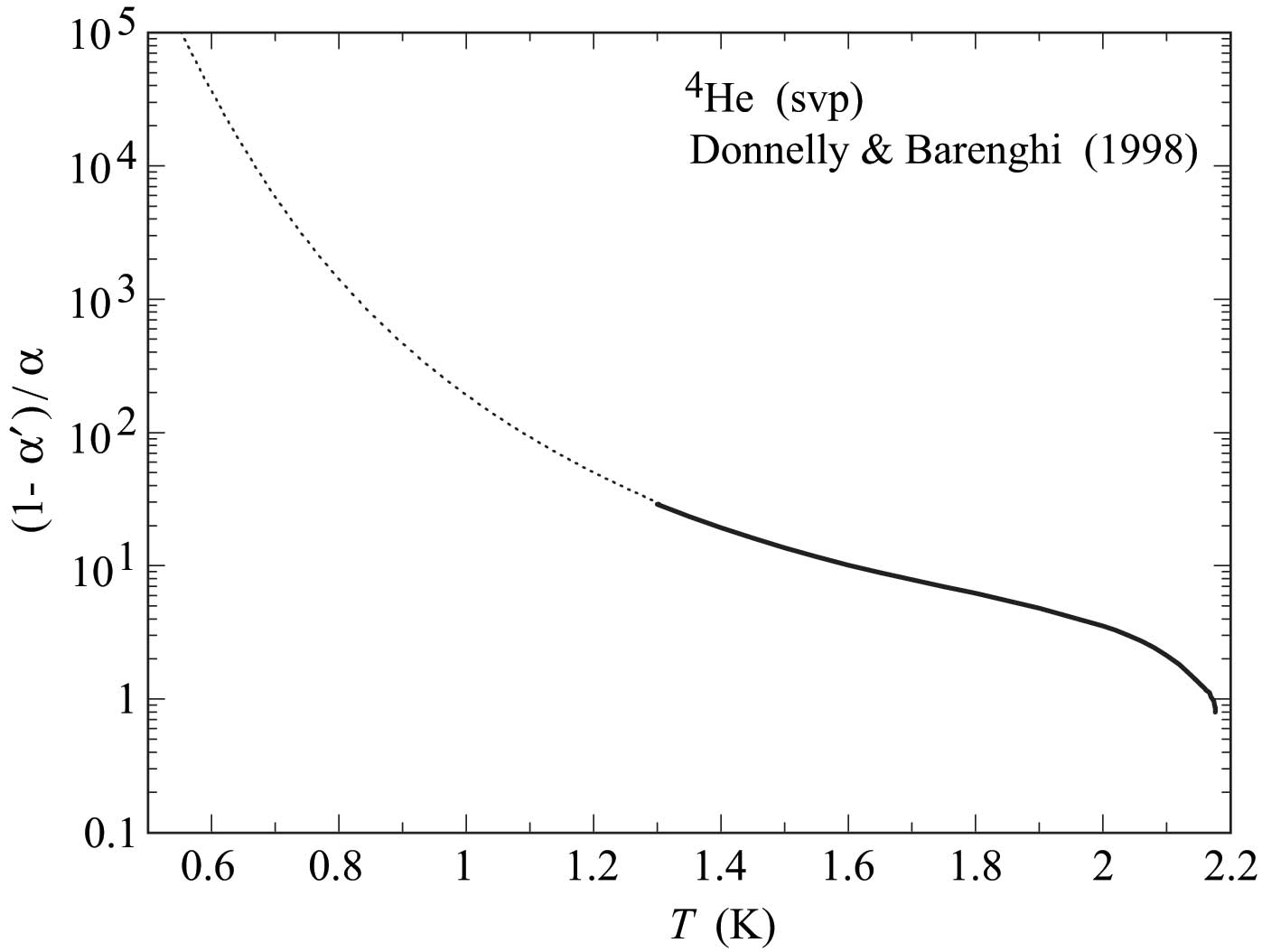}}
\caption{Mutual friction parameter $\zeta=(1 - \alpha^{\prime})/
\alpha$ as a function of temperature. In superfluid dynamics this
parameter, composed of the dissipative mutual friction $\alpha
(T)$ and the reactive mutual friction $\alpha^{\prime} (T)$
corresponds to the Reynolds number $Re$ of viscous hydrodynamics.
Typically, when $Re > 1$, turbulence becomes possible in the bulk
volume between interacting evolving vortices. This transition to
turbulence as a function of temperature can readily be observed in
$^3$He-B (at $0.59\,T_{\rm c}$), while in $^4$He II it would be
within $\sim 0.01\,$K from the lambda temperature and has not been
identified yet. } \label{MutualFriction}
\end{center}
\end{figure}

Turbulent flow in superfluid $^3$He-B and $^4$He is generally
described by the same two-fluid hydrodynamics of an inviscid
superfluid component with singly-quantized vortex lines and a
viscous normal component. The two components interact via mutual
friction. There are generic properties of turbulence that are
expected to be common for both superfluids. However, there are
also interesting differences which extend the range of the
different dynamic phenomena which can be studied in the He
superfluids:
\begin{itemize}
\item{In typical experiments with $^3$He-B, unlike with $^4$He, the mutual
friction parameter $\alpha$ can be both greater and smaller than
unity (Fig.~\ref{MutualFriction}) -- this  allows the study of the
critical limit for the onset of turbulence at $\alpha \sim 1$
(Sec.~\ref{VortexInstability});}
\item{The viscosity of the normal component in $^3$He-B is four
orders of magnitude higher than in $^4$He, hence the normal
component in $^3$He-B is rarely turbulent, which amounts to a
major simplification at finite temperatures (but not in the $T=0$
limit with a vanishing normal component);}
\item{While the critical velocity $v_{\rm c}$ for vortex nucleation is much
smaller in $^3$He-B, pinning on wall roughness is also weaker;
this makes it possible to create vortex-free samples, which are
instrumental in the transitional processes studied in
Secs.~\ref{VortexInstability} and \ref{s:FrontMotion}; on the
other hand, the ever-present remanent vortices in superfluid
$^4$He are expected to ease the production of new vortices, which
becomes important in such experiments as spin-up from rest;}
\item{The vortex core diameter in both liquids is small
(which allows us to use the model of one-dimensional line
filaments), but in $^3$He-B it is up to three orders of magnitude
larger than in $^4$He; hence the dissipation mechanisms in the $T
\rightarrow 0$ regime, which ultimately rely on the emission of
excitations, are expected to work in $^3$He-B at larger length
scales and to lead to more significant energy loss in vortex
reconnections. }
\end{itemize}

A comparison of the turbulent dynamics in these two superfluids
allows one to identify generic properties that are common for both
superfluids, and also to pinpoint specific reasons when there are
differences. The main quantity controlling dissipation is the
mutual friction dissipation $\alpha(T)$, which dominates the
temperature dependence of the dynamic mutual friction parameter
$\zeta=(1 - \alpha^{\prime})/ \alpha$, shown in
Fig.~\ref{MutualFriction}. Experimental values are plotted with
filled symbols for $^3$He-B \citep{Bevan:1997} and with a solid
line for $^4$He \citep{Donnelly:1998}. At low temperatures, the
following extrapolations are used (shown as dashed lines): For
$^3$He-B at a pressure $P=29$~bar, we use $\alpha = 37.21
\exp(-1.968 T_{\rm c}/T)$, where the value for the superfluid gap
$\Delta=1.968 T_{\rm c} k_{\rm B}$ is a linear interpolation as a
function of density $\rho$ between the weak coupling value at zero
pressure and that measured by \cite{delta} at melting pressure.
For $^4$He at saturated vapour pressure (svp), we follow
\cite{Svistunov:2008b} and use $\alpha = 25.3 \exp(-8.5/T)T^{-1/2}
+ 5.78\cdot 10^{-5} T^{5}$, where $T$ is in K.

\section{Dynamic instability -- precursor to turbulence}
\label{VortexInstability}

\subsection{Introduction}

In practice, superfluid flow remains dissipation-free only as long
as there are no quantized vortices (or the existing vortices do
not move, which is more difficult to arrange). The classic
question in superfluid hydrodynamics is therefore: How is the
quantized vortex formed \citep{Feynman:1955,Vinen:1963}? In flow
measurements with bulk liquid the understanding about the origin
of the first vortex has been improving in recent times. Whether it
is created in an intrinsic nucleation process \citep{Ruutu:1997}
or from remanent vortices \citep{Solntsev:2007}, which were
created earlier in the presence of flow or while cooling the
sample to the superfluid state \citep{Yano:2007}, these questions
we are not going to address here. Instead we assume that the first
vortex is already there, for instance as a remanent vortex. We
then ask the question: How is turbulence started when the flow
velocity is suddenly increased by external means? After all a
turbulent vortex tangle is created through the interaction of many
vortices: So how can turbulence start from a single seed vortex?

In rotating $^3$He-B one can create reliably a meta-stable state
of vortex-free flow. It is then possible to inject a single vortex
ring into the flow with neutron irradiation. When a slow thermal
neutron undergoes a capture reaction in liquid $^3$He with a
$^3$He nucleus, a vortex ring may escape from the overheated
reaction bubble into the flow if the flow velocity is above a
critical value \citep{Eltsov:2005}. Making use of this phenomenon
one can inject a single vortex ring in vortex-free flow at
different temperatures. At low temperatures it is observed that a
turbulent vortex tangle is spontaneously formed from the injected
ring \citep{Finne:2004a}, while at high temperatures only a single
vortex line results \citep{Ruutu:1998a}. What is the explanation?

\begin{figure}[t]
\begin{center}
\centerline{\includegraphics[width=0.9\linewidth]{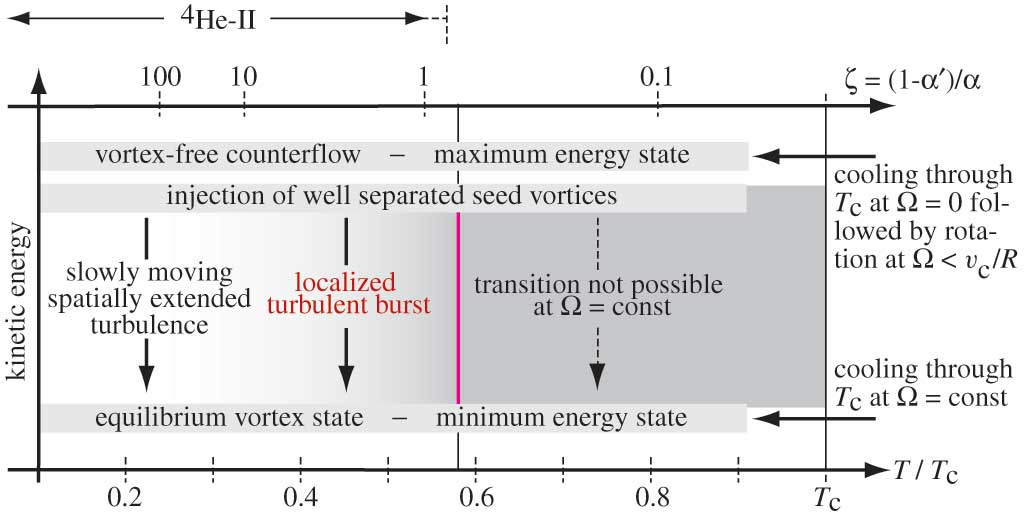}}
\caption{Principle of measurements on seed vortex injection at
constant rotation $\Omega$ and temperature $T$. Well separated
isolated seed vortices are introduced in rotating vortex-free
counterflow. The initial high-energy state may then relax to the
equilibrium vortex state via vortex generation processes which
become possible at temperatures below the hydrodynamic transition
at $0.59\,T_{\rm c}$. The Kelvin-wave instability of a single
evolving seed vortex is the first step in this process. It is then
followed by a turbulent burst which is started if the density of
newly created vortices grows sufficiently. The combined process
depends on the dynamic mutual friction parameter $\zeta = (1-
\alpha^{\prime})/\alpha $ which is shown on the top. On the very
top the range of variation for this parameter in $^4$He-II  is
indicated, \textit{i.e.} the temperature regime of conventional
$^4$He-II measurements. } \label{RotStates+Injection}
\end{center}
\vspace{-3mm}
\end{figure}

This demonstration in $^3$He-B shows that in addition to the
applied flow velocity also mutual friction matters importantly in
the formation of new vortices, in their expansion, and in the
onset of turbulence.  In $^3$He-B, mutual friction dissipation
$\alpha (T)$ is strongly temperature dependent
(Fig.~\ref{MutualFriction}) and it so happens that $\alpha (T)$
drops to sufficiently low value for the onset of turbulence in the
middle of the accessible temperature range. The principle of seed
vortex injection experiments is summarized in
Fig.~\ref{RotStates+Injection}. In these measurements the number
and configuration of the injected vortex loops can be varied. It
turns out that the highest transition temperature is observed when
turbulence starts in bulk volume from many small vortex loops in
close proximity of each other. This transition has been found to
be at $T_{\rm on}^{\rm bulk} \sim 0.6\,T_{\rm c}$
\citep{Finne:2004b} and to be independent of flow velocity over a
range of velocities (3 -- 6\,mm/s). In viscous hydrodynamics the
Reynolds number is defined as $Re = U D/\nu_{\rm cl}$, where $U$
is the characteristic flow velocity, $D$ the relevant length scale
of the flow geometry, and $\nu_{\rm cl} = \eta/\rho$ the kinematic
viscosity. In an isotropic superfluid the equivalent of the
Reynolds number proves to be $\zeta = (1-\alpha^{\prime})/\alpha$.
It defines the boundary between laminar and turbulent flow as a
function of dissipation and is independent of flow velocity or
geometry \citep{Finne:2006b}.

\begin{figure}[t]
\begin{center}
\includegraphics[width=1\linewidth,keepaspectratio]{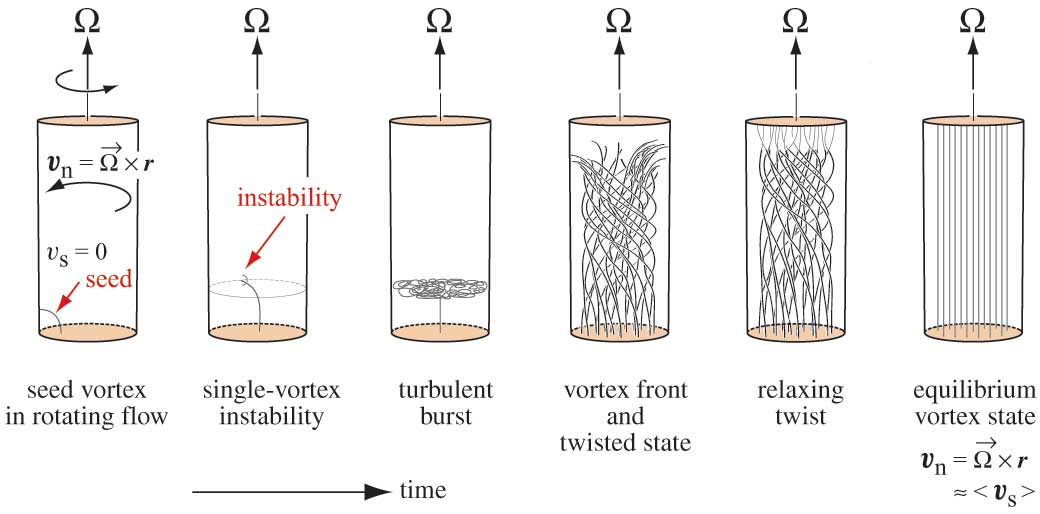}
\caption{Vortex instability and turbulence in a rotating column of
$^3$He-B in the turbulent temperature regime, $T < T_{\rm on}^{\rm
bulk}$. A seed vortex loop is injected in applied vortex-free flow
and the subsequent evolution is depicted. Different transient
states are traversed, until the stable rotating equilibrium vortex
state is reached.  } \label{RotColumnEvolution}
\end{center}
\end{figure}

However, if we inject instead of several closely packed vortex
loops only one single seed loop in vortex-free flow (or several
loops but so far apart that they do not immediately interact) then
the transition to turbulence is found to be at a lower temperature
and to depend on the flow velocity. Thus the onset of turbulence
must also have a velocity dependent boundary.

To explain all these observations, one has to assume that an
independent precursor mechanism exists which creates more vortices
from a seed vortex evolving in externally applied flow. The
characterization of this instability is the topic of this section.
It turns out that this can be done in $^3$He-B in the temperature
regime close to the onset $T_{\rm on}^{\rm bulk}$ of turbulence in
the bulk volume. Here the precursor often progresses sufficiently
slowly so that it can be captured with present measuring
techniques, while at lower temperatures turbulence starts too
rapidly. This latter case is exactly what happens in $^4$He II:
mutual friction dissipation is always so low in the usual
experimental temperature range that the instability has not been
explicitly identified.

The central question is the reduced stability of the evolving seed
vortex loop when mutual friction dissipation is decreasing on
cooling to lower temperature: At sufficiently low $\alpha (T)$ an
evolving vortex becomes unstable with respect to loop formation,
so that one or more new vortex loops are split off, before the
seed vortex has managed to reach its stable state as a rectilinear
line parallel to the rotation axis. The evolution during this
entire process, from injection to the final state, is depicted in
Fig.~\ref{RotColumnEvolution} in a rotating cylindrical sample.
The final state is the equilibrium vortex state, with an array of
rectilinear vortex lines, where their areal density $n_{\rm v}$ in
the transverse plane is given by the rotation velocity $\Omega$:
$n_{\rm v} = 2 \Omega/\kappa$. Here $\kappa = 2 \pi \hbar /2m_3 =
0.066\,$mm$^2$/s is the superfluid circulation quantum of the
condensate with Cooper-pairs of mass $2 m_3$. In this equilibrium
state at constant rotation the superfluid component is locked to
solid-body rotation with the normal component, when averaged over
lengths exceeding the inter-vortex distance $\ell \sim
1/\sqrt{n_{\rm v}}$. In the ideal case all vortices are here
rectilinear, while in the transient states in
Fig.~\ref{RotColumnEvolution} vortices can exist in many different
configurations.

\begin{figure}[t]
\begin{center}
\centerline{\includegraphics[width=0.9\linewidth]{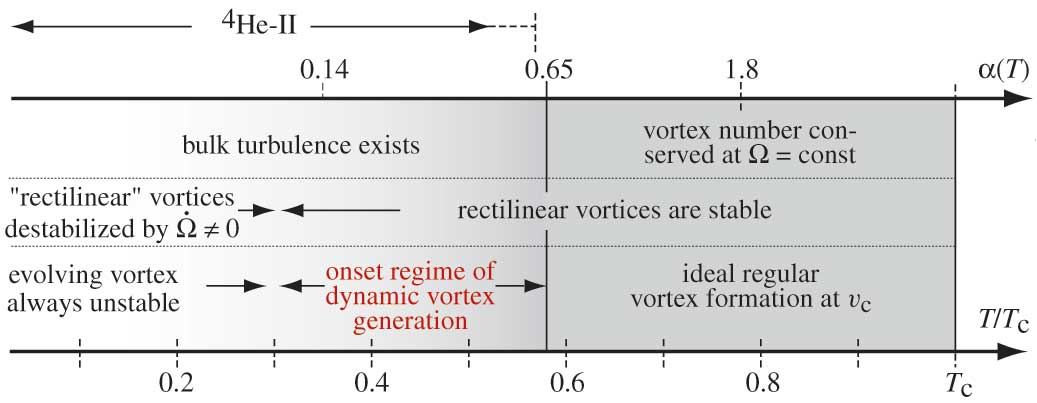}}
\caption{Summary of vortex stability in rotating counterflow of
$^3$He-B, as a function of temperature. {\it (Top row)} The
hydrodynamic transition at $T_{\rm on}^{\rm bulk} \approx 0.59 \,
T_{\rm c}$ (at $P = 29\,$bar pressure) separates regular and
turbulent vortex dynamics. Above the transition vortices are
stable in all situations which have been studied, while below
turbulence becomes possible. {\it (Middle row)} In rotation at
constant $\Omega$ rectilinear vortices are stable. In
time-dependent rotation ($|\dot{\Omega}| \neq 0$) the
``rectilinear" vortex turns out to be an idealization, presumably
because of the experimentally inevitable slight misalignment
between the rotation and the sample cylinder axes and because of
surface interactions. In practice the ``rectilinear" vortices are
found to be stable above $\sim 0.3\,T_{\rm c}$ in time-dependent
rotation, while at lower temperatures they tend to transform to
increasingly turbulent configurations with increasing
$|\dot{\Omega}|$. {\it (Bottom row)} Dynamically evolving vortices
are stable above the transition, but at lower temperatures an
evolving vortex may become unstable, generate a new vortex, and
eventually bulk turbulence. The conditions at seed vortex
injection determine the onset temperature $T_{\rm on}$ below which
turbulence follows. The onset temperatures have been found to
concentrate in the regime $0.35\,T_{\rm c} < T_{\rm on} <
0.59\,T_{\rm c}$. The very low temperatures below $0.3\,T_{\rm c}$
display consistently turbulent response. } \label{VorStability}
\end{center}
\end{figure}

In Fig.~\ref{VorStability} a rough classification is provided of
the stability of vortices as a function of temperature (or more
exactly mutual friction) in different configurations and rotating
situations. The lowest temperatures below $0.3\,T_{\rm c}$ are in
the focus of current research and have by now been probed with a
few different types of measurements. The most extensive work has
been performed by the Lancaster group. They create with various
vibrating oscillators in a quiescent $^3$He-B bath a vortex tangle
and then monitor the decay of the tangle with a vibrating wire
resonator \citep{Bradley:2006}. The total turbulent dissipation in
a vortex front propagating along a rotating column (see
Fig.~\ref{RotColumnEvolution}) has recently been measured
\citep{Eltsov:2007} and will be discussed in
Sec.~\ref{s:FrontMotion}. Also the response of the superfluid
component has been studied to rapid step-like changes in rotation,
when $\Omega$ is changed from one constant value to another. This
type of measurement is commonly known as spin up or spin down of
the superfluid fraction and will be extensively described for spin
down in the case of superfluid $^4$He later in this review.

\begin{figure}[t]
\begin{center}
\includegraphics[width=.7\linewidth,keepaspectratio]{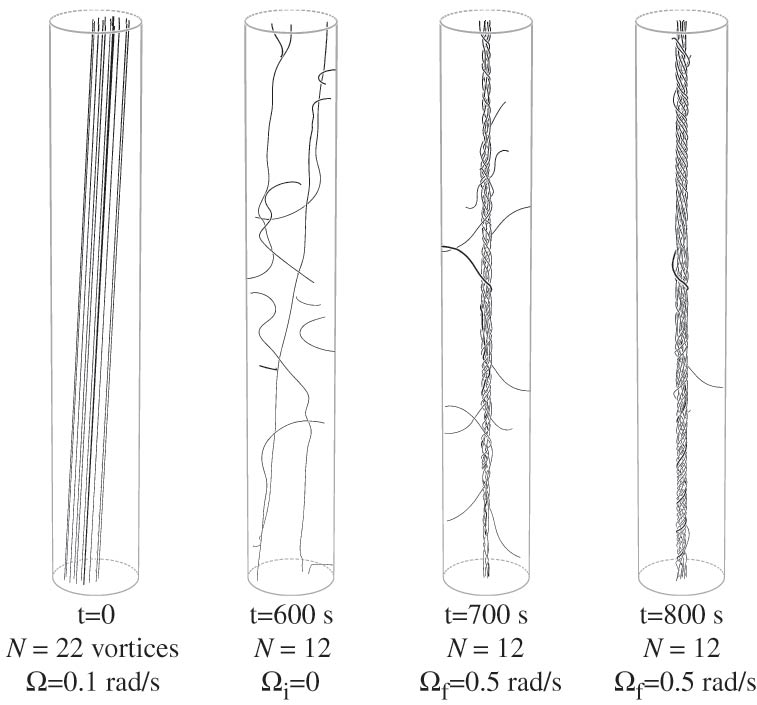}
\caption{Numerical calculation of the evolution of remanent
vortices in rotating flow \citep{de Graaf:2007}. $(t \leq 0)$
Initial state with 22 vortices at 0.1\,rad/s rotation. The
vortices have been artificially tilted by $1.4^{\circ}$, by
displacing their end points uniformly by 1\,mm at both end plates
of the cylinder, to break cylindrical symmetry. $(t = 0)$ Rotation
is abruptly reduced to zero, to allow vortices to annihilate. $(t
= 600\,$s) After a waiting period $\Delta t = 600\,$s at zero
rotation, 12 remanent vortices remain in dynamic state. Rotation
is then suddenly increased to $\Omega_{\rm f} = 0.5\,$rad/s. $(t
\geq 600\,$s) The 12 remnants start evolving towards rectilinear
lines. This requires that the vortex ends on the cylindrical wall
travel in spiral motion to the respective end plates. The
parameters are: radius $R=3\,$mm and length $h=80\,$mm of
cylinder, $T=0.4\,T_{\rm c}$, $P = 29.0\,$bar, $\alpha = 0.18$ and
$\alpha^{\prime} = 0.16$ \citep{Bevan:1997}. In the figure, the
radial lengths have been expanded by two compared to axial
distances. } \label{RemnantCalculation}
\end{center}
\end{figure}

\subsection{Seed vortex evolution in rotating column}
\label{SeedEvolution}

\begin{figure}[t]
\begin{center}
\centerline{\includegraphics[width=.7\linewidth]{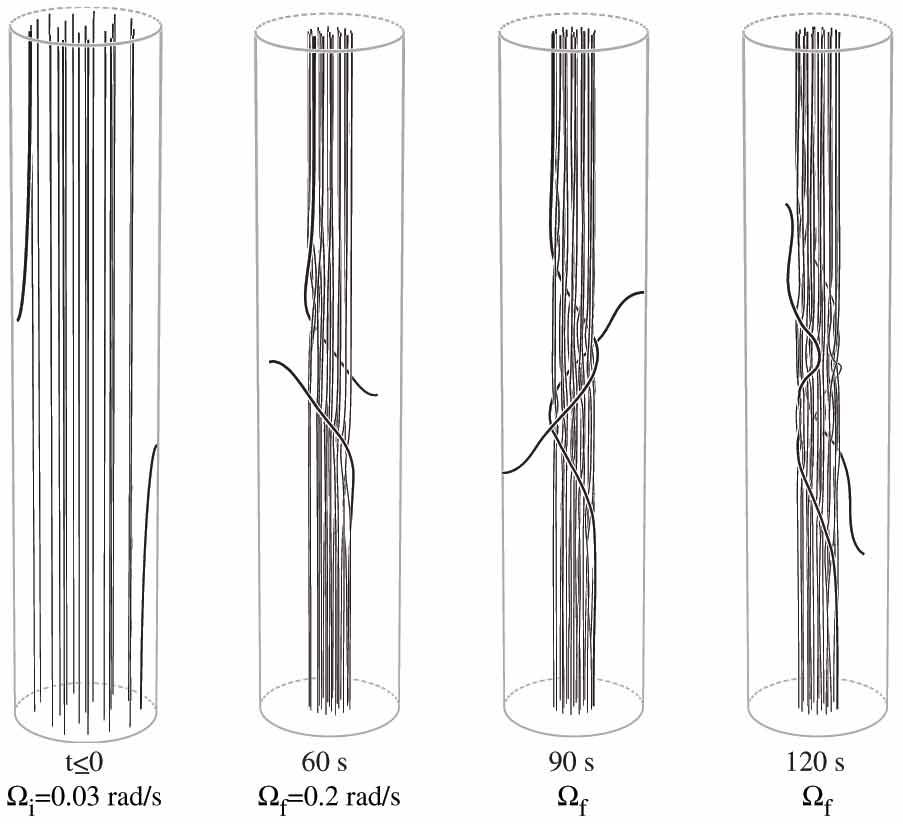}}
\caption{Numerical calculation of the evolution of two curved
peripheral vortices in an inclined rotating cylinder, when
rotation is suddenly increased at $t = 0$ from the equilibrium
vortex state at $\Omega_{\rm i} \approx 0.03\,$rad/s to
$\Omega_{\rm f} = 0.2\,$rad/s \citep{Hanninen:2007a}. There are 22
vortices in this sample, of which two in the outermost ring (lying
opposite to each other) have been initially bent to the
cylindrical wall, to mimic an inclined cylinder. In the later
snapshots at $\Omega_{\rm f}$, the two short vortices expand
towards the top and bottom end plates of the cylinder, to reach
their final stable state as rectilinear lines. Parameters:
$R=3\,$mm, $h=30\,$mm, $P = 29.0\,$bar, and $T=0.4\,T_{\rm c}$
(which corresponds to $\alpha = 0.18$ and $\alpha^{\prime} =
0.16$).} \label{EquilVorStateCalculation}
\end{center}
\end{figure}

The motion of a seed vortex follows a distinctive pattern, while
it expands in a rotating cylinder. Numerically calculated
illustrations are shown in Figs.~\ref{RemnantCalculation} and
\ref{EquilVorStateCalculation} which depict the evolution of the
seeds to stable rectilinear lines. In
Fig.~\ref{RemnantCalculation} an example with remanent vortices
 is examined, while in Fig.~\ref{EquilVorStateCalculation}
the initial configuration is an equilibrium vortex state in the
usual situation that the rotation and cylinder axes are inclined
by some small residual angle $\sim 1^{\circ}$.  These calculations
(in a rotating circular cylinder with radius $R$ and length $h$)
describe the situation at intermediate temperatures when the
vortex instability does not yet occur. The purpose is to focus on
the motion of the expanding vortices and the transient
configurations which thereby evolve. The characteristic property
is the spiral trajectory of a vortex end along the cylindrical
wall and the strong polarization on an average along the rotation
axis. The calculations have been performed using the numerical
techniques described in Sec.~\ref{Simulation} \citep{de
Graaf:2007}.

In Fig.~\ref{RemnantCalculation} the remnants are obtained from an
equilibrium vortex state rotating at 0.1\,rad/s, by reducing
rotation to zero in a sudden step-like deceleration. The vortices
are then allowed to annihilate at zero rotation for a period
$\Delta t$. Some left-over remnants, which have not yet managed to
annihilate, still remain after this annihilation time. Ideally
smooth walls are assumed without pinning.  By suddenly increasing
rotation from zero to a steady value $\Omega_{\rm f} =
0.5\,$rad/s, the remaining remnants are forced to expand. The
configuration 100\,s later shows how the spiral vortex motion has
created a twisted vortex cluster in the center, with a few vortex
ends still traveling in circular motion around the cluster. This
motion thus winds the evolving vortex around the straighter
vortices in the center. On the far right 200\,s after the start of
the expansion the cluster is almost completed. Nevertheless, this
state is still evolving since ultimately also the helical twist
relaxes to rectilinear lines, while the vortex ends slide along
the end plates of the container.

\begin{figure}[t]
\begin{center}
\centerline{\includegraphics[width=.65\linewidth]{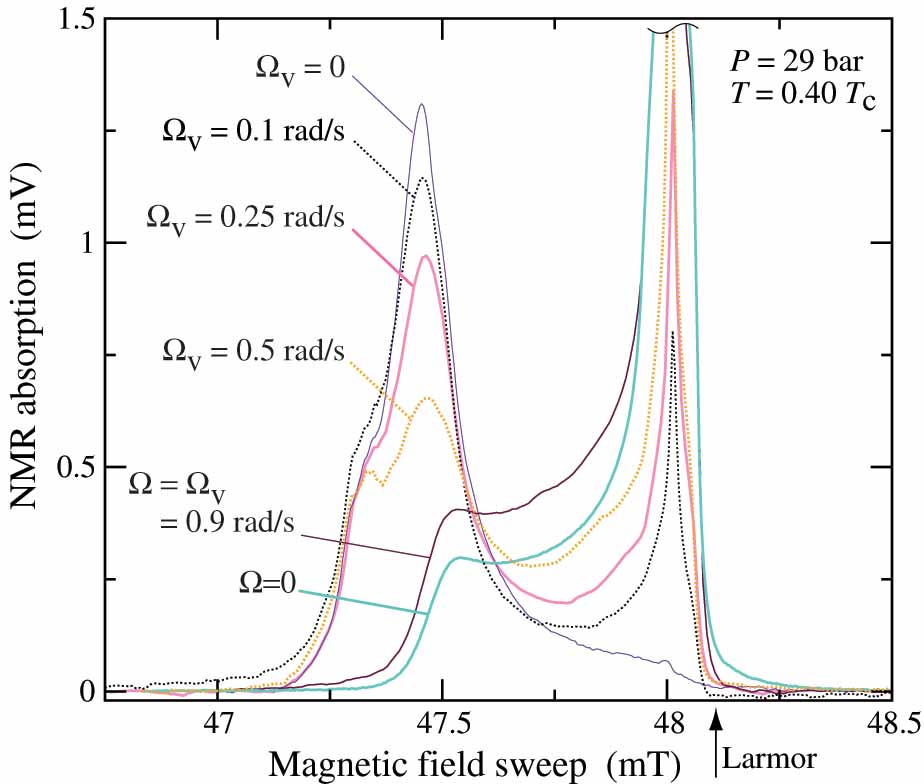}
\includegraphics[width=.35\linewidth]{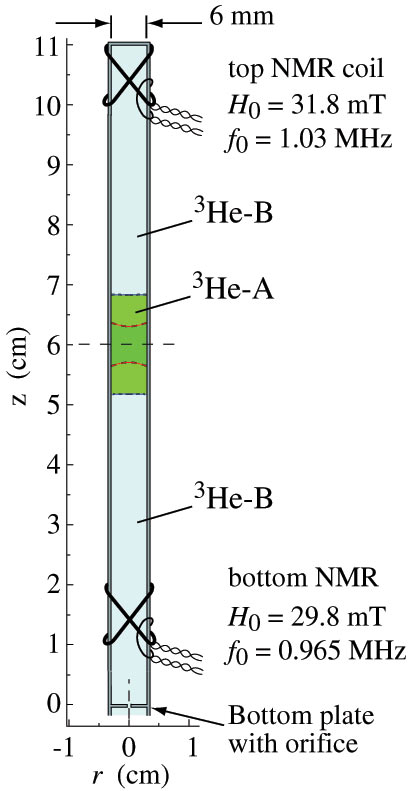}}
\caption{\textit{On the left} the NMR absorption spectra
constitute an image of the  ``flare-out" order parameter texture
in the long rotating cylinder \citep{Kopu:2000}. The Larmor field,
around which the NMR absorption is centered in the normal phase,
is here at 48.1\,mT. In the B phase the Larmor value becomes a
sharp edge beyond which at higher fields the absorption vanishes,
while most of the absorption is shifted to lower fields. The
dominant absorption maximum on the left is the counterflow peak.
Its height depends on the number of vortices $N$ in the central
cluster. $N$ can be defined in terms of the rotation velocity
$\Omega_{\rm v}$ at which the vortices form the equilibrium state:
$N = N_{\rm eq}(\Omega_{\rm v})$. Here all counterflow peaks have
been recorded at the counterflow velocity $\Omega - \Omega_{\rm v}
= 0.8\,$rad/s. The conversion from peak height to $N$ in given
conditions $(T, \Omega, P)$ can be obtained experimentally or from
a numerical calculation of the order parameter texture and the
corresponding line shape \citep{Kopu:2006}. The two line shapes
without a counterflow peak, but with a large truncated maximum
bordering to the Larmor edge, are for the non-rotating sample
$(\Omega = 0)$ and for the equilibrium vortex state $(N=N_{\rm
eq})$ at $\Omega_{\rm v} = 0.9\,$rad/s. For both of them the
absorption at the site of the counterflow peak is close to zero.
\textit{On the right} a NMR setup is shown which was used to
study the onset temperature of turbulence (\textit{cf.} also
Fig.~\ref{ExpSetup}). Two different contours of the AB interface
are shown when the A-phase barrier layer is present at
$0.6\,T_{\rm c}$. The contours correspond to a current of 4\,A
(narrow A-phase sliver with curved concave interfaces) and 8\,A
(wider A-phase layer with flat interfaces) in the superconducting
A-phase stabilization solenoid
 \citep{Finne:2004b}.} \label{NMR-LineShapes}
\end{center}
\end{figure}

In Fig.~\ref{EquilVorStateCalculation} a calculation is presented
with 20 rectilinear vortex lines and two short vortices which
connect at one end to the cylindrical wall. This configuration
mimics the equilibrium vortex state in a real rotating experiment
where there exists some residual misalignment between the rotation
and sample cylinder axes. Depending on the angle of misalignment
and the angular velocity of rotation $\Omega_{\rm i}$, some of the
peripheral vortices may then end on the cylindrical side wall in
the equilibrium vortex state, as shown on the far left.  At $t=0$
rotation is increased in step-like manner from $\Omega_{\rm i}$ to
a higher value $\Omega_{\rm f} $. Two types of vortex motion are
started by the rotation increase. First, the $N=20$ rectilinear
vortices are compressed to a central cluster with an areal density
$n_{\rm v} = 2 \Omega_{\rm f} /\kappa$ by the surrounding
azimuthally flowing counterflow. Outside the vortex cluster the
counterflow has the velocity
\begin{equation} v(\Omega_{\rm f},r,N) = v_{\rm n} - v_{\rm s} = \Omega_{\rm f} r -
\kappa N/(2\pi r)\;, \label{cf}
\end{equation} The normal excitations are in solid-body
rotation and thus $v_{\rm n} = \Omega_{\rm f} r$, while the
superflow velocity around a cluster of $\kappa N$ circulation
quanta is decaying as $v_{\rm s} = \kappa N/(2\pi r)$, where $r
\geq R_{\rm v}$ and the cluster radius $R_{\rm v} \approx R
\sqrt{\Omega_{\rm i}/\Omega_{\rm f}}$. Experimentally it is
convenient to define the number of vortices $N$ via the initial
equilibrium vortex state: $N = N_{\rm eq}(\Omega_{\rm i})$, where
the externally adjusted rotation velocity $\Omega_{\rm i} \sim
\kappa N_{\rm eq} /(2\pi R^2)$ uniquely defines $N_{\rm eq}$ in a
given experimental setup. It is customary to denote this specially
prepared calibration value with $\Omega_{\rm v} = \Omega_{\rm i}$.
The maximum counterflow velocity at $\Omega_{\rm f}$ at the
cylinder wall is then given by $v \approx (\Omega_{\rm f} -
\Omega_{\rm v})R$. This definition has been used in
Fig.~\ref{NMR-LineShapes} to characterize the number of vortices
in the central cluster.

The second type of vortex motion, which is enforced by the
increased rotation in Fig.~\ref{EquilVorStateCalculation}, is the
spiral motion of the two short vortices, as they become mobile and
start expanding towards the top and bottom end plates,
respectively. Let us now examine this motion in more detail.

A vortex moving with respect to the superfluid component is
subject to the influence from the Magnus lift force. This force
can be written as \citep{Donnelly:1991} ($\rho_{\rm s}$ is the
density of the superfluid fraction)
\begin{equation} \mathbf{f}_{\rm M} = \rho_{\rm
s} \kappa \; \hat{\mathbf{s}} \times (\mathbf{v}_{\rm L} -
\mathbf{v}_{\rm s} )\;,\label{Magnus}
\end{equation}
which acts on a vortex element $\mathbf{s}(\xi, t)$ with a tangent
unit vector $\hat{\mathbf{s}} = d\mathbf{s}/d\xi$ moving with
velocity ${\bf v}_{\rm L} = d\mathbf{s}/dt$ with respect to the
superfluid component, which locally has the velocity
$\mathbf{v}_{\rm s}$. The motion from the Magnus force is damped
by mutual friction which arises when the vortex moves with respect
to the surrounding cloud of normal excitations:
\begin{equation}
\mathbf{f}_{\rm mf} = - \gamma_0 \rho_{\rm s} \kappa \;
\hat{\mathbf{s}} \times [\hat{\mathbf{s}} \times (\mathbf{v}_{\rm
n} - \mathbf{v}_{\rm L} )] + \gamma_0^{\prime} \rho_{\rm s} \kappa
\; \hat{\mathbf{s}} \times (\mathbf{v}_{\rm n} - \mathbf{v}_{\rm
L} )\;.\label{MutFrictionForce} \end{equation} The mutual friction
force has dissipative and reactive components, which here are
expressed in terms of the two parameters $\gamma_0$ and
$\gamma_0^{\prime}$. Balancing the two hydrodynamic forces,
$\mathbf{f}_{\rm M} + \mathbf{f}_{\rm mf} =0$, one gets the
equation of motion for the vortex line element, which when
expressed in terms of the superfluid counterflow velocity
$\mathbf{v} = \mathbf{v}_{\rm n} - \mathbf{v}_{\rm s}$ has the
form
\begin{equation} {\bf v}_{\rm L}= \mathbf{v}_{\rm s} +\alpha
\hat{\mathbf{s}} \times (\mathbf{v}_{\rm n}-\mathbf{v}_{\rm s})
-\alpha' \hat{\mathbf{s}} \times
[\hat{\mathbf{s}}\times(\mathbf{v}_{\rm n}-\mathbf{v}_{\rm s}
)]\:.\label{vl}
\end{equation}
Here the dissipative and reactive mutual friction coefficients
$\alpha$ and $\alpha^{\prime}$ appear. Conversion formulae between
different sets of friction parameters are listed by
\cite{Donnelly:1991}. Evidently solutions of the equation of
motion can be classified according to the ratio of the two
components. The important parameter proves to be $\zeta =
(1-\alpha^{\prime})/\alpha$ (Fig.~\ref{MutualFriction}), which is the equivalent of the
Reynolds number of viscous fluid flow \citep{Finne:2003}.

Two elementary examples are useful to consider. A single
rectilinear vortex in rotating counterflow at $v =\Omega r$ moves
such that its velocity components in the transverse plane consist
of the radially oriented dissipative part $\alpha \Omega r$ and
the azimuthally oriented reactive part $-(1-\alpha^{\prime})\,
\Omega r$, when expressed in the rotating coordinate frame. The
former is responsible for the contraction of the rectilinear
vortices to a central cluster in Figs.~\ref{RemnantCalculation}
and \ref{EquilVorStateCalculation}. The latter causes the
rectilinear vortex to rotate with the azimuthal flow with respect
to the cylindrical wall.

The second simple consideration concerns the end point motion of
an evolving vortex along the cylindrical wall. Since the vortex
end is perpendicular to the cylindrical wall, it has from
Eq.~(\ref{vl}) a longitudinal velocity $v_{{\rm L}z} = \alpha
v(\Omega_{\rm f},R,N)$ and an azimuthal component $v_{{\rm
L}{\phi}} = -(1-\alpha^{\prime}) v(\Omega_{\rm f},R,N)$. Evidently
other parts of the vortex also contribute to its motion, in
particular its curvature where it connects to the cylindrical
wall. However, it turns out that the end point velocity is an
approximate guide for the expansion of a single vortex in
vortex-free rotation. For comparison, the calculated velocities of
the two vortex ends in Fig.~\ref{EquilVorStateCalculation} are
$v_{{\rm L}z} \approx 0.84 \, \alpha \Omega R \approx 0.96 \,
\alpha v(R)$ and $v_{{\rm L}{\phi}} \approx 0.73
(1-\alpha^{\prime}) \Omega R \approx 0.83(1-\alpha^{\prime})
v(R)$. The wave length of the spiral trajectory is $\lambda = $
$2\pi R \, v_{{\rm L}z}/v_{{\rm L}{\phi}} \approx 5\,$mm and the
period $p = 2\pi R/v_{{\rm L}{\phi}} \approx \,50\,$s. Thus the
end point motion can be used to construct a simplified model of
the motion of the two short vortices in
Fig.~\ref{EquilVorStateCalculation}.

As seen in Fig.~\ref{EquilVorStateCalculation}, the spiral motion
of the vortex end point along the cylindrical wall winds the rest
of the evolving vortex around the central vortex cluster with a
wave vector $Q$, such that $QR = v_{{\rm L}{\phi}}/v_{{\rm L}z} =
\zeta$. The other almost straight end of the evolving vortex is
fixed to a flat end plate of the cylinder and resides there at the
edge of the vortex cluster, where the counterflow velocity is
close to zero. Therefore the helical twist is removed only by a
slow sliding of the vortex end along the end plate. As seen in
Fig.~\ref{EquilVorStateCalculation}, occasional reconnections
between the twisted evolving vortex and a straight outer vortex in
the cluster or with a second oppositely twisted vortex can help to
reduce the twist. Finally, we see in
Fig.~\ref{EquilVorStateCalculation} that while the evolving vortex
is wound tightly around the cluster this induces Kelvin wave
oscillations which propagate vertically along the vortices in the
cluster \citep{Hanninen:2007a}.

\subsection{Onset temperature of turbulence} \label{OnsetTemp}

When the dynamics calculated in Figs.~\ref{RemnantCalculation} and
\ref{EquilVorStateCalculation} is probed with measurements, the
final state proves to depend crucially on temperature: At
temperatures above the transition to turbulence, $T > T_{\rm
on}^{\rm bulk}$, the calculations are confirmed and the number of
vortices remains constant. At low temperatures, in contrast,
evolving vortices may become unstable in applied counterflow and
trigger a turbulent burst. The evolution after the burst continues
as illustrated in Fig.~\ref{RotColumnEvolution}. The final state
is then consistently close to the equilibrium vortex state.
Interestingly, it turns out that for each initial configuration,
such as those in Figs.~\ref{RemnantCalculation} or
\ref{EquilVorStateCalculation}, there exists a specific
temperature $T_{\rm on}$, which characterizes the onset of
turbulence: well above $T_{\rm on}$ no turbulent burst occurs
while well below $T_{\rm on}$ a burst always occurs.

In these onset measurements only two different types of final
states are observed: sufficiently far above $T_{\rm on}$ the
number of vortices is conserved, while well below $ T_{\rm on}$
close to the equilibrium number of vortices is formed. The change
over takes place within a narrow temperature interval around
$T_{\rm on}$, typically within $\pm 0.02\,T_{\rm c}$. Within this
interval either of the two final states can emerge. The reason for
the narrow width is the strong, nearly exponential temperature
dependence of the mutual-friction parameter $\zeta = (1 -
\alpha^{\prime})/ \alpha$, which controls the dynamic instability
of seed vortices evolving in the applied counterflow
(Fig.~\ref{MutualFriction}). As sketched in
Fig.~\ref{RotColumnEvolution},  two sequential processes are
needed to start the low-temperature evolution: First the
single-vortex instability \citep{Finne:2006a}, the precursory
process which becomes possible only at temperatures below $T_{\rm
on}^{\rm bulk}$ and which is responsible for generating a bunch of
new evolving vortices. Secondly, the turbulent burst has to be
triggered as a collective process in which several evolving
vortices interact and generate a sudden localized event of
turbulence, which expands across the entire cross section of the
rotating column, but only over a short section of its length (of
order $\sim R$).

The experimental confirmation of this scenario is obtained by
examining the final state as a function of temperature. By
recording the line shape of the NMR absorption profile, when the
magnitude of the magnetic polarization field is swept across the
resonance region, the number of vortices in the final state can be
determined. The line shapes of the two types of final states
differ drastically, as seen in Fig.~\ref{NMR-LineShapes}, where
all the spectra have been recorded at the same temperature of
$0.40\,T_{\rm c}$ and where thus the integrated area under each
line shape is the same. The characteristic feature are the large
NMR shifts. These are controlled by the temperature and pressure
dependent spin-orbit coupling. If the central vortex cluster is
surrounded by applied counterflow at some sizeable velocity, a
large sharp peak is formed which is shifted down field from the
Larmor value. The number of vortices in the central cluster can be
determined from the height of this so-called counterflow peak. At
small vortex numbers ($N \ll N_{\rm eq}$), the reduction in the
counterflow peak height is directly proportional to the number of
rectilinear vortices $N$ in the cluster. At larger $N$ the
dependence becomes nonlinear and ultimately the peak height drops
to zero well before $N_{\rm eq}$ is reached, in practice around
$v(R) \lesssim 1\,$mm/s \citep{Kopu:2006}. Accordingly, in the
equilibrium vortex state the line shape is radically different and
easily distinguished from a state with sizeable counterflow.

A measurement of $T_{\rm on}$ for any particular initial setup,
such as in Fig.~\ref{RemnantCalculation} for remanent vortices,
requires repeating the measurement at different temperatures and
recording the line shape in the final state. Surprisingly, it
turns out that in the final state the vortex number is either
preserved or it has increased close to that in the equilibrium
vortex state. Practically no intermediate cases are observed. As a
result, in practice a measurement of $T_{\rm on}$ requires simply
a visual check of the measured line shape in the final state. This
feature about the turbulent burst is similar to recent
observations from measurements with a closely spaced pair of
vibrating wires in superfluid $^4$He \citep{Goto:2008}. One of the
wires is driven at high oscillation amplitude as generator while
the second is operated at low amplitude as detector. Once
turbulence has been switched on by running the generator at high
drive, the generator can be switched off and turbulent flow will
still be maintained around the detector. Only if the detector
drive is reduced to sufficiently low level, turbulence ceases and
the flow around the detector returns to the laminar state. This
shows that once turbulence has been switched on it can be
sustained at much lower flow velocities. Similarly, once the
turbulent burst is started in the rotating column, turbulent
vortex formation will continue until the counterflow velocity has
dropped close to zero in a section of the column of height $\sim
R$.

\begin{figure}[t]
\begin{center}
\includegraphics[width=1\linewidth]{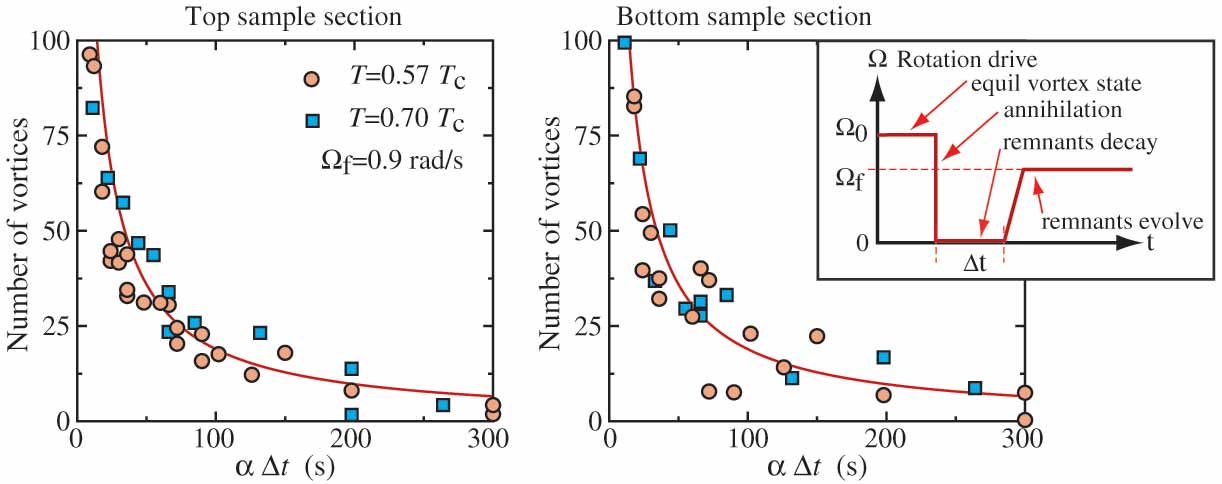}
\caption{Number of remanent vortices $\cal{N}$$(\Delta t)$ for
A-phase separated top and bottom sample sections, measured as a
function of the annihilation time $\Delta t$ in the temperature
regime of laminar vortex motion. The results of these two
independent measurements can be fitted in both cases with the
solid curve $\cal{N}$$(\Delta t) \approx 2 \cdot 10^3 /(\alpha \,
\Delta t + b )$ [$\Delta t$ in sec, $b \approx 7\,$s].
\textit{(Inset)} In the upper right corner the sequence of
rotations $\Omega (t)$ is shown which was used to perform the
measurement. The data are for $0.57\,T_{\rm c}$ with $\alpha =
0.60$ and $0.7\,T_{\rm c}$ with $\alpha = 1.1$. Parameters:
$\Omega_{\rm f}=0.9\,$rad/s, $R= 3\,$mm, length of top sample
section $h_{\rm t} = 44\,$mm (41\,mm) and $h_{\rm b} =54\,$mm
(51\,mm) for the bottom section at $0.57 \, T_{\rm c}$ ($0.70 \,
T_{\rm c}$).} \label{RemnantVorWaitTime}
\end{center}
\end{figure}

The situation at temperatures above $T_{\rm on}$ is illustrated by
the measurements on vortex remanence in
Fig.~\ref{RemnantVorWaitTime} \citep{Solntsev:2007}. These
measurements have been performed at two different temperatures
above onset, where no increase in the number of vortices is
expected. This is confirmed by extracting from the counterflow
peak height the number of vortex lines in the final state. The
measurement proceeds as sketched in the inset on the top right of
Fig.~\ref{RemnantVorWaitTime}: An equilibrium vortex state is
decelerated to zero and the vortices are allowed to annihilate for
a time interval $\Delta t$ before rotation is turned back on. The
measurement is repeated many times by varying the annihilation
time $\Delta t$ at zero rotation. The annihilation time is found
to govern the number of rectilinear vortex lines in the final
state and thus the number of remnants at the end of the
annihilation period: $N(\Delta t) \propto (1+\Delta t /\tau_{\rm
mf})^{-1}$, where the mutual-friction-controlled time constant is
$\tau_{\rm mf} = [2 \, \alpha \, \Omega_0]^{-1}$ and $\Omega_0 =
\Omega (t < 0)$. This is exactly as expected for the
mutual-friction damped motion of vortices in the radial direction,
when straight vortices move outward to annihilate on the
cylindrical wall at zero rotation. For this to apply, the vortices
have to be polarized along the cylindrical symmetry axis
\citep{Krusius:1993}. As seen in Fig.~\ref{RemnantCalculation}
(second from left, at $t \leq 600\,$s) this is the case: The
polarization remains at high level even in the remanent state at
zero rotation. Consequently, the measurements in
Fig.~\ref{RemnantVorWaitTime} confirm that at constant temperature
above $T_{\rm on}$ the number of vortices in this experiment is
controlled by the annihilation period $\Delta t$ and no
uncontrolled increase occurs.

It is useful to note some additional features about vortex
remanence in the measurements of Fig.~\ref{RemnantVorWaitTime}.
Let us denote the number of remnants after the annihilation period
with $\cal{N}$$(\Delta t)$ and the number of rectilinear lines in
the final state with $N$. Although the annihilation time $\Delta
t$ controls the number of remnants $\cal{N}$$(\Delta t)$, the
result $N = \cal{N}$ is independent of $\Delta t$. It is also to a
large extent independent of how the measurement is performed,
\textit{i.e.} what the rotation velocity $\Omega_{\rm f}$ in the
final state is or what acceleration $\dot{\Omega}$ is employed to
reach $\Omega_{\rm f}$ (as long as $\cal{N}$$(\Delta t) < N_{\rm
eq} (\Omega_{\rm f})$ or the critical velocity for vortex
formation, $\Omega_{\rm f} - \Omega_{\rm v} < v_{\rm c}/R$, is not
exceeded). Furthermore, the result $N = \cal{N}$ is established
separately both for the top and bottom sections of the cylinder,
when these are separated by a magnetic-field stabilized A-phase
barrier layer (\textit{cf.} Fig.~\ref{NMR-LineShapes}), and for
the entire cylinder without A-phase barrier.  The A-phase barrier
layer prevents vortices from traversing across the AB interfaces
at low counterflow velocity \citep{Blaauwgeers:2002}. In this way
the number of remanent vortices $\cal{N}$$(\Delta t)$ has been
found to be proportional to the length $h$ of the cylinder (as
long as $h \gg R$). All these properties are consistent with the
conclusions that when $T > T_{\rm on}$, the vortex number is
conserved in dynamical processes, the annihilation decay of
remnants is a laminar process regulated by mutual friction
damping, and that pinning is weak. For simplicity, we neglect
vortex pinning altogether and assume ideal wall properties
throughout this review.

\begin{figure}[t]
\begin{center}
\centerline{\includegraphics[width=1\linewidth]{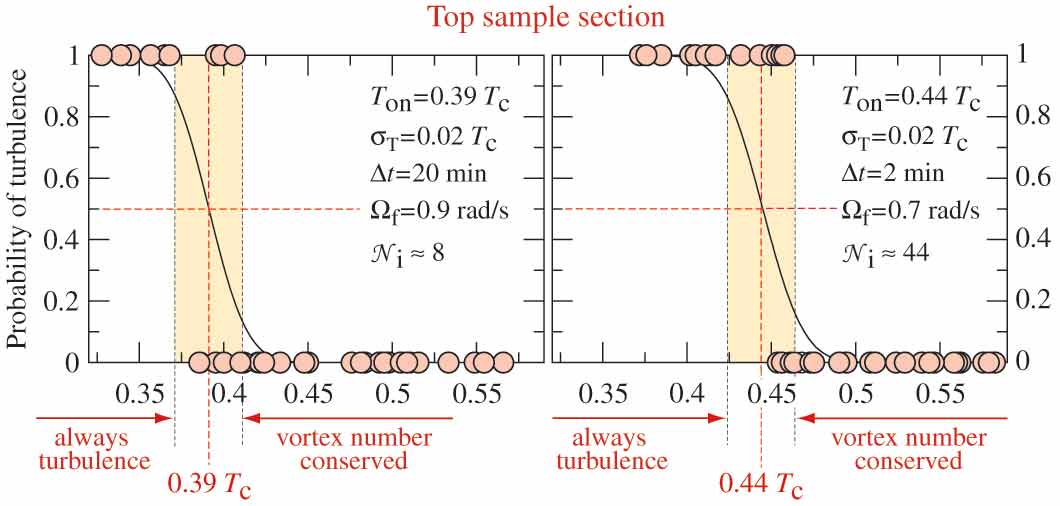}}
\caption{Measurements on the onset temperature $T_{\rm on}$ of the
transition to turbulence. The measurements are performed similar
to those in Fig.~\protect\ref{RemnantVorWaitTime} and start from
an initial state which is obtained by decelerating an equilibrium
vortex state at 1.7\,rad/s to zero at a rate 0.01\,rad/s$^2$. The
remaining vortices are left to annihilate for a period $\Delta t$
at $\Omega = 0$. Rotation is then increased to $\Omega_{\rm f}$ at
a rate 0.02\,rad/s$^2$. When all transients have decayed the
number of vortices is measured in the final steady state at
$\Omega_{\rm f}$. The result is plotted as a function of
temperature with 30 -- 40 data points per panel.  The solid curve
is a gaussian fit which represents the probability for turbulence
with a half width $\sigma_{\rm T} = 0.02\,T_{\rm c}$, centered
around $T_{\rm on}$. Comparing results in the two panels for
$\Delta t = 20\,$min and 2\,min, we see that $T_{\rm on}$
decreases with increasing $\Delta t$, since the number, average
size, and density of remnants is reduced as $\Delta t$ increases.
Both panels have been measured for the upper sample section which
is separated from the bottom half with an A-phase barrier layer.
Parameters: $R = 3\,$mm, $h = 45 \, $mm, and $P = 29.0\,$bar.}
\label{RemnantTonTop}
\end{center}
\end{figure}

The situation at temperatures around $T_{\rm on}$ is illustrated
by the measurements in Fig.~\ref{RemnantTonTop} which determine
$T_{\rm on}$ for this particular choice of initial state
\citep{{Solntsev:2007},{de Graaf:2007}}. The probability of the
turbulent burst is plotted as a function of temperature, when the
annihilation time $\Delta t = 20\,$min \textit{(on the left)} and
$\Delta t = 2\,$min \textit{(on the right)}. The striking feature
is the abrupt change over from the laminar behaviour, where the
vortex number is conserved, to turbulence, where the vortex number
surges close to $N_{\rm eq}$ and the system relaxes to its minimum
energy state. The center of the narrow transition defines the
onset temperature of turbulence $T_{\rm on}$, which proves to be
different for the two cases studied in Fig.~\ref{RemnantTonTop}.

As seen in Fig.~\ref{RemnantTonTop}, $T_{\rm on}$ depends on the
annihilation time $\Delta t$ and thus on the initial number and
configuration of evolving remnants $\cal{N}$$(\Delta t)$ at the
moment when the step increase in rotation from zero to
$\Omega_{\rm f}$ is applied. Calculating from the results in
Fig.~\ref{RemnantVorWaitTime} one finds that the number of
remnants  at the start of acceleration is $\cal{N}$$_{\rm i}
\approx 40$, when $\Delta t =2\,$min and $T \approx T_{\rm on} =
0.44\,T_{\rm c}$, and $\cal{N}_{\rm i}$ $\approx 10$, when $\Delta
t =20\,$min and $T = 0.39\,T_{\rm c}$. Thus at a higher
temperature a larger number of remnants is needed to achieve the
turbulent burst. Both panels in Fig.~\ref{RemnantTonTop} refer to
the top sample half where, with no orifice, there is no preferred
site for the remnants to accumulate and the turbulent burst occurs
randomly at any height $z$ in the column \citep{de Graaf:2007}.
Similar measurements at different values of $\Delta t$ and final
rotation velocity $\Omega_{\rm f}$ show that the onset temperature
depends weakly on both the initial number of remnants
$\cal{N}_{\rm i}$ and the applied flow velocity. These dependences
can be summarized in the form
\begin{equation} \zeta(T_{\rm on})^{-1} \propto
\protect{\cal{N}}_{\rm i}^{0.3} \Omega_{\rm f}^{1.3}\;.
\label{PowerLaw}
\end{equation} Thus the onset temperature $T_{\rm on}$ depends primarily
on the mutual friction parameter $\zeta (T)$, but also weakly on
other factors which influence the likelihood of achieving locally
somewhere in the maximum available counterflow velocity a density
of evolving vortices which allows to trigger the turbulent burst.
Among these additional factors most important are (i) the applied
counterflow velocity $v = v_{\rm n} - v_{\rm s}$,  (ii) the number
and configuration of the injected seed vortices, and (iii) the
sample geometry. In Figs.~\ref{RemnantVorWaitTime} and
\ref{RemnantTonTop} we examined the response of remanent vortices
to a step-like increase in rotation. The same measurements can
also be performed by starting from the equilibrium vortex state at
finite rotation, as discussed in the context of
Fig.~\ref{EquilVorStateCalculation}. In fact, the most extensive
study of the scaling law in Eq.~(\ref{PowerLaw}) was performed
using this approach.

Finally we note that in the rotating container all measured onset
temperatures, which depend on the presence of the precursor, are
found to be below the transition to turbulence in the bulk:
$T_{\rm on} < T_{\rm on}^{\rm bulk}$. Furthermore, since the onset
also depends on the applied counterflow velocity in
Eq.~(\ref{PowerLaw}), the instability is expected to occur first
close to the cylinder wall, where the applied velocity $v=\Omega
\, r$ reaches its maximum value at $r=R$. Thus the reconnection of
the expanding loop will most likely occur with the wall.
Surprisingly it is also found that once the instability is
triggered, the turbulent burst essentially always follows next,
since little if no increase in the vortex number is detected at $T
~\sim T_{\rm on}$ in such cases where the turbulent burst does not
switch on (\textit{cf.} Fig.~\ref{RemnantTonTop}). To provide more
understanding on the role of the single-vortex instability as the
precursor mechanism to turbulence, we next examine it in the onset
temperature regime, $ T \sim T_{\rm on}$, where the instability
proceeds sufficiently slowly in time so that it can be monitored
with continuous-wave NMR measurement.

\subsection{Single-vortex instability in applied flow}
\label{SingleVorInstability}

Since the time when it was first understood that superfluid
turbulence is made up of tangled quantized vortices
\citep{{Vinen:1956},{Vinen:1961}}, the most basic question has
been its onset as a function of applied counterflow velocity: How
is turbulence started and what defines its critical velocity? An
important clue was provided by the rotating experiments of
\cite{Cheng:1973} and \cite{Swanson:1983}, who found that
rectilinear vortex lines in rotation are broken up in a turbulent
tangle if a heat current is applied parallel to the rotation axis.
The thermal current is transported as a counterflow of the normal
and superfluid components along the rotation axis. Rectilinear
vortices become unstable in this parallel flow and above a low
critical velocity transform to a tangle which tends to be aligned
in the plane transverse to the heat current.

This phenomenon was explained by Glaberson and his coworkers
\citep{Glaberson:1974} who showed that an array of rectilinear
vortices becomes unstable in longitudinal counterflow above the
critical velocity $v = 2 (2 \Omega \kappa_{\rm e})^{1/2}$, where
$\kappa_{\rm e} \approx \kappa$ is an effective circulation
quantum (supplemented with the logarithmic cutoff term:
$\kappa_{\rm e} = (\kappa/4 \pi)\; \ln{(\ell/a_0)}$, where the
average inter-vortex distance is $\ell \sim (\kappa/2
\Omega)^{1/2}$ and the vortex core radius $a_0$). The instability
appears when a Kelvin-wave mode with wave vector $k = (2 \Omega/
\kappa_{\rm e})^{1/2}$ starts to build up, whose amplitude then
grows exponentially in time. The Glaberson instability has also
been examined in numerical calculations which qualitatively
confirm the instability and the vortex tangle in the transverse
plane which starts to form above a first critical velocity
\citep{Tsubota:2003}.

In general, the dispersion relation of a helical Kelvin wave
disturbance $\propto$ exp[-i$(\omega_{\rm k} t -k z)$] can be
written as \citep{{Donnelly:1991},{Finne:2006b}}
\begin{equation}
\omega_{\rm k} (k) = \kappa_{\rm e} k^2 - \alpha^{\prime}
(\kappa_{\rm e} k^2 - k v) - {\rm i} \alpha (\kappa_{\rm e} k^2 -
k v) \;. \label{KelvinModeDisp}
\end{equation}
In the absence of flow $(v = 0)$, these Kelvin modes are always
damped, at high temperatures they are actually overdamped, but at
low temperatures $(\alpha < 1-\alpha^{\prime})$ this is not the
case. In applied flow $(v > 0)$, the long wave length modes with
$0 < k < v/\kappa_{\rm e}$ become exponentially unstable. If an
evolving vortex accumulates enough length $L_{\parallel}$ parallel
to the applied flow, then a disturbance with wave length
$\lambda_{\rm min} \sim L_{\parallel} \sim \kappa_{\rm e} / v \sim
1/k_{\rm max}$ may start to grow. The expanding loop may
reconnect, either with the wall of the container, with itself, or
with another vortex. This leads to a growing number and density of
evolving vortices, which ultimately start interacting and trigger
the onset of turbulence in the bulk.

\begin{figure}[t]
\centerline{\includegraphics[width=0.9\linewidth]{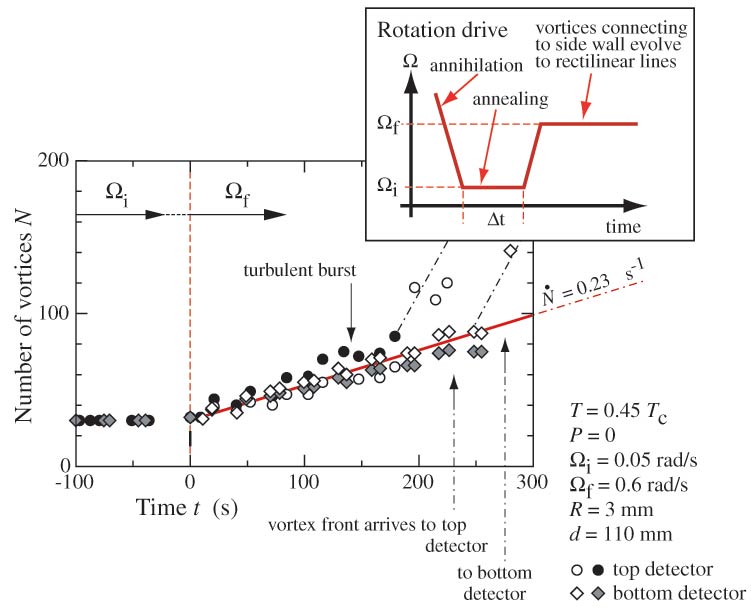}}
\caption{Experimental illustration of single-vortex instability as
precursor of bulk turbulence.  The number of vortices $N(t)$ is
recorded with NMR coils at the top and bottom ends of the sample.
As seen \textit{in the inset}, initially the sample is in the
equilibrium vortex state at $\Omega_{\rm i}= 0.05\,$rad/s with $N
\approx 30$ vortices, of which close to one half connect to the
cylindrical side wall. Rotation is then increased to a new stable
value $\Omega_{\rm f} = 0.6\,$rad/s, which is reached at $t = 0$.
During the ramp to $\Omega_{\rm f}$ the counterflow builds up,
compresses the rectilinear sections of all vortices to a central
cluster, and starts the spiral motion of the vortex ends
connecting to the side wall. Eventually in the increased applied
flow at $\Omega_{\rm f}$ the instability starts to generate new
vortices which contribute to the average slow rate ${\dot N}$ of
vortex formation, shown by the solid straight line. After about
140\,s the turbulent burst occurs 63\,mm above the bottom end
plate. Vortex fronts traveling up and down along the column then
approach the two detector coils and reach their closer ends as
indicated by vertical arrows (at 230\,s and 275\,s). The filled
data symbols are derived from experimentally calibrated
counterflow peak heights and the open symbols from order parameter
texture calculations fitted to the NMR signal in the non-rotating
state. } \label{SlowVorFormation}
\end{figure}

No rigorous analytical calculation has been presented of the
single-vortex instability in the rotating container, but a simple
scaling model illustrates the problem. Consider a vortex ring in
vortex-free counterflow, which is initially perpendicular to the
plane of the ring in a rotating cylinder of radius $R$. If the
ring is large enough, then it expands until it reaches the
container size $R$. The time needed for this expansion is of order
$\delta t\sim R/\alpha v$, where $v$ is the average normal
velocity through the ring. The ring also has a self-induced
velocity component $v_{\rm r} \sim \kappa_{\rm e}/R$, which arises
from its own curvature and is directed along the normal of the
plane of the ring. Because of this velocity component, the plane
of the ring is rotated away from being perpendicular to the
azimuthal flow in the cylinder, while it drifts in the flow.
During the time $\delta t$, the vortex length parallel to the flow
becomes of order $(1-\alpha') v_{\rm r} \, \delta t$. Equating
this to $L_{\parallel}$, it is seen that the instability condition
$L_{\parallel} \gtrsim \lambda_{\rm min}$ leads to the requirement
$\zeta = (1-\alpha')/\alpha \gtrsim 1$. This condition is
virtually independent of velocity; the only restriction is imposed
by the finite container radius, $L_\parallel < R$, which defines a
critical velocity $v_c \sim \kappa_{\rm e}/R$. Typically the time
spent by a vortex in radial motion before reaching the sample
boundary is of order $R/(\alpha\Omega R)=(\alpha\Omega)^{-1}$. At
$0.45\,T_{\rm c}$ and $\Omega=0.6$\,rad/s, the inverse of this
quantity equals 0.20\,s$^{-1}$, which fits with the measured
vortex generation rate of $dN/dt = \dot{N} = 0.23\,$s$^{-1}$ in
Fig.~\ref{SlowVorFormation}. However, numerical calculations
confirm that the presence of surfaces is required to demonstrate
the single-vortex instability in usual experimentally relevant
flow conditions \citep{Finne:2006a}. Also the calculations
demonstrate that the instability is not characterized by a unique
critical velocity, since it depends on the relative orientation of
the flow with respect to the vortex, while the vortex expands in
helical motion in the rotating cylinder. Thus the above model is
incomplete.

Let us now examine direct observations of the single-vortex
instability in rotating flow. In the onset temperature regime, $T
\sim T_{\rm on}$, in about half of the measured cases, which lead
to a turbulent burst, a slow increase in the number of vortex
lines $N(t)$ can be observed to precede the turbulent burst. If
present, the increase is invariably followed by a turbulent burst.
Thus it appears reasonable to associate the slow increase with the
single-vortex instability. In Fig.~\ref{SlowVorFormation} the
number of vortex lines $N(t)$ is plotted as a function of time
while the precursor generates new vortices at slow rate. In this
example the increase in $N(t)$ is almost linear (solid line),
until the turbulent burst sets in and starts the vortex front
motion along the rotating column both upwards and downwards from
the site of turbulence. At $\Omega_{\rm f} = 0.6\,$rad/s the slow
increase lasts in this example for about 140\,s, generating
approximately one vortex every five seconds, until some 30 new
vortices have been created and the turbulent burst manages to
switch on. The time interval from $t=0$ to the turbulent burst is
called the burst time which here is $t_{\rm b} = 140\,$s. At
larger $\Omega_{\rm f}$ the burst time is shorter in duration,
{\it eg.} in a repetition of the measurements in
Fig.~\ref{SlowVorFormation} at $\Omega_{\rm f} = 1\,$rad/s the
turbulent burst was found to start in less than 30\,s.

Two further observations about the precursor can be made from
Fig.~\ref{SlowVorFormation}. First, vortex formation proceeds
independently in different parts of the sample. At $\Omega_{\rm f}
= 0.6\,$rad/s it takes more than 300\,s for a vortex created at
one end of the sample to reach the other end. Still, vortex
formation at the top and bottom ends is observed to proceed at
roughly the same rate. Thus vortex generation by the single-vortex
instability is not localized, in contrast to the turbulent burst.
The random occurrence of the single-vortex instability agrees with
the notion of ideal walls (or at least weak pinning), as opposed
to a vortex mill localized at a surface defect on the cylinder
wall.

Secondly, in Fig.~\ref{SlowVorFormation} the equilibrium vortex
state at low initial rotation $\Omega_{\rm i} = 0.05\,$rad/s has
been used to introduce evolving vortices in the applied flow. This
approach provides a more reproducible initial vortex configuration
than remanent vortices, since the number of those vortices, which
connect to the cylindrical side wall, is primarily determined by
the misalignment between the cylinder and rotation axes
(Fig.~\ref{EquilVorStateCalculation}). In a given experiment the
residual angle between the two axes is generally a constant.

To appreciate the influence of the vortices curving to the side
wall, the experiment was repeated differently. A cluster with only
rectilinear vortices $(N < N_{\rm eq})$ was prepared at higher
temperatures and was then cooled below $0.5\,T_{\rm c}$. As long
as this cluster is separated by a sufficiently wide vortex-free
counterflow annulus from the cylindrical boundary, $\Omega$ can be
increased or decreased without change in $N$ at any temperature
down to $0.35\,T_{\rm c}$ (which is the lower limit of the so far
measured onset temperatures $T_{\rm on}$). If $\Omega$ is reduced
too much, the cluster makes contact with the cylindrical side
wall, some outermost vortices become curved, and during a
subsequent increase of $\Omega$, while $T < 0.5\,T_{\rm c}$, the
behavior in Fig.~\ref{SlowVorFormation} is reproduced. Therefore
we are led to assume that, to observe the vortex instability, at
least one curved vortex connecting to the cylindrical side wall
needs to be present. At temperatures below $0.35\,T_{\rm c}$ this
may not be the case, since in rapid changes of rotation even
rectilinear vortices seem to be destabilized
(Fig.~\ref{VorStability}).

More statistics on the properties of the precursor has been
collected from measurements similar to that in
Fig.~\ref{SlowVorFormation} by \cite{de Graaf:2007}. Important
characteristics are the initial rate of vortex generation
$\dot{N}(t=0)$ and the burst time $t_{\rm b}$. These can be
examined for events with sufficiently long burst times $t_{\rm b}
\gtrsim 20\,$s, so that the rate of the counterflow peak height
decrease with time can be adequately resolved. In general it is
found that $\dot{N}$ increases and $t_{\rm b}$ decreases rapidly
with decreasing temperature below $T_{\rm on}$. To find events
with well-resolved pre-turbulent vortex generation and long burst
time one has to scan for data (i) in the onset temperature regime,
$T \approx T_{\rm on}$, (ii) with a low initial formation rate
$\dot{N} \lesssim 1\,$vortex/s, and (iii) by starting from a state
with a small number of seed vortices (iv) at low applied flow
velocity.

These measurements demonstrate that the precursor generates new
independent vortex loops which start to evolve along spiral
trajectories towards the final state of a rectilinear vortex line.
When the density of evolving vortices rises sufficiently, so that
interactions between them in the bulk volume become possible, then
the process is terminated in a turbulent burst. The burst is a
localized event which from one measurement to the next happens
randomly at different heights $z$ of the sample
(Fig.~\ref{NMR-LineShapes}). The measured properties of the
precursor are consistent with those expected for a single-vortex
instability based on the excitation of Kelvin-wave modes of
sufficiently long wave length. Overall, measurements in the onset
regime reveal the precursor mechanism, owing to the strongly
temperature dependent mutual friction of $^3$He-B, which makes the
precursor observable within a narrow temperature interval around
the onset temperature. At lower temperatures  the turbulent burst
develops so rapidly that the measuring techniques, which have been
employed so far, are not fast enough to capture the details. The
latter case is the typical situation in superfluid $^4$He
experiments.

\subsection{Numerical calculation of dynamic vortex
generation}\label{Simulation}

Numerical calculations on vortex dynamics are carried out with the
vortex filament model introduced by \cite{Schwarz:1988}. With
today's computing power, one uses Biot-Savart integration along
all vortex lines, so that the superfluid velocity field from
vortices is obtained from \citep{Hanninen:2005}
\begin{equation}
\mathbf{v}_{s,\omega}({\bf r},t) = \frac{\kappa}{4\pi} \int
\frac{(\mathbf{s-r}) \times d\mathbf{s}}{|\mathbf{s-r}|^3}\;.
\label{Biot-Savart}
\end{equation}
The line integral is taken along all vortices in the system, ${\bf
s}(\xi,t)$ denotes the location of the vortex core at time $t$,
and $\xi$ is measured along the arc length of the vortex core. In
the presence of solid boundaries the total superfluid velocity
field, $\mathbf{v}_\mathrm{s} = \mathbf{v}_{\mathrm{s},
\omega}+\mathbf{v}_\mathrm{b}$, is modified by the boundary
induced velocity $\mathbf{v}_\mathrm{b}$. At a plane boundary one
can use image vortices to satisfy the requirement of zero flow
through the boundary, $\hat{\bm{n}} \cdot \mathbf{v}_\mathrm{s} =
0$, where $\hat{\bm{n}}$ is the unit vector along the surface
normal. More generally we obtain $\mathbf{v}_\mathrm{b} =
\nabla\Phi$ by solving the Laplace equation $\nabla^2\Phi = 0$
combined with the requirement that at the boundary $\hat{\bm{n}}
\cdot \nabla \Phi = - \hat{\bm{n}} \cdot \mathbf{v}_{\mathrm{s},
\omega}$.

No surface pinning or even surface friction is generally included,
the boundaries are assumed ideal, as indicated so far by
measurements on $^3$He-B in smooth-walled simple cylindrical
containers. Mutual friction in the bulk superfluid is included
using the equation of motion (\ref{vl}) for the vortex element at
${\bf s}(\xi,t)$, which moves with the velocity ${\bf v}_{\rm L} =
d\mathbf{s}/dt $. For the mutual friction parameters $\alpha
(T,P)$ and $\alpha'(T,P)$ one uses the $^3$He-B data measured by
\cite{Bevan:1997} at 10 and 29\,bar pressures. A reconnection
between two vortex segments is enforced if they have drifted
within a distance from each other which is less than the minimum
spatial resolution of the calculation (usually $\sim 0.05\,$mm).
The configuration after reconnection should correspond to shorter
overall vortex length than the initial state. In practice, the
computing time limits severely what can be calculated and what
becomes too time consuming. Therefore the practical implementation
becomes of great importance, how the Biot-Savart integration and
the proper solution for the boundary conditions are worked out.
For details we refer to \cite{de Graaf:2007}.

\begin{figure}[t]
\begin{center}
\centerline{\includegraphics[width=0.6\linewidth]{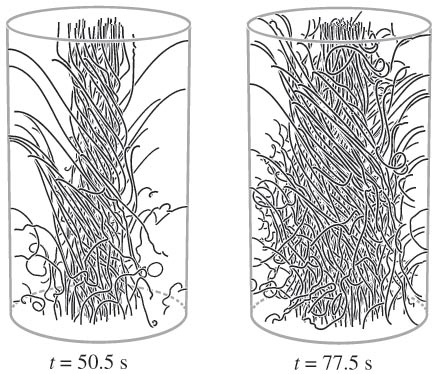}}
\caption{Two snapshots from a calculation of vortex generation in
a rotating cylinder. The summary of these calculations with
results accumulated over more than 100\,s is shown in
Fig.~\protect\ref{VorReconRotCyl}. } \label{VorConfigRotCyl}
\end{center}
\end{figure}

In Fig.~\ref{VorConfigRotCyl} two snapshots are shown from
calculations on vortex formation and the configurations which
evolve in a rotating cylinder \citep{de Graaf:2007}. Recently
formed younger vortices are here in helical configurations on the
outer circumference closer to the cylindrical wall. There in the
outer regions one can see loops of Kelvin waves, small separated
loops with both ends of the vortex on the cylindrical wall, and
even closed vortex rings (lower right corner at $t=50\,$s). Since
it is primarily surface reconnections at the cylindrical wall
which contribute to the formation of new vortices in the early
stages of the calculation (at $t < 100\,$s), the many newly formed
short loops are still close to the side wall. Further inside the
cluster one can see older and straighter vortices which congregate
within the central parts.

\begin{figure}[t]
\begin{center}
\centerline{\includegraphics[width=0.9\linewidth]{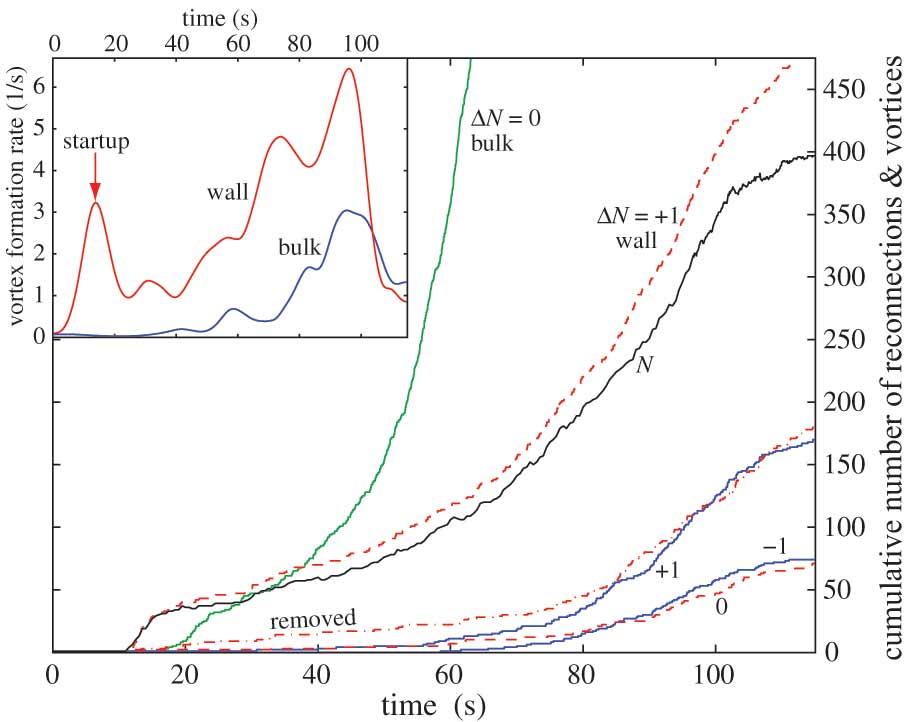}}
\caption{Calculation of the cumulative number of reconnections and
vortices in a rotating cylinder. The different curves denote:
($\Delta N = 0$, solid curve) reconnections in the bulk which do
not change $N$; (+1, dashed) reconnections with the cylindrical
wall which add one new vortex loop; ($N$, solid) total number of
vortices; (removed, dash-dotted) small loops which form in
reconnections mainly close to the cylindrical wall, but which are
contracting and are therefore removed; (+1, solid) reconnections
in the bulk which add one vortex and (-1, solid) which remove one
vortex; (0, dashed) reconnections at the cylindrical wall which do
not change $N$. {\it (Inset)} Averaged rate of increase in $N$
owing to reconnections on the cylindrical wall and in the bulk.
The large initial peak in the boundary rate represents the
starting burst, which is used to start vortex formation.
Parameters: $R = 3\,$mm, $h = 10\,$mm, $\Omega = 0.9\,$rad/s, and
$T= 0.35\,T_{\rm c}$ (where $\alpha = 0.095$ and $\alpha^{\prime}
= 0.082$).} \label{VorReconRotCyl}
\end{center}
\end{figure}

The general observation from these calculations is that evolving
vortices in a rotating sample are more stable in the numerical
experiment than in measurements. For instance, in
Fig.~\ref{VorConfigRotCyl} vortex formation has to be started from
an artificial initial configuration \citep{Finne:2006a}. This
consists from an initial single vortex ring which is placed in the
plane perpendicular to the rotation axis at height $0.2 \, h$
slightly off center, to break cylindrical symmetry \citep{de
Graaf:2007}. This is an unstable configuration where Kelvin waves
of large amplitude immediately form and then reconnect at the
cylindrical wall. The end result is the sudden formation of
roughly 30 vortices which have one end on the bottom end plate and
the other moving in spiral trajectory along the cylindrical wall.
After the initial burst the later evolution is followed as a
function of time $t$, the number of vortices $N(t)$ is listed, and
the reconnections of different type are classified. The results
are shown in Fig.~\ref{VorReconRotCyl}.

In Fig.~\ref{VorReconRotCyl} one keeps account of all reconnection
processes which occur in the rotating sample as a function of time
while it is evolving towards its final stable state with an array
of rectilinear vortices and $N \rightarrow N_{\rm eq}$. After the
initial burst of the first $\sim 30$ vortices $N$ increases first
gradually, but after about 50\,s the rate $\dot{N}$ picks up.
During the first 50\,s reconnections in the bulk do not contribute
to the generation of new vortices, but later such processes also
start to appear. However, even during the later phase a
reconnection of a single vortex at the cylindrical wall, while
Kelvin waves expand along this vortex, remains the dominant
mechanism of vortex generation. This is seen from the fact that
the curve for $N$ follows closely that of the successful surface
reconnections (dashed curve marked as ``$\Delta N =+1$").  The
most frequent reconnections after the first 40\,s are denoted by
the solid ``$\Delta N = 0$" curve and occur in the bulk between
two different vortices.  These inter-vortex reconnections do not
lead to changes in $N$ and are primarily associated with processes
occurring between the twisted vortices in the bundle further away
from the wall.

The inset in Fig.~\ref{VorReconRotCyl} compares the rates of
vortex generation from reconnections at the wall and in the bulk:
The reconnection of a single vortex at the cylindrical wall is
clearly the most important mechanism for the generation of new
independent vortex loops in the early stages of the calculation.
The dominant role of such wall reconnections is compelling. A
second important consideration is correspondence with measurement.
The obvious difference is the higher stability of evolving
vortices in the calculation as compared to experiment. In
Fig.~\ref{VorReconRotCyl} the rate of vortex generation remains
modest, no clearly identifiable turbulent burst can be
distinguished, and the vortex number approaches the equilibrium
value from below. After 115\,s of evolution the vortex number has
progressed to $N \approx 400\,$, where the increase is almost
stopped, well below the saturation value of $N_{\rm eq} \approx
780\,$ \citep{de Graaf:2007}. This is the general experience from
calculations on an ideal rotating cylinder, with smooth surfaces
and no surface friction or pinning. The calculations become more
and more time consuming with decreasing temperature, which limits
the possibilities to obtain a more comprehensive understanding of
their predictions and of the origin of the differences with
measurement. The low probability of the single-vortex instability
in the calculations appears to be a particular property of
rotating flow in a circular cylinder, since linear pipe flow, for
instance, displays a steady rate of vortex generation \citep{de
Graaf:2007}.

Clearly numerical calculations provide important illustrations and
guidance in situations where measurements answer only specific
limited questions. The calculations take full account of
interactions between vortices and between a vortex and the ideal
container wall. Nevertheless, the correspondence between
calculation and measurement is not satisfactory at present, when
we speak about the single-vortex instability and the onset of
turbulence in a rotating cylinder. It appears that some mechanism,
which makes vortices more unstable and adds to the vortex generation rate,
is missing from the calculations. The difficulty is
likely to reside on the cylindrical wall, where the assumption of
ideal conditions should be examined closer. Attempts in this
direction have so far not produced more clarification. However,
these uncertainties about the mechanisms behind the single-vortex
instability in rotating flow do not change the fact that at low
vortex density Kelvin-wave formation on a single vortex, followed
by a reconnection at the surface, is the only efficient mechanism
for generating new vortices.

\subsection{Summary: onset of turbulence}

Since the advent of $^3$He-B new possibilities have appeared to
study turbulence. Firstly, it has become possible to
distinguish and characterize, in measurements with large samples,
vortex formation at a stable reproducible critical velocity,
vortex remanence and turbulent proliferation of vortices.
Secondly, the mutual friction dissipation $\alpha (T)$ with strong
temperature dependence around $\alpha \sim 1$ has made it possible
to evaluate the role of mutual friction in the onset of
turbulence. The important dynamic parameter proves to be $\zeta =
(1 - \alpha^{\prime})/ \alpha$. It controls the onset of the
single-vortex instability, where an evolving vortex becomes
unstable and generates, during a reconnection at the wall, a new
vortex loop. After several such events the density of evolving
vortices is sufficient to produce a turbulent burst. The necessary
condition is $\zeta \gtrsim 1$, to start the cascade of the
single-vortex instability followed by the turbulent burst.

The single-vortex instability becomes possible only at
temperatures below the turbulent transition in the bulk volume and
thus $T_{\rm on} \leq T_{\rm on}^{\rm bulk}$. The onset
temperature $T_{\rm on}$ of these two series-coupled processes has
been found to obey a power-law dependence which relates the mutual
friction parameter $\zeta$ to the magnitude of the "flow
perturbations" in Eq.~(\ref{PowerLaw}). Well above $T_{\rm on}$ no
new vortices are detected (with a resolution $< 10$ new vortices),
while well below $T_{\rm on}$ all final states are found to be
equilibrium vortex states with close to the equilibrium number of
vortices, $N \lesssim N_{\rm eq}$. In the onset regime itself, $T
\sim T_{\rm on}$, one finds events with and without turbulent
burst,  but surprisingly practically no incomplete transitions
with $N_{\rm i} < N \ll N_{\rm eq}$.

In the intermediate temperature regime $0.3 \, T_{\rm c} < T < 0.6
\, T_{\rm c}$, the equilibrium vortex state is reached after a
single turbulent burst. In fact, in the measurements with the
sample setup of Fig.~\ref{NMR-LineShapes} no case of two or more
almost simultaneous bursts was identified above $0.35\,T_{\rm c}$.
Apparently the probability of the single vortex instability to
start a turbulent burst is still low at these temperatures.
Secondly, after the burst the vortex front moves rapidly and
removes the vortex-free flow. At these intermediate temperatures
the burst is both spatially and temporally a localized event in a
short section (of length $\sim  R$) of the column. From one
measurement to the next, it occurs randomly at different heights
of the column. Below $ 0.3 \, T_{\rm c}$ the longitudinal
propagation velocity of vortices becomes slow and evolving
vortices go rapidly unstable everywhere. As a result turbulence
tends to be both spatially and temporally more extended, filling
larger sections of the column. The later events, the evolution
after the turbulent burst, are the subject of the next section.


\newcommand{\BSE}[1]{\begin{subequations}\label{#1}}
\def\ese{\end{subequations}}
 \newcommand{\BE}[1]{\begin{equation}\label{#1}}
 \newcommand{\BEA}[1]{\begin{eqnarray}\label{#1}}
\def\ee{\end{equation}} \def\eea{\end{eqnarray}}

\renewcommand{\sb}[1]{_{\text {#1}}}  
\renewcommand{\sp}[1]{^{\text {#1}}}  
\newcommand{\Sp}[1]{^{^{\text {#1}}}} 
\def\Sb#1{_{\scriptscriptstyle\rm{#1}}}

   \newcommand{\eq}[1]{(\ref{#1})}
  \newcommand{\Eq}[1]{Eq.~(\ref{#1})}
  \newcommand{\Eqs}[1]{Eqs.~(\ref{#1})}
  \newcommand{\Fig}[1]{Fig.~\ref{#1}}
  \newcommand{\Figs}[1]{Figs.~\ref{#1}}
 \newcommand{\Sec}[1]{Sec.~\ref{#1}}
 \newcommand{\Secs}[1]{Secs.~\ref{#1}}
  \newcommand{\Ref}[1]{Ref.~\citep{#1}}
 \newcommand{\Refs}[1]{Refs.~\citep{#1}}

\newcommand{\B}[1]{{\bm{#1}}}
\newcommand{\C}[1]{{\mathcal{#1}}}    
\newcommand{\BC}[1]{\bm{\mathcal{#1}}}
\newcommand{\F}[1]{{\mathfrak{#1}}}
\newcommand{\BF}[1]{{\bm{\F {#1}}}}

\renewcommand{\a}{\alpha} \renewcommand{\b}{\beta}
\newcommand{\g}{\gamma}\newcommand{\G}{\Gamma}
\renewcommand{\O}{\Omega} \renewcommand{\o}{\omega}
\newcommand{\D}{\Delta}\def\r{\rho} \def\ve{\varepsilon}
\renewcommand{\L}{\Lambda} \renewcommand{\k}{\kappa}

\let \= \equiv \let\*\cdot \let\~\widetilde \let\^\widehat \let\-\overline
\let\p\partial \def\pp {\perp} \def\pl {\parallel}

\def\<{\left\langle}    \def\>{\right\rangle}
\def\({\left(}          \def\){\right)}
 \def \[ {\left [} \def \] {\right ]}

 \let \nn  \nonumber  \newcommand{\br}{\\ \nn}
\newcommand{\BR}[1]{\\ \label{#1}}
\def\hf{\frac{1}{2}}

\section{Propagating vortex front in rotating
flow}\label{s:FrontMotion}

\subsection{\label{s:FrontMotionIntro}Introduction }

In rotation at constant angular velocity the steady state
superfluid response is generally not turbulent. Nevertheless,
transient states of turbulence can be formed by rapidly changing
the rotation velocity, especially if the sample container does not
have circular cross section or its symmetry axis is inclined by a
larger angle from the rotation axis. The decay of turbulence and
the approach to equilibrium can then be monitored at constant
$\Omega$. The normal component relaxes back to solid body rotation
by means of viscous interactions, while the superfluid component
adjusts much slower, coupled only by mutual friction dissipation
from vortex motion with respect to the normal component and (if
any) by the deviations of the container walls from being
axially-symmetric around the rotation axis. Such measurements on
transient turbulence are generally known as spin up or spin down
of the superfluid component. This used to be an important topic in
superfluid $^4$He work in the fifties and sixties
\citep{Androkashvili:1967}, but was then replaced (with few
exceptions \citep{Adams:1985}) by other methods which one expected
to lead to results with more straightforward interpretation.

The turbulent burst, which suddenly starts the motion of $N
\approx N_{\rm eq}$ vortices along the rotating column at
temperatures $T \lesssim T_{\rm on}$, as discussed in
Sec.~\ref{VortexInstability} (\textit{cf.}
Fig.~\ref{RotColumnEvolution}), provides a novel technique to
investigate transient turbulence in rotation. Originally it was
assumed that this motion would take place as a tangle of vortices,
which spreads longitudinally along the rotating column. It was
soon realized from NMR measurements \citep{Eltsov:2006b} that this
could not be the case; rather the propagating vortices were highly
polarized and had to be coiled in a helical configuration owing to
their spirally winding motion. This recognition presented a new
problem: Is there any room at all for turbulence in this kind of
motion and if there is, how is it expressed? Or perhaps the nature
of the motion changes on approaching the zero temperature limit,
when mutual friction dissipation vanishes $\alpha \propto$
exp$(-\Delta/T)$? These questions provided the incentive to
examine the propagation more closely and to measure its velocity
as a function of temperature. The results demonstrate that
turbulent losses depend crucially on the type of flow, flow
geometry, external conditions, the physical properties of the
superfluid itself, \textit{etc.}

A measurement of the front propagation in the laminar and turbulent
temperature regimes allows one to determine the rate of kinetic
energy dissipation. The measurement proceeds as follows: The
initial starting state is the rotating vortex-free state, the
so-called Landau state, which is metastable with much larger
free energy than the stable equilibrium vortex state. The latter
consists of rigidly co-rotating normal and superfluid components,
owing to the presence of a regular array of rectilinear vortices,
while in the vortex-free state the superfluid component is not
rotating at all: it is at rest in the laboratory frame of
reference. When the turbulent burst is triggered in the Landau
state, a rapid evolution towards the equilibrium vortex state is
started, where a boundary between the vortex-free and the vortex
states propagates along the rotating column and displaces the
metastable vortex-free counterflow. Particularly at temperatures
below $0.4\,T_{\rm c}$ the boundary has the form of a sharp thin
vortex front which travels at a steady velocity $V\sb f$. The
dissipation rate of the total kinetic energy, $\C E(t)$, is
related to $V\sb f$ as
\begin{equation} {d \C E}\big/ {d t} = - \pi \r\sb s V\sb f\, \O^2 R^4/4 \, .
\label{KineticEnergy}
\end{equation}
 By
measuring $V\sb f$ one determines directly the energy dissipation
${d \C E} / {d t}$ as a function of temperature.

At high temperatures the motion is laminar and the front velocity
is determined by mutual friction dissipation between the normal
and superfluid components, $V_{\rm f}(T) \approx \alpha(T)\,
\Omega R $. Below $0.4\,T_{\rm c}$, $V_{\rm f}(T)$ deviates more
and more above the laminar extrapolation \citep{Eltsov:2007}, in
other words the dissipation becomes larger than expected from
mutual friction in a laminar flow. At the very lowest temperatures
a striking anomaly becomes apparent: ${d \C E(t)}\big/ {d t}$ does
not go to zero, but the measured velocity $V_{\rm f}(T)$ appears
to level off at a constant value which corresponds to an effective
friction $\alpha_{\rm eff} \sim 0.1$,  even though
$\alpha(T)\rightarrow 0$, when $T\rightarrow0$. Evidence for a
similar conclusion has been offered by the Lancaster group
\citep{Bradley:2006}, who measured the density of the vortex
tangle created by an oscillating grid and found that this kind of
turbulence decays at a temperature-independent finite rate below
$0.2\,T_{\rm c}$.

When mutual friction decreases and turbulent motions in the vortex
front cascade downward to progressively smaller length scales,
eventually individual quantized vortex lines must become
important. This is the quantum regime of superfluid hydrodynamics.
The energy cascade on length scales smaller than the inter-vortex
distance and the nature of dissipation on these scales are
currently central questions in superfluid turbulence
\citep{Vinen:2002}. Theoretical predictions exist on the role of
non-linear interactions of Kelvin waves and the resulting
Kelvin-wave cascade, which is ultimately terminated in
quasiparticle emission
\citep{Svistunov:2004,Svistunov:2005a,Svistunov:2008a,Vinen:2000,Vinen:2003},
or on the importance of reconnections which could rapidly
redistribute energy over a range of scales and also lead to
dissipation \citep{Svistunov:1995}. From their front propagation
measurements \cite{Eltsov:2007} conclude that the Kelvin-wave
cascade accounts for an important part in the increased
dissipation below $0.3\,T_{\rm c}$. The different sources of
dissipation in this analysis are discussed in Secs.~\ref{NumFront}
and \ref{s:basicEq}.

It is worth noting that a propagating turbulent vortex front has
many interesting analogues in physics \citep{Saarloos:2003}. For
instance, it is similar to the propagation of a flame front in
premixed fuel. Flame front propagation can also proceed in laminar
or in turbulent regimes. In the latter case the effective area of
the front increases and its propagation speed becomes higher than
in the laminar regime. This property finds its practical use in
combustion engines, but has also been used by \cite{SN} to
describe intensity curves of type Ia supernovae. In all such cases
a metastable state of matter is converted to stable state in the
front and $V\sb f$ is determined by the rate of dissipation of the
released energy.

\begin{figure}[t]
\begin{center}
\centerline{\includegraphics[width=0.7\linewidth]{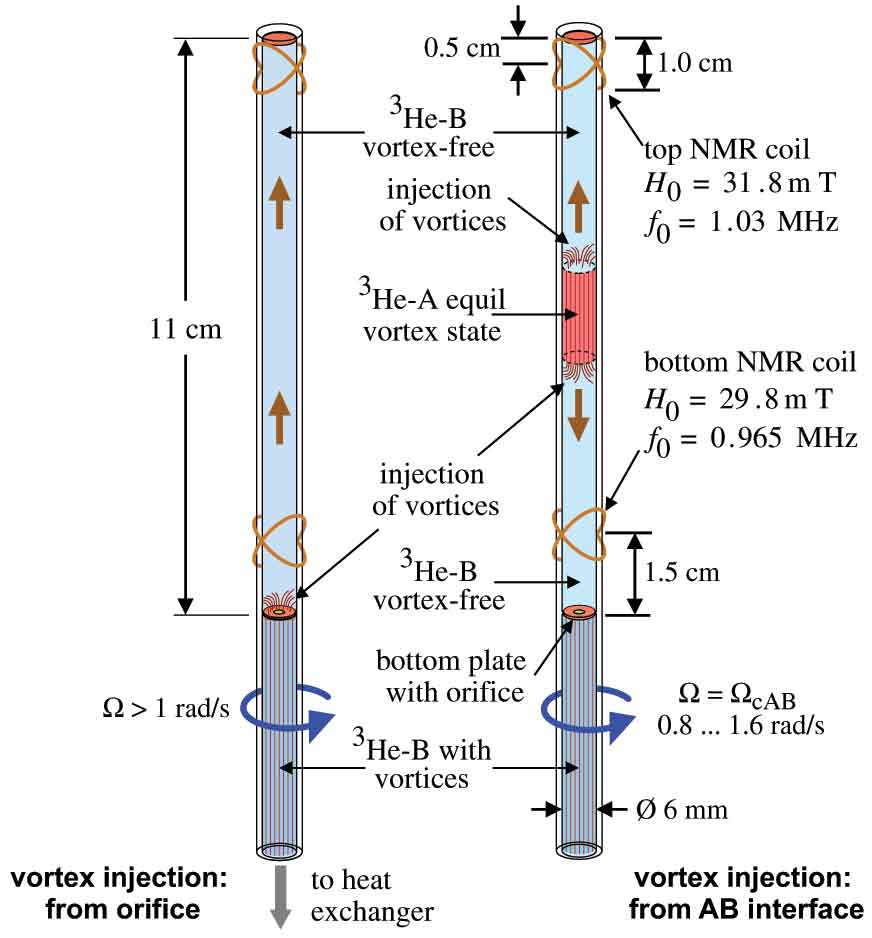}}
\caption{Experimental setup for measuring the propagation velocity
of the vortex front in the rotating column. Two methods are shown
for measuring the front motion across different flight lengths in
one single experimental setup. \textit{On the left} the seed
vortices are tiny remnants at the orifice. In increasing rotation
at $\Omega \gtrsim 1\,$rad/s they produce a turbulent burst in the
volume around the orifice below the bottom detector coil. A single
vortex front is then observed to pass first through the bottom
coil and later through the top coil. The time difference
separating the signals from the passing front over the flight path
of 90\,mm defines the front velocity $V_{\rm f}$. \textit{On the
right} the seed vortices are injected via the Kelvin-Helmholtz
shear flow instability of the two AB interfaces. The injection
event is followed instantaneously by a turbulent burst close to
the AB interface on the B-phase side. A vortex front is then
observed to propagate independently both up and down along the
cylinder. The lengths of the flight paths are equal for the upper
and lower halves. In Fig.~\ref{CF+LarmorResponse} it is explained
how the flight time is determined in this case.} \label{ExpSetup}
\end{center}
\end{figure}

\subsection{\label{ExpVorFront}Measurement of vortex front propagation}

The velocity of the vortex front was measured by
\cite{Eltsov:2007} with the setup in Fig.~\ref{ExpSetup}. The
initial vortex-free state was prepared by warming the sample above
0.7\,$T_{\rm c}$, where remanent vortices annihilate rapidly, and
by then cooling it in the vortex-free state at constant rotation
to the target temperature. Two different procedures were used to
trigger the turbulent burst at the target temperature. These are
sketched in Fig.~\ref{ExpSetup}. In both cases the front velocity
is determined by dividing the flight distance by the flight time,
assuming that the front propagates in steady-state configuration.
Although this is not exactly true, for instance owing to initial
equilibration processes which follow injection, it is assumed for
now that this simplification is justified. Especially, since the
two injection techniques for different propagation lengths give
the same result.

\begin{figure}[t]
\begin{center}
\centerline{\includegraphics[width=.8\linewidth]{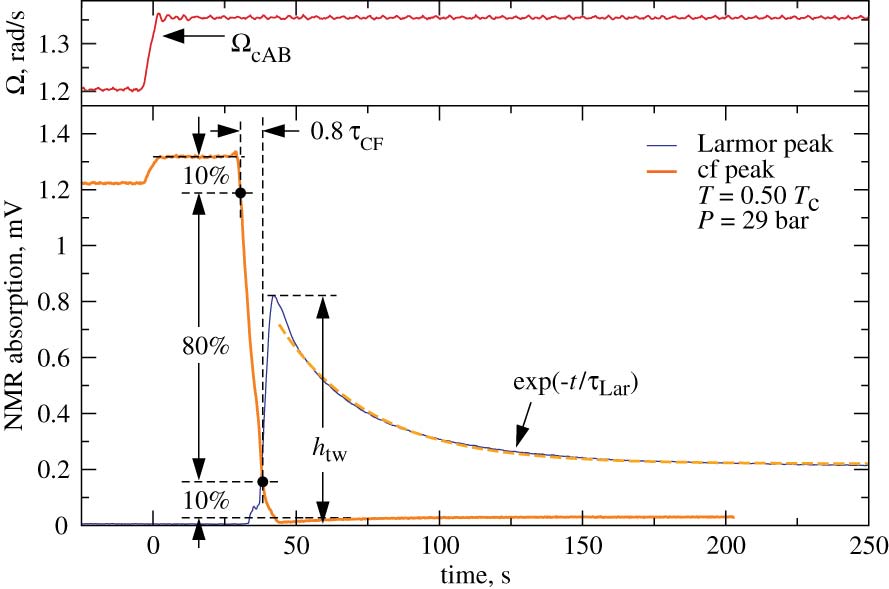}}
\caption{NMR signal responses of the propagating vortex front and
the twisted vortex state behind it. Here the turbulent burst is
started with the Kelvin-Helmholtz instability of the AB interface.
\textit{(Top panel)} It is triggered by increasing $\Omega$ by a
small amount $\Delta \Omega$ across the critical value
$\Omega_{\rm cAB}$ at 1.3\,rad/s. \textit{(Main panel)} Two
absorption responses are shown, which have been recorded with the
bottom detector coil at constant, but different values of magnetic
field. The responses are from two consecutive identical
measurements, to allow direct comparison of signal amplitudes. The
counterflow peak height \textit{(thick line)} shows the KH trigger
$\Delta \Omega$ and a rapid collapse when the front moves through
the coil. The time interval from $t=0$ (when $\Omega (t) =
\Omega_{\rm cAB}$) to the start of the collapse measures the
flight time of the front from the AB interface to the closer end
of the detector. The moment when the peak height reaches zero
corresponds to the point when the front has passed through the
rear end of the coil. The time required for the collapse, $\sim
\tau_{\rm CF}$, measures the width of the vortex front. The second
signal \textit{(thin line)} is recorded close to the Larmor edge
and is sensitive to the longitudinal velocity $v_{{\rm s}z}$ which
is generated by the twisted vortex state
(Fig.~\ref{AxialVel+Fluct}). Its sudden steep rise at $t \approx
30\,$s is caused by the passage of the first helical sections of
the twisted state through the coil. Its later exponential decay
reflects the unwinding of the twist, which starts when the front
has reached the end plate of the cylinder and the vortex ends
begin to slip along the flat surface. } \label{CF+LarmorResponse}
\end{center}
\end{figure}

The first injection technique (depicted on the left,
Fig.~\ref{ExpSetup}) makes use of remanent vortices
\citep{Solntsev:2007}. By trial and error it was found that one or
more remnants can be freed with a small step increase in rotation
from the region around the orifice on the bottom of the sample
cylinder. This was done by increasing $\Omega$ in small steps,
until at some point usually above 1\,rad/s a remnant starts
expanding which below $0.35\,T_{\rm c}$ immediately gives rise to
a turbulent burst. The ensuing vortex front then propagates
upwards along the entire column through both pick-up coils in
succession.

The second injection method (depicted on the right,
Fig.~\ref{ExpSetup}) relies on the superfluid Kelvin-Helmholtz
(KH) instability of the interface between the A and B phases of
superfluid $^3$He \citep{{Blaauwgeers:2002}}. Two stable AB
interfaces are formed by applying a specially configured magnetic
field which stabilizes a narrow A-phase barrier layer over the mid
section of the sample cylinder. The shear flow instability of
these two AB interfaces is controlled by rotation velocity,
temperature, and the stabilization field. At the target
temperature the instability can be triggered with a step increase
of the rotation velocity or of the stabilization field
\citep{Finne:2004b}. The instability causes vortices from the A
phase to escape across the AB interface into the vortex-free
B-phase flow in the form of a bunch of small closely packed loops.
Once in the B phase, at temperatures below $0.59\,T_{\rm c}$ the
loops immediately interact and generate a turbulent burst. Two
vortex fronts then propagate independently up and down from the AB
interfaces, arriving to the top and bottom pick-up coils
practically simultaneously (since the setup is symmetric with
respect to the mid plane of the stabilization field). An example
of the NMR readout as a function of time is shown in
Fig.~\ref{CF+LarmorResponse}. The KH shear flow instability and
the associated vortex leak across the AB interface have been
extensively described in the review by \cite{Finne:2006b}.

In Fig.~\ref{CF+LarmorResponse} one of the signal traces records
the absorption at the counterflow peak (\textit{cf.}
Fig.~\ref{NMR-LineShapes}). It is at maximum in the initial
vortex-free flow at $v = v_{{\rm n}\phi} - v_{{\rm s}\phi}$ (where
$v_{{\rm n}\phi} = \Omega \, r$  and $v_{{\rm s}\phi} = 0$ in the
laboratory frame). To trigger the KH instability,  $\Omega$ is
increased by a small increment across the critical rotation
velocity $\Omega_{\rm cAB}$, which is instantaneously registered
as a small increase in absorption level, owing to the increased
counterflow velocity. Following the instability and the turbulent
burst, the vortices subsequently propagate along the column, but a
response in the counterflow peak height is not observed until they
reach the closer end of the NMR coil. From thereon the absorption
in the peak rapidly decreases and drops to zero. Keeping in mind
that the peak height measures the azimuthally circulating flow in
the transverse plane (\textit{cf.} Eq.~(\ref{cf})), its sudden
removal requires that the passage of the vortices through the coil
must occur as an organized sharp front, followed by a highly
polarized state behind the front. The passage is characterized by
the time $\tau_{\rm CF}$, which is defined in
Fig.~\ref{CF+LarmorResponse}. The propagation velocity $V_{\rm f}$
of the front is determined from its flight time, measured from the
AB interface instability (when $\Omega(t) = \Omega_{\rm cAB}$) to
the arrival of the front at the closer edge of the pick-up coil,
\textit{i.e.} where the rapid drop in the counterflow peak height
starts. Compared to the flight time, the AB interface instability
and the turbulent burst can be considered as instantaneous.

The second signal trace in Fig.~\ref{CF+LarmorResponse} records
the absorption in the Larmor peak. It is near zero in the initial
vortex-free state, displays a sharp maximum after the collapse of
the counterflow peak, and then decays to a small, but finite
value, which is a characteristic of the final equilibrium vortex
state. The transient maximum is the new feature, which arises from
flow in the axial direction, created by a helically twisted vortex
bundle \citep{Eltsov:2006b}. With increasing wave vector $Q$ of
the helix, the axial flow at a velocity $v_{{\rm s}z}$ and the
absorption in the Larmor peak increase monotonically
\citep{Kopu:2006}. Thus the maximum Larmor peak height $h_{\rm
tw}$ in Fig.~\ref{CF+LarmorResponse} is reached when $v_{{\rm
s}z}$ reaches its largest value inside the detector coil. This
happens when the most tightly spiralled section of the twisted
cluster (which is just behind the front, as seen in
Fig.~\ref{AxialVel+Fluct}) passes through the middle of the
detector coil. When the front arrives at the bottom end plate of
the cylinder the twist starts to relax, since the vortices have to
obey the boundary condition on the flat end plate, where they slip
to reduce their length and winding. The unwinding produces the
exponentially relaxing absorption with time constant $\tau_{\rm
Lar}$. The signal decay continues down to the absorption level
characteristic of the equilibrium vortex state, with an array of
rectilinear vortex lines. At their other end the vortices, after
crossing the AB interface, have a continuation as doubly quantized
A-phase vortices \citep{Hanninen:2003}.

To interpret measurements on the front velocity it is of
importance that the structure of the twisted state and the front
itself are known. Information on these characteristics can be
obtained by measuring quantitatively the various features denoted
in Fig.~\ref{CF+LarmorResponse} as a function of temperature.

\subsection{Velocity of vortex front} \label{FrontVelocityMeasurement}

A striking consequence from the twisted state is the appearance of
superflow directed along the helically spiralling vortex cores.
This situation is reminiscent of a "force-free" vortex
configuration, where all the flow is directed along the vortex
core. Such a structure is expected to be stable up to some
instability limit, similar to the Glaberson limit of a rectilinear
vortex array in parallel flow (Sec.~\ref{SingleVorInstability}).
In the twisted state, which is uniform in the axial and azimuthal
directions, the superflow has both an azimuthal component at the
velocity $v_{{\rm s}\phi}$ and an axial component at $v_{{\rm
s}z}$, which depend on the radial coordinate $r$ and are described
by the expressions \citep{Eltsov:2006b}:
\begin{equation}
v_{{\rm s}\phi}(r)=\frac{(\Omega+Qv_0)r}{1+Q^2r^2},\ \ v_{{\rm
s}z}(r)=\frac{v_0-Q\Omega r^2}{1+Q^2r^2}\,. \label{uniftw}
\end{equation}
Since the net flow through the cross section of the cylindrical
container should vanish, from this condition for $v_{{\rm s}z}$
one finds that $v_0=(\Omega/Q)[Q^2R^2/\ln(1+Q^2R^2)-1]$. The axial
flow is directed along the vortex expansion direction close to the
cylindrical wall and in the opposite direction closer to the
center. In practice, there has to exist also a radial velocity
$v_{{\rm s}r}$, since any laboratory example of the twisted state
is nonuniform.   In the case of a propagating vortex front, the
wave vector $Q$ has its maximum value close to the rear end of the
front and decreases to zero at the bottom and top end plates of
the sample. As seen in Fig.~\ref{CF+LarmorResponse}, the twisted
state prominently changes the line shape of the NMR spectrum and
it is the axial superflow which here has the strongest influence.

\begin{figure}
\begin{center}
\centerline{\includegraphics[width=0.52\textwidth]{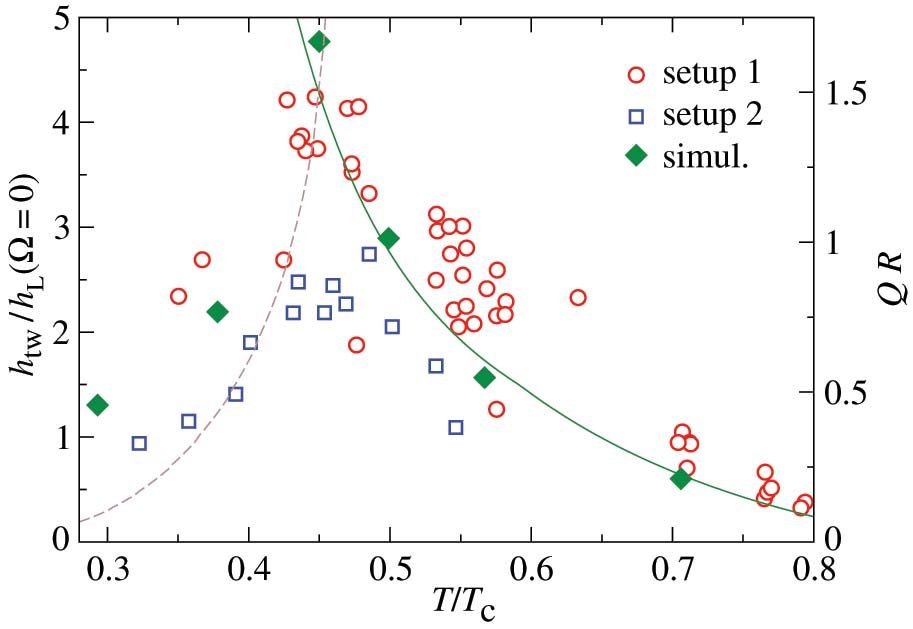}
\includegraphics[width=0.45\textwidth]{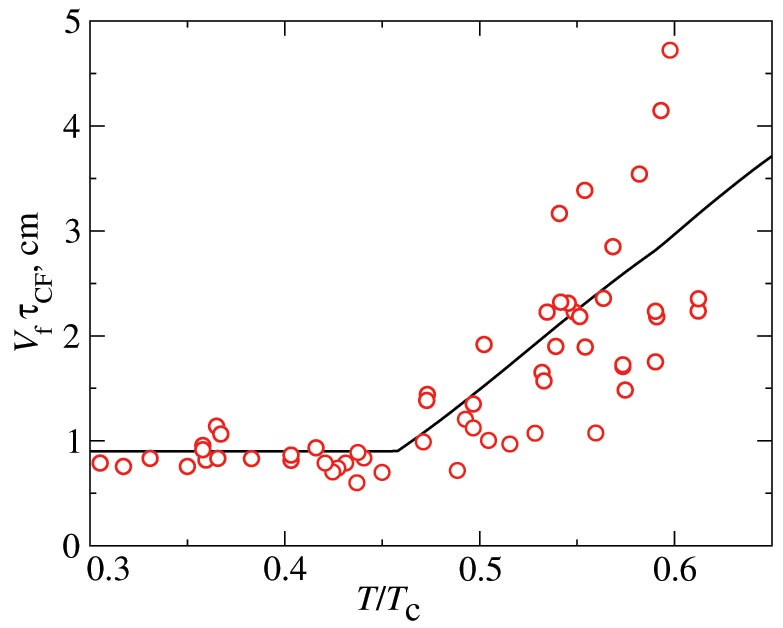}}
\caption{\textit{(Left)} Magnitude of the twist as a function of
temperature.  The measurements are performed using the bottom
spectrometer.  Setups 1 and 2 refer to measurements with the
detector coils positioned in two different sets of positions along
the sample cylinder; in Fig.~\ref{ExpSetup} setup 2 is depicted.
$(\circ, \Box)$: The measured ratio of the maximum amplitude
$h_{\rm tw}$ of the Larmor peak in the twisted state to the
amplitude $h_{\rm L}(\Omega = 0)$ of the Larmor peak in the
nonrotating sample is plotted on the left axis. ($\blacklozenge)$:
The maximum value of the twist wave vector $Q$, obtained from
simulation calculations, is plotted on the right axis. The solid
curve shows the fit $Q\, R = 0.7 (1-\alpha')/\alpha$. The dashed
curve shows the minimum $Q$ at which the vortex front still
propagates in a thin steady-state configuration. \textit{(Right)}
Apparent thickness of the vortex front $\tau_{\rm CF} V_{\rm f}$
as a function of temperature, as determined from measurements
triggered with the Kelvin-Helmholtz instability
(Fig.~\ref{ExpSetup}). The solid line is the prediction of the
model in Eq.~(\ref{thickmodel}). } \label{overshoot}
\end{center}
\end{figure}


As discussed in Sec.~\ref{SeedEvolution}, in vortex-free
counterflow the end point of a single vortex moves on the
cylindrical side wall roughly with the longitudinal velocity
$v_{{\rm L}z} \approx \alpha \Omega R$, while its azimuthal
velocity is  $v_{{\rm L}\phi} \approx - (1-\alpha') \Omega R$ (in
the rotating frame). Thus the wave vector of the spiral trajectory
is $Q = |v_{{\rm L}\phi}/(R v_{{\rm L}z})| \approx (1-\alpha')/(R
\alpha) = \zeta/R$. If this value is used as an estimate for the
wave vector of the twisted state, it follows that the helical
winding of the vortex bundle becomes tighter with decreasing
temperature. The tighter twist increases the flow velocities in
Eq.~(\ref{uniftw}), \textit{i.e.} flow parallel to the vortex
cores is enhanced, which will ultimately destabilize the
``force-free" twisted-cluster configuration. Nevertheless,
twisted-cluster propagation appears to persist even below
$0.2\,T_{\rm c}$. The twist is removed by the slip of the vortex
ends along the flat end plates of the cylinder, which generates
the exponentially relaxing absorption in
Fig.~\ref{CF+LarmorResponse} with the time constant $\tau_{\rm
Lar}$.

In Fig.~\ref{overshoot} \textit{(left panel)} the measured
temperature dependence of the magnitude of the twist is plotted in
terms of the maximum height of the Larmor peak $h_{\rm tw}$,
normalized to the height of the Larmor peak at $\Omega=0$ at the
same temperature \textit{(left vertical axis)}. Two experimental
setups with slightly different specifications were used in these
measurements. The line shape of the NMR absorption spectrum in the
Larmor region depends both on the magnitude and homogeneity of the
magnetic polarization field. In the two setups the homogeneities
varied by a factor of two, which is believed to explain the
differences in the absolute values between the two data sets  (for
details we refer to \cite{Eltsov:2008}). Nevertheless, it is seen
here that the twist increases towards low temperatures as
expected, but only until a maximum at $0.45T_{\rm c}$, whereas
below $0.45\,T_{\rm c}$ it abruptly starts to decrease.

The non-monotonic temperature dependence of the twist is confirmed
in numerical calculations: The value of the twist wave vector
behind the front, as determined from a fit of the calculated
velocity profiles to Eq.~(\ref{uniftw}) \citep{Eltsov:2008} and
plotted in Fig.~\ref{overshoot} \textit{(left panel, right
vertical axis)}, also peaks at $0.45T_{\rm c}$. Two reasons can be
suggested for the change in temperature dependence at
$0.45\,T_{\rm c}$. First, the twist can relax via reconnections
between neighboring vortices in the bundle, which become more
frequent with decreasing temperature below $0.4\,T_{\rm c}$
(\textit{cf.} Fig.~\ref{Reconnections}). Secondly, the source of
the twist is at the vortex front, while the sink is at the end
plate of the cylinder, where the twist vanishes because of the
boundary conditions. From there the relaxation of the twist
advances in a diffusive manner along the twisted bundle. The
effective diffusion coefficient increases as the temperature
decreases \citep{Eltsov:2006b} and thus the faster diffusion
limits the maximum twist in a finite-size sample at low
temperatures. However, overall the stability of twisted-cluster
propagation appears to be a complicated question at temperatures
below $0.3\,T_{\rm c}$, where it controls the average number of
vortices threading through each cross section of the column behind
the front.

The properties of the twist in Fig.~\ref{overshoot} roughly agree
with the estimate that the front velocity can be approximated with
the longitudinal velocity of a single vortex expanding in
vortex-free rotation, $v_{{\rm L}z} \approx \alpha \Omega R$.
However, this simplification suffers from the following
difficulty: Ahead of the front the vortex-free superfluid
component is at rest and the effective counterflow velocity might
really be approximated with $v = \Omega r$, but behind the front
the density of vortices is close to equilibrium and $v_{{\rm
s}\phi} \approx v_{{\rm n}\phi}$. In this case a vortex, which has
fallen behind in the motion, feels a much reduced counterflow and
continues to fall more behind. Therefore the thickness of the
front should increase with time. The explanation to this dilemma
is that behind the front the superflow induced by the twisted
vortex bundle has to be taken into account. The longitudinal
expansion velocity should now be modified to $v_{{\rm L}z} =
\alpha\,[v_{{\rm n}\phi}(R) - v_{{\rm s}\phi}(R)] +
(1-\alpha')v_{{\rm s}z}(R)$. Since here $v_{{\rm s}z}(R)$ is
oriented in the direction of the front propagation and $v_{{\rm
s}\phi}(R) < v_{{\rm n}\phi}(R)$ in the twisted state, the
longitudinal expansion velocity $V_{\rm t}$ of the vortices in the
tail of the front is enhanced. This velocity can be estimated
taking $v_{{\rm s}z}(R)$ and $v_{{\rm s}\phi}(R)$ from
Eq.~(\ref{uniftw}):
\begin{equation}
V_{\rm t} = \alpha \Omega R \left[ 1 +
\frac{1-\alpha'}{\alpha}\,\frac{1}{Q\,R} \right]\, \left[ 1 -
\frac{Q^2R^2}{(1 + Q^2R^2)\, \log(1 + Q^2R^2)} \right]\;.
\label{Vt}
\end{equation}
Eq.~(\ref{Vt}) has a maximum as a function of the $Q$ vector. If
$(1-\alpha')/\alpha < 1.9$ (which corresponds to $T>0.46\,T_{\rm
c}$ according to the measurements of \cite{Bevan:1997}), the
maximum value of $V_{\rm t}$ is less than the velocity of the
foremost vortices $V_{\rm f} \approx \alpha \Omega R$. In these
conditions the thickness of the front increases while it
propagates. When $T<0.46\,T_{\rm c}$, a wide range of $Q$ values
exists for which formally $V_{\rm t} \geqslant V_{\rm f}$. The
minimum possible value of $Q$ is shown in Fig.~\ref{overshoot}
\textit{(left panel)} as the dashed curve. In these conditions the
front propagates in a steady-state ``thin'' configuration.


Experimentally, the decay time of the counterflow peak $\tau_{\rm
CF}$ in Fig.~\ref{CF+LarmorResponse} can be used to extract the
front thickness. The decay starts when the head of the front
arrives at the closer edge of the detector coil and it is over
when that part of the front leaves the far edge of the detector
coil where the counterflow is not sufficient to generate a
non-zero absorption response. The product $\tau_{\rm CF} V_{\rm
f}$ has the dimension of length and can be called the apparent
thickness of the front. At higher temperatures the actual
thickness of the front grows with time. Here the apparent
thickness depends on the distance of the observation point from
the site of the turbulent burst and on the rate at which the
thickness increases and vortices fall behind. With decreasing
temperature the front starts to propagate as a thin steady-state
structure and ultimately its apparent thickness decreases to equal
the height of the pick-up coil ($h_{\rm c} = 9\,$mm,
Fig.~\ref{NMR-LineShapes}) and remains thereafter approximately
constant.

\begin{figure}[t]
\begin{center}
\centerline{\includegraphics[width=0.6\linewidth]{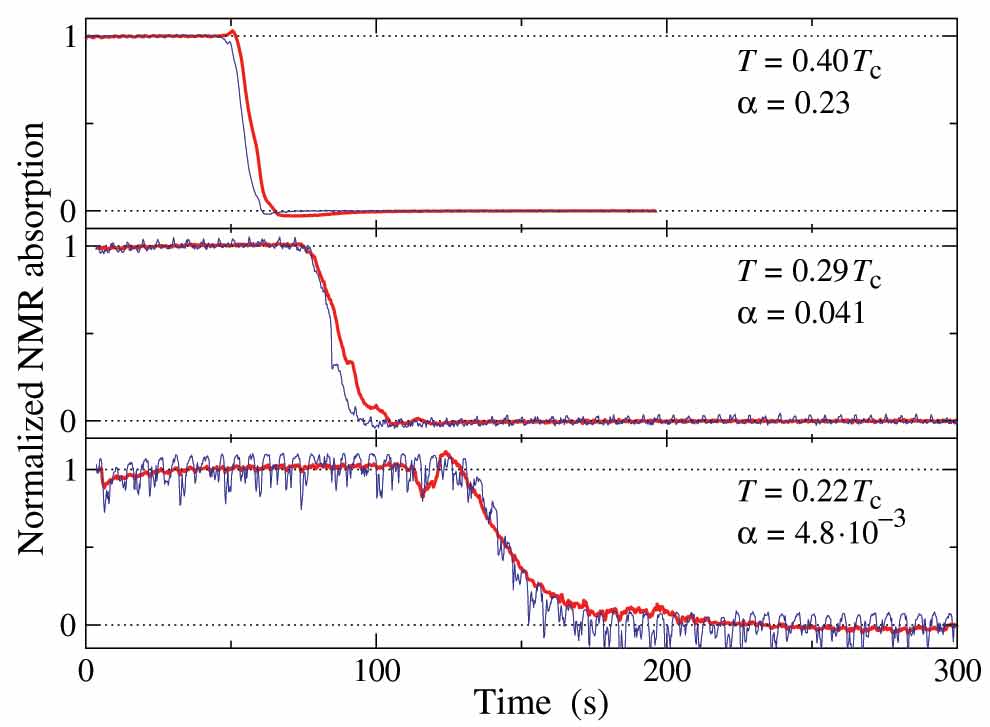}}
\caption{Measurement of vortex front propagation. The NMR
absorption in the counterflow peak is monitored as a function of
time, after triggering the Kelvin-Helmholtz instability at $t = 0$
(\textit{cf.} Fig.~\ref{CF+LarmorResponse}). The instability is
started in constant conditions in the vortex-free state at
1.2\,rad/s, by increasing $\Omega$ in one small step above the
critical value $\Omega_{\rm cAB}$, when a magnetic field
stabilized A-phase layer is present \textit{(topmost panel)}, or
by increasing the magnetic field stepwise above $H_{\rm AB}$ at
constant $\Omega$ \textit{(two lower panels)}. The two signal
traces denote the top \textit{(thin noisy line)} and the bottom
\textit{(thick line)} detector. Since the flight paths for the
upper and lower sample sections are almost equal, the two traces
display almost identical flight times.  } \label{FrontVelExp}
\end{center}
\end{figure}

Measurements of the apparent thickness of the front are presented
in the right panel of Fig.~\ref{overshoot}. At $T>0.45T_{\rm c}$,
$\tau_{\rm CF} V_{\rm f} > h_{\rm c}$ and the apparent thickness
increases with increasing temperature. At $0.45\,T_{\rm c}$ the
apparent front thickness becomes comparable with the height of the
detector coil and thereafter at lower temperatures remains at that
value.  Assuming that initially at the turbulent burst the front
is infinitely thin we can write
\begin{equation}
\tau_{\rm CF} = \frac{h_{\rm b} + h_{\rm c}}{V_{\rm t}^*} -
\frac{h_{\rm b}}{V_{\rm f}}, \label{thickmodel}
\end{equation}
where $h_{\rm b} $ is the distance from the site of the turbulent
burst to the nearest edge of the pick-up coil and $V_{\rm t}^*$ is
the expansion velocity at the position in the front where the NMR
signal from the counterflow vanishes. Given that the latter
condition roughly corresponds to $v_{{\rm s}\phi} \sim (1/2)
v_{{\rm n}\phi}$ we take $V_{\rm
  t}^* = (V_{\rm t}+V_{\rm f})/2$ if $V_{\rm t} < V_{\rm f}$ and simply
$V_{\rm t}^* = V_{\rm f}$ otherwise. Using $V_{\rm t}$ from
Eq.~(\ref{Vt}) and the simple estimates $QR = (1-\alpha')/\alpha$
and $V_{\rm f} = \alpha \Omega R$, we get from
Eq.~(\ref{thickmodel}) the solid line in Fig.~\ref{overshoot}
\textit{(right panel)}, which is in reasonable agreement with
experiment.


The rapid change in the counterflow peak height during the passage
of the vortex front through the detector coil provides a
convenient signal for measuring the propagation velocity $V_{\rm
f}$. Examples of these signals are shown in
Fig.~\ref{FrontVelExp}. They have been measured at different
temperatures to illustrate how the temperature dependence of
$V_{\rm f}$ is expressed in the practical measurement. All three
examples have been measured using the externally triggered
Kelvin-Helmholtz instability to start the turbulent burst. In
Fig.~\ref{FrontVelExp} the vortex fronts have traveled a distance
of $\sim 4\,$cm, before they pass through the detector coil, and
thus have already acquired their steady-state thin-front
configuration.

\begin{figure}[t]
\begin{center}
\centerline{\includegraphics[width=1\linewidth]{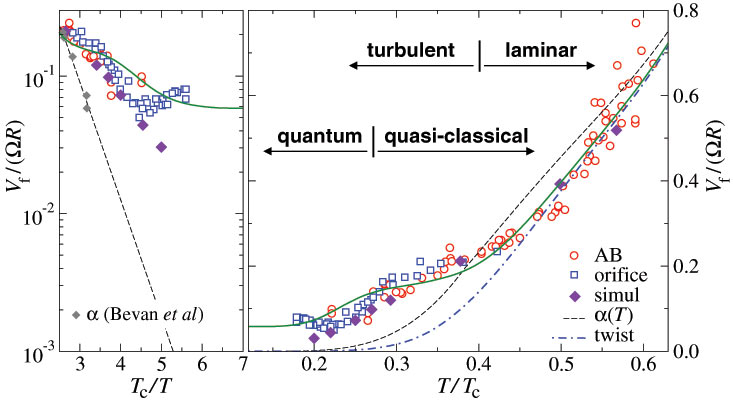}}
\caption{Normalized velocity $V_{\rm f}/ \Omega R$ of vortex front
propagation in a rotating column. All externally controlled
variables are kept constant during this measurement where the
initial state is vortex-free rotation and the final state an
equilibrium array of rectilinear vortices.  \textit {(Main panel)}
The assignments of different hydrodynamic regimes refer to the
dynamics in the front motion. The open circles denote measurements
in which the turbulent burst is started from the AB interfaces in
the middle of the column. The squares refer to the case where the
front is started at the orifice and then moves upward through the
entire column. The large filled diamonds mark results from
numerical calculations. The dashed line represents the mutual
friction dissipation $\alpha (T)$ measured by \cite{Bevan:1997}
and extrapolated below $0.35\,T_{\rm c}$ with exp$(- \Delta /T)$
\citep{delta}. The dash-dotted curve takes into account the
twisted vortex state behind the propagating front
(Eq.~(\ref{Vlam})). The solid curve displays the theoretical model
from Sec.~\ref{s:basicEq} which includes corrections from the
twisted vortex state, turbulent energy transfer, and quantum
bottleneck. \textit {(Left panel)} This semi-log plot shows that
the analytically calculated model provides a reasonable fit to the
low temperature data.} \label{FrontVel}
\end{center}
\end{figure}

Measurements on the front velocity $V_{\rm f}$ are shown in
Fig.~\ref{FrontVel}. As a function of temperature two different
regimes of front propagation can be distinguished, the laminar and
turbulent regimes. The crossover between them is gradual and
smooth. This is in sharp contrast to the sudden onset of bulk
turbulence (as a function of temperature around $\zeta \sim 1$) in
injection measurements in the same circular column, when a bundle
of closely spaced seed vortex loops escapes across the AB
interface in a Kelvin-Helmholts instability event
(Fig.~\ref{RotStates+Injection}). A sharp transition with a
clearly defined critical velocity is the usual case, for instance
in all measurements with mechanical vibrating objects (in the
regime $\zeta > 1$) as a function of drive and flow velocity
\citep{VinenSkrbekThisVolume}. The smooth crossover here in front
propagation may be a special property of the circular column where
vortex polarization along the rotation axis is always $\gtrsim
90\,$\%. Nevertheless, smooth crossovers have been observed
before, for instance from linear to turbulent wave acoustics in
second sound propagation in a circular cylinder as a function of
driving amplitude \citep{Kolmakov:2006}.

Above $ 0.4 T_{\rm c}$ in the laminar regime the results in
Fig.~\ref{FrontVel} are consistent with the earlier measurements
of \cite{old-flight-meas}. Here the single-vortex dynamics apply,
when inter-vortex interactions can be neglected and $V_{\rm f}
\approx \alpha \Omega R$ (Sec.~\ref{SeedEvolution}). At closer
inspection it is noticed that the data for the normalized front
velocity $v_{\rm f} = V_{\rm f}/\Omega R$ lie on average below
$\alpha (T)$ (dashed curve). The reason is that behind the front
the vortices are in the twisted state and not as rectilinear
vortex lines in an equilibrium array. If we integrate the kinetic
energy stored in the twist-induced flow (Eq.~\ref{uniftw}) over
the sample cross section, the result is found to be
\begin{equation}
v_{\rm {f,lam}} = 2[1/\log(1+ \zeta^2) -  \zeta^2]\, \alpha\,.
\label{Vlam}
\end{equation}
After including this reduction from the twisted state, the
agreement is improved (dash-dotted curve). In the turbulent regime
below $ 0.4 T_{\rm c}$, the data deviate with decreasing
temperature more and more above the extrapolations from the
laminar regime. Eventually at the lowest temperatures the
measurements become temperature independent, with a peculiar
transition from one plateau to another at around $0.25\,T_{\rm
c}$. These features are attributed to turbulent dynamics and are
analyzed in the next sections in more detail.

\begin{figure}[t]
\begin{center}
\centerline{\includegraphics[width=0.8\linewidth]{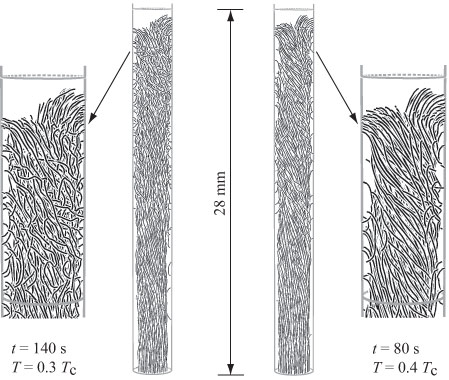}}
\caption{Calculated vortex propagation at $0.3\,T_{\rm c}$ and
$0.4\,T_{\rm c}$. The motion is started from the bottom end of the
cylinder, by placing the equilibrium number of vortices in the
form of quarter loops between the bottom end plate and the
cylindrical side wall.  \textit{On the right} at $0.4\,T_{\rm c}$
the front has traveled for 80\,s to a height $ z \approx 28\,$mm
above the bottom end plate in a cylinder of 3\,mm diameter
rotating at 1\,rad/s. The zoom on the far right shows the vortices
in the front and immediately below in more detail. \textit{On the
left} at $0.3\,T_{\rm c}$ the same distance is covered in 140\,s.
Here the vortices appear more wrinkled, owing to short-wavelength
Kelvin wave excitations. The vortices are tightly twisted below
the front, but become straighter and smoother on approaching the
bottom end plate. Many vortices can be seen to connect to the
cylindrical side wall also below the front. Their average length
is shorter than the distance from the front to the bottom end
plate, although the number of vortices $N(z)$ threading through
each cross section of the cylinder below the front is roughly
constant and comparable to that in the equilibrium vortex state:
$N(z) \lesssim N_{\rm eq}$. The average polarization along the
vertical axis is high, $\sim 90\,$\%.  }
\label{TurbFront03Tc+04Tc}
\end{center}
\end{figure}

The measured properties of the propagating vortex front are
confirmed qualitatively in numerical calculations. They show that
in the laminar regime the thickness of the front grows with time,
but with decreasing temperature the twist increases
(Fig.~\ref{overshoot}) and finally at about $ 0.45\,T_{\rm c}$ the
thin steady-state front configuration is established, with a
time-independent thickness roughly equal to the radius of the
sample. Below $ 0.45\,T_{\rm c}$ the twist decreases with
decreasing temperature, but remains within the limits where the
twist-induced superflow is sufficient to maintain a thin
time-invariant front configuration. The calculated front velocity
in Fig.~\ref{FrontVel} approximately agrees with the measurements
down to about $0.22\,T_{\rm c}$. It is therefore instructive to
analyze the calculations, to identify where and by what mechanisms
turbulent losses occur in the rotating column. This will be
discussed in the next section.

\begin{figure}[t]
\begin{center}
\centerline{\includegraphics[width=0.8\linewidth]{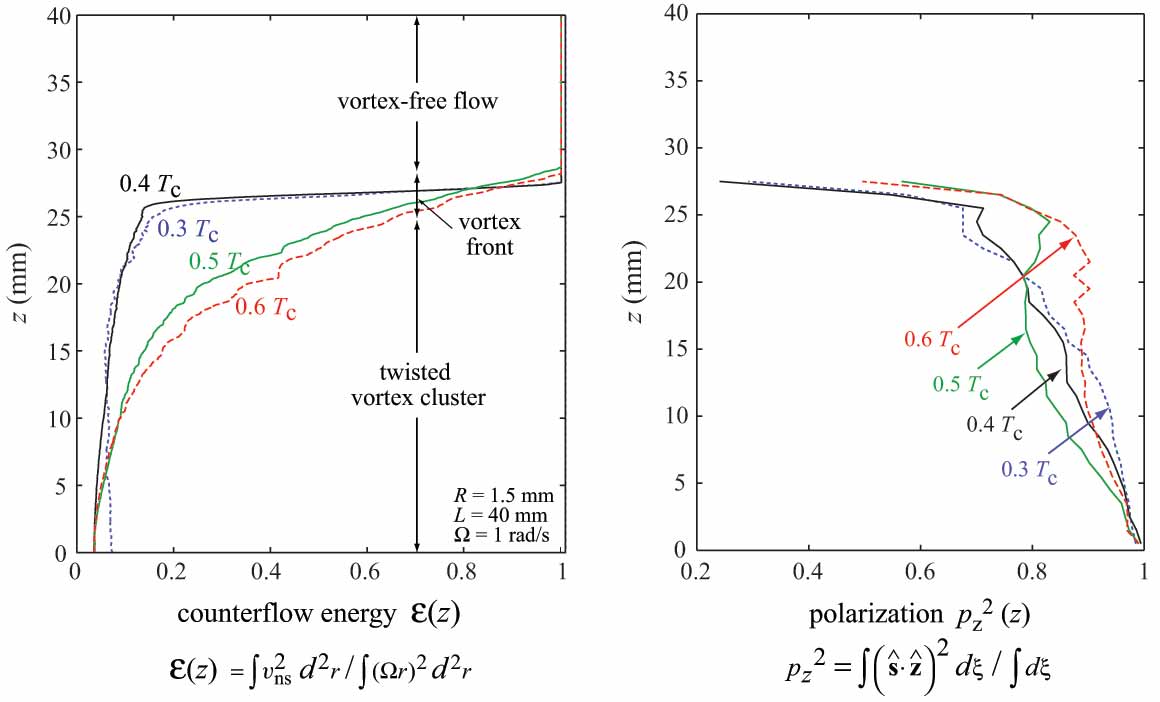}}
\caption{Axial distributions of counterflow energy $\mathcal{E}
(z)$ (see Eq.~(\ref{KineticEnergy})) and polarization $p_z (z)$
(both in normalized units), calculated for the vortex propagation
in Fig.~\ref{TurbFront03Tc+04Tc}. On the left $\mathcal{E} (z)$ is
seen to drop steeply within the narrow vortex front at the
temperatures of 0.4 and $0.3\,T_{\rm c}$. Here the front is in
steady-state time-invariant motion and its velocity provides a
direct measure of dissipation. Above $0.45\,T_{\rm c}$ some
vortices tend to fall more and more behind during the motion and
the shape of the front becomes more extended with time. As seen in
Fig.~\ref{TurbFront03Tc+04Tc}, small-scale structure from Kelvin
waves accumulates increasingly on the vortices below $0.45\,T_{\rm
c}$, but as shown in the panel on the right, the polarization
$p_z(z)$ remains always high behind the front. }
\label{FrontEnergy+Polarization}
\end{center}
\end{figure}

\subsection{\label{NumFront}Numerical calculation of turbulence in
vortex front propagation}

\begin{table}
  \centering

 \begin{tabular}{||c||c|c||}
   \hline
    $T/T_c$   & 0.3 & 0.4 \\   \hline
    Mutual  friction parameter $\alpha$& 0.040 & 0.18 \\
   Mutual  friction parameter $\alpha'$& 0.030 & 0.16 \\
Front velocity $V\sb{f}$  & ~$0.12\,\Omega R$~ & ~$0.22\, \Omega R$~ \\
Front velocity $V\sb{f}$ (mm/s) & ~$0.18$~ & ~$0.33$~ \\
 ~Total Reconnection rate   (event/s)~~ & $~300$ & $~130 $ \\  
 Number of vortex lines/cross section & ~$125$~ & ~$150$~ \\
 Reconnection rate   per line $1/\tau$ [events/(line s)] & 2.4 & 0.87 \\  
     Front shift $\delta$ during $\tau$ (mm) & 0.075 & 0.38 \\ 
     Interline separation $\ell $ (mm) & 0.20& 0.19 \\ 
     Ratio $\delta /\ell  $ & 0.36  & 2.0 \\
 \hline
 \end{tabular}
 \vspace{5mm}
\caption{Comparison of vortex front propagation at $0.3\,T_{\rm
c}$ and $0.4\,T_{\rm c}$. The bottom line of the table shows that
below $0.4\,T_{\rm c}$ front propagation is rapidly moving into
the quantum regime. The calculations are for a cylinder of radius
$R = 1.5\,$mm and length $h = 40\,$mm, which is filled with
$^3$He-B at a liquid pressure of $P = 29\,$bar and rotates at an
angular velocity $\Omega = 1\,$rad/s. The number of vortex lines
is the average through each cross section of the cylinder over the
length of the twisted cluster. The maximum resolution of the
calculations is $0.05$\,mm.}\label{t:1}
\end{table}

\begin{figure}[t]
\begin{center}
\centerline{\includegraphics[width=0.6\linewidth]{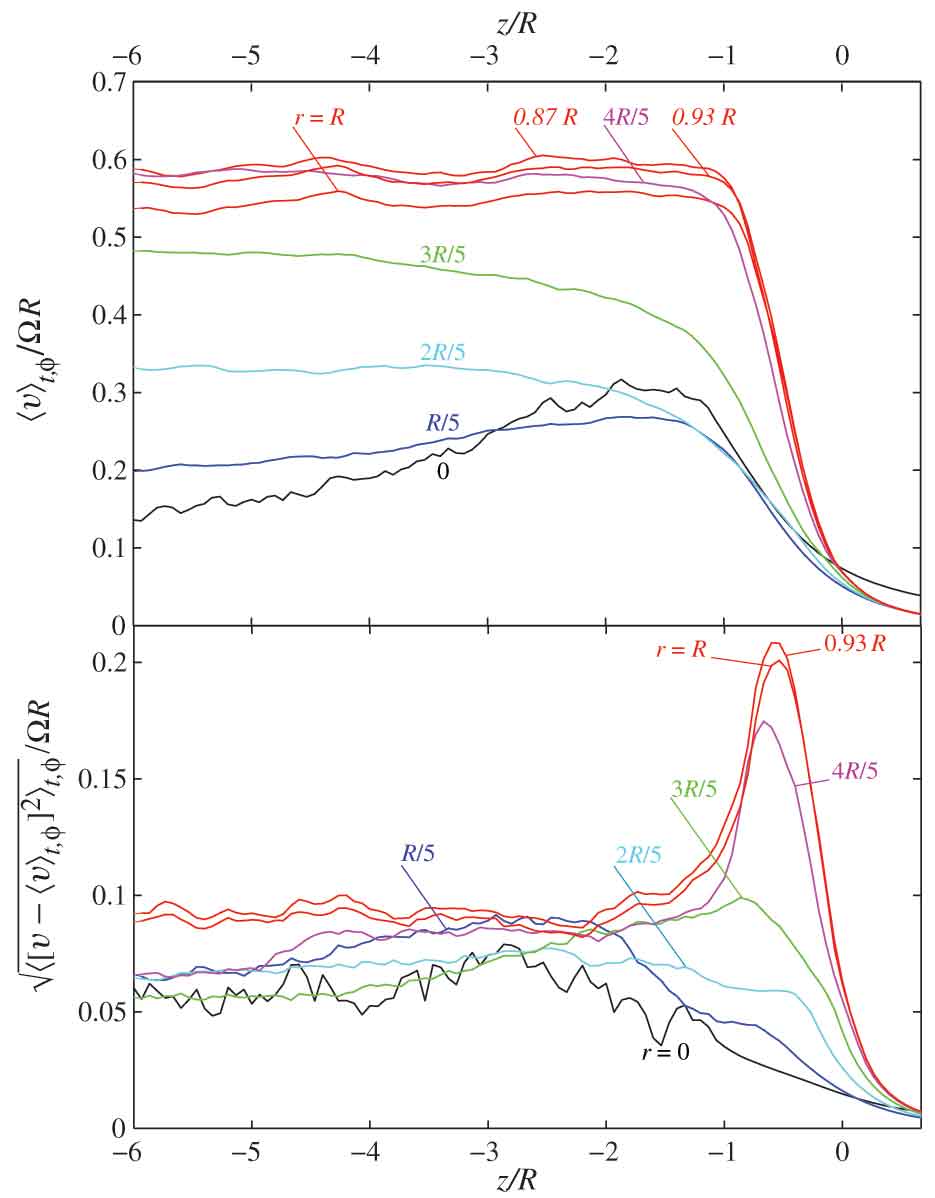}}
\caption{Axial and radial distributions of the velocity $\mid \!\B
{v}_{\rm s} (r, \phi, z, t)\! \mid$ and its fluctuations in vortex
front propagation. Both quantities are normalized with respect to
$\Omega R$ and are expressed in the laboratory frame in terms of
their averages, $\langle v_{\rm s}\rangle_{t,\phi}$ \textit{(top
panel)} and $[\langle (v_{\rm s} - \langle v_{\rm
s}\rangle_{t,\phi})^2\rangle_{t,\phi}]^{1 \over 2}$
\textit{(bottom panel)}, by integrating over the azimuthal
coordinate $\phi$ and time $t$. The different contours are plotted
as a function of $z$ at fixed radial value, $r =0, \, R/5, \,
2\,R/5, ...$. They are calculated for the example at $0.3\,T_{\rm
c}$ in Fig.~\ref{TurbFront03Tc+04Tc},  over the time interval from
60 to 80\,s, after starting the front motion. At $t = 80\,$s  the
front has climbed to a height $z(80\,{\rm s}) \approx 16\,$mm
(here placed at $z/R = 0$). Both plots are generated from vortex
configurations, which are saved every 0.5\,s. The steep change in
the top panel in the interval $-1 < z/R < 0$ signifies the front
with its narrow thickness $\Delta z \sim R$. Similarly in the
bottom panel the amplitude of fluctuations (sampled at 2\,Hz)
reaches a sharp maximum in the front region $-1 < z/R < 0$ at
large radii $r \gtrsim 3\,R/5$. This peak is twice larger than the
flat values from behind the front in the region of the twisted
cluster.  At $0.4\,T_{\rm c}$ (the case of
Fig.~\ref{TurbFront03Tc+04Tc} right) the fluctuation peak from the
front is 14\,\% lower. } \label{TotalVel+Fluct}
\end{center}
\vspace{-4mm}
\end{figure}

\begin{figure}[t]
\begin{center}
\centerline{\includegraphics[width=0.6\linewidth]{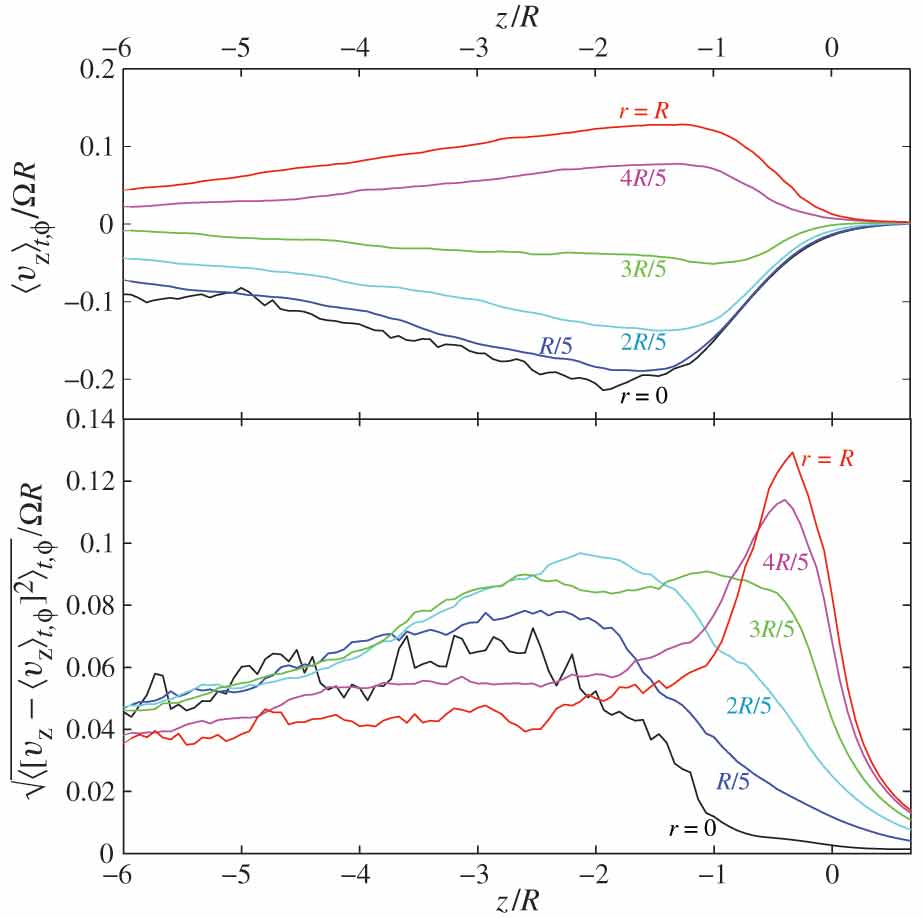}}
\caption{Axial and radial distributions of the axial velocity
component $v_{{\rm s}z} (r,\phi,z,t)$ and its fluctuations. The
profiles depict the same calculations at $0.3\,T_{\rm c}$ as
Fig.~\ref{TotalVel+Fluct}. This component is generated by the
helically wound sections in the twisted cluster. The slowly
unwinding twist at the bottom end plate (on the left of the
figure) causes the characteristic linear increase in $\mid \!
v_{{\rm s}z} \! \mid$ towards the right where new twist is
continuously formed by the spirally winding motion of the vortex
front. Note that $v_{{\rm s}z}$ changes sign, since it is directed
antiparallel to the propagation direction of the front in the
central parts $r \lesssim 2\,R/3$ and  parallel at larger radii.
Its maximum magnitude is in the center just behind the front where
it is the dominant component in $\mid \!\B {v}_{\rm s} \! \mid$.
Its largest fluctuations are at large radii within the front
region.} \label{AxialVel+Fluct}
\end{center}
\vspace{-4mm}
\end{figure}

\begin{figure}[t]
\begin{center}
\centerline{\includegraphics[width=0.6\linewidth]{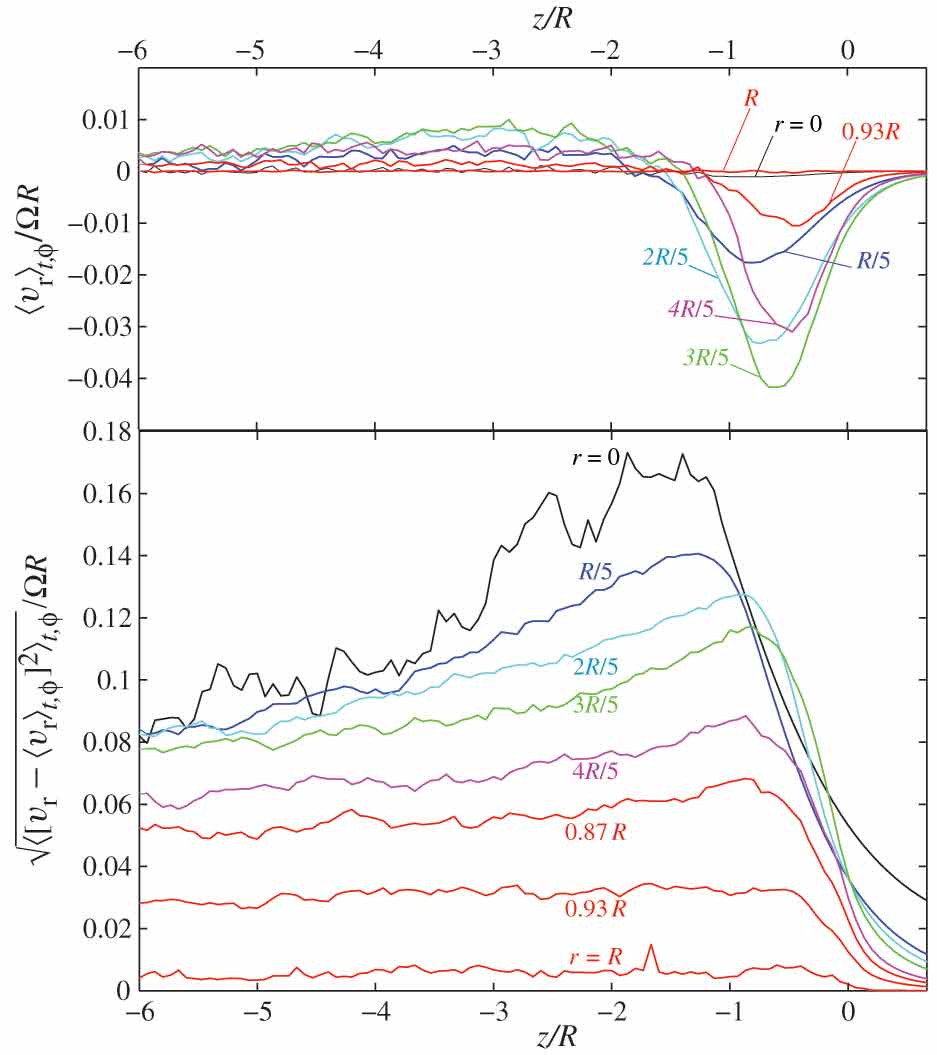}}
\caption{Axial and radial distributions of the radial velocity
component $v_{{\rm s}r} (r,\phi,z,t)$ and its fluctuations in
vortex front propagation at $0.3\,T_{\rm c}$ in
Fig.~\ref{TurbFront03Tc+04Tc}. This component provides the return
currents to the axial component $v_{{\rm s}z}$. It is an order of
magnitude smaller than either the azimuthal $v_{{\rm s} \phi}$ or
the axial $v_{{\rm s} z}$ component. It is directed inward towards
the central axis in the narrow front region, while the outward
flowing radial return currents are distributed more evenly along
the length of the twisted cluster. The fluctuations in $v_{{\rm s}
r}$ are largest  close to the cylinder axis and steadily decrease
as a function of $r$. As a function of $z$ the maximum
fluctuations are just behind the front where $v_{{\rm s} r}$
changes sign. } \label{RadialVel+Fluct}
\end{center}
\vspace{-4mm}
\end{figure}

\begin{figure}[t]
\begin{center}
\centerline{\includegraphics[width=0.6\linewidth]{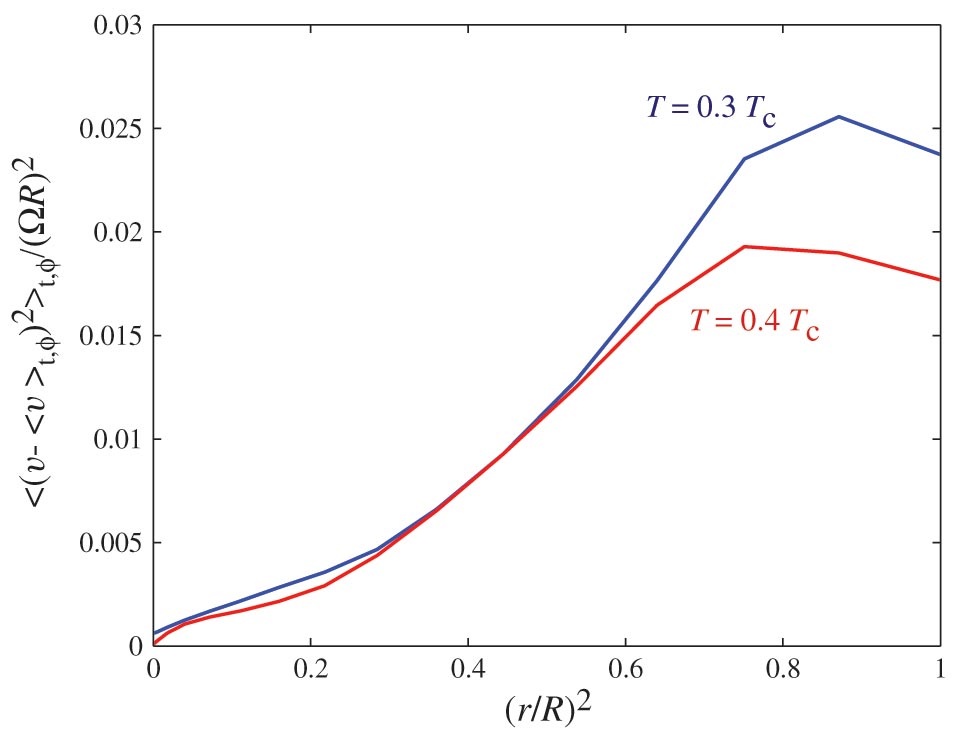}}
\caption{Radial distribution of velocity fluctuations within the
propagating vortex front from Fig.~\ref{TotalVel+Fluct}.  The
normalized square of the velocity deviation from the average is
plotted on the vertical scale, $\langle (v_{\rm s} - \langle
v_{\rm s} \rangle_{t, \phi})^2 \rangle_{t, \phi} /(\Omega R)^2$,
versus the square of the normalized radial position, $(r/R)^2$.
The averages in the angled brackets, besides having been
integrated over time $t$ and azimuthal coordinate $\phi$ (or front
motion), represent an average over the interval $ -1 < z/R < 0$.
The fluctuations are seen to increase rapidly towards large radii,
up to where the vortex-free annulus starts at $R_{\rm v} \approx
0.87 \,R$. } \label{RadVelFluctuation}
\end{center}
\end{figure}

\begin{figure}[t]
\begin{center}
\centerline{\includegraphics[width=0.5\linewidth]{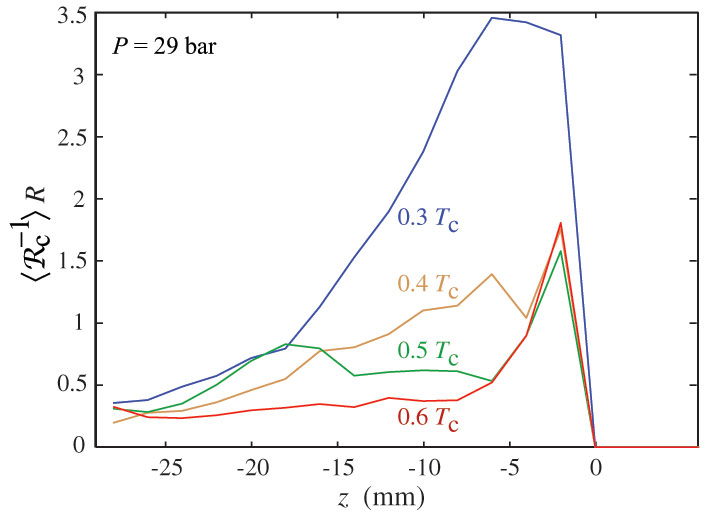}}
\caption{Distribution of curvature radii along vortices during
front propagation, calculated for the setup in
Fig.~\protect\ref{TurbFront03Tc+04Tc} \textit. The curvature is
defined as the inverse of the local vortical curvature radius
$\cal{R}_{\rm c}$, \textit{i.e.} $\cal{R}$$_{\rm c}^{-1} = |d^2
s(\xi,t)/d\xi^2|$, where $\xi$ is the coordinate along the vortex
core. The curvature is calculated on an equidistant linear grid
along each line vortex. On the vertical axis the average curvature
$\langle {\cal{R}}^{-1}_{\rm c} \rangle $ is plotted as a
dimensionless quantity $\langle {\cal{R}_{\rm c}}^{-1} \rangle
\,R$ as a function of $z$ at different temperatures. The averages
are taken over the cross section of the cylinder and over an
interval $\Delta z = 2\,$mm in length centered at $z$.  The sharp
peak at the very front ($z \approx -2\,$mm) at temperatures $T
\geq 0.4\,T_{\rm c}$ represents the vortices curving to the
cylindrical side wall with $\cal{R}_{\rm c}$ $\approx R$.  In
contrast, at $0.3\,T_{\rm c}$ small-scale curvature dominates, the
average radius of curvature has dropped to $\cal{R}_{\rm c}$
$\approx 0.4\,$mm and extends from the front well inside the
twisted cluster. } \label{CurvatureDistribution}
\end{center}
\end{figure}

\begin{figure}[t]
\begin{center}
\centerline{\includegraphics[width=0.6\linewidth]{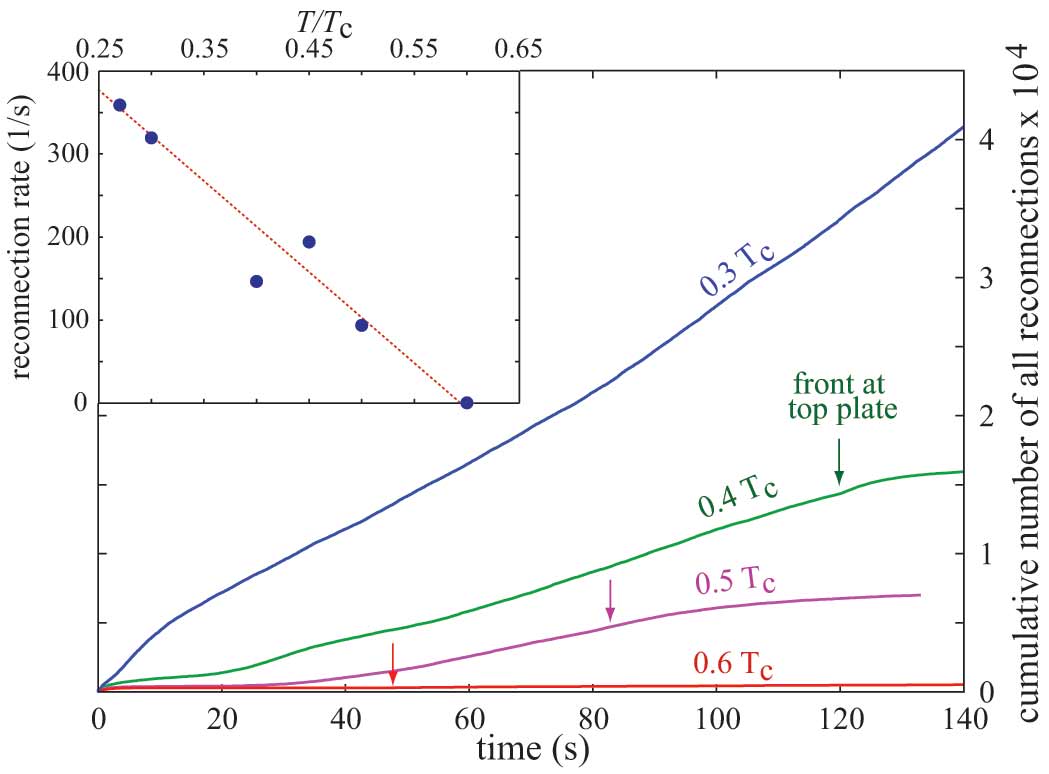}}
\caption{\textit{(Main panel)} Cumulative number of all vortex
reconnection events during vortex front propagation in the setup
of Fig.~\protect\ref{TurbFront03Tc+04Tc}, as a function of time.
The reconnections occur between two different vortices and mainly
in the twisted cluster where they reach a maximum close behind the
front. This process does not change the number of vortices. The
vertical arrows indicate the moment when the front reaches the top
end plate (at height $h = 40\,$mm). The reconnections are here
seen to increase rapidly at low temperatures, reaching a rate of
300\,s$^{-1}$ or $\sim 1$ reconnection/(vortex sec) at
$0.3\,T_{\rm c}$. \textit{(Inset)} The roughly constant
reconnection rate from the main panel (when the front is in
steady-state propagation before the end plate is reached), plotted
as a function of temperature. } \label{Reconnections}
\end{center}
\end{figure}

The promising agreement of the calculated vortex front velocity
$v_{\rm f}$ with the measurements in Fig.~\ref{FrontVel} suggests
that numerical calculations could provide useful guidance for the
interpretation of the measurements in
Sec.~\ref{FrontVelocityMeasurement} and for constructing the
analytic theory in Sec.~\ref{s:basicEq}. Our main question is the
following: Why are the numerical results on $v_{\rm f}$ deviating
with decreasing temperature more and more above the
mutual-friction controlled extrapolation from the laminar regime?
For simplicity we split the discussion of vortex motions to three
different length scales: (i) large scales of order $\sim R$, where
turbulent fluctuations become visible as variations in the number
and distribution of vortices, (ii) the inter-vortex scale $\sim
\ell$, where the presence of Kelvin-waves on individual vortex
lines can be seen to grow, and (iii) small scales, where
reconnections between neighboring vortices might occur and excite
turbulent fluctuations on individual vortex lines. In contrast to
the more usual studies of turbulent tangles, which generally
monitor their free decay as a function of time when the external
pumping is switched off, we are here dealing with highly polarized
steady-state motion along the rotating column. This motion appears
to remain intact down to the lowest temperatures, but becomes
dressed with turbulent fluctuations in growing amounts towards
decreasing temperatures.

Two examples are shown in Fig.~\ref{TurbFront03Tc+04Tc} which
illustrate the development with decreasing temperature. These
configurations at $0.4\,T_{\rm c}$ and $0.3\,T_{\rm c}$ in an
ideal column of radius $R = 1.5\,$mm rotating at $\Omega =
1\,$rad/s have been calculated using the techniques described in
Sec.~\ref{Simulation}. Similar results at $0.4\,T_{\rm c}$ and
higher have been reported in the review by \cite{Finne:2006b}. The
calculations are started from an initial configuration with
roughly the equilibrium number of vortices placed  as
quarter-loops between the bottom end plate and the side wall of
the cylinder.  During the subsequent evolution the propagating
vortex front is formed and the twisted cluster starts to acquire
its shape ~\citep{Hanninen:2006}.

Above $ 0.45 \, T\sb c$  the motion is laminar with relatively
smooth vortices, which only twist at large length scales. Below
$0.45 \, T\sb c$ the thickness of the front (in the axial
direction) settles at $\Delta (r)\simeq r \,d$, where the
parameter $d\sim 1$. The large-scale characteristics as a function
of temperature from 0.3 to $0.6\,T_{\rm c}$ are displayed in
Fig.~\ref{FrontEnergy+Polarization} for the setup of
Fig.~\ref{TurbFront03Tc+04Tc}. Here the counterflow energy
$\mathcal{E}(z)$ is obtained by integrating the momentary
distribution of counterflow $\mathbf{v} (r, \phi,z,t)$ over each
cross section $z$ of the column, while the polarization $p_z (z)$
is similarly derived by integrating over the vortices threading
through this cross section. With decreasing temperature the front
acquires more and more turbulent features which become visible as
increasing small-scale structure: Kelvin waves, kinks, and
inter-vortex reconnections. These small scale fluctuations exist
on top of a strongly polarized vortex orientation, which is
preserved even at $0.2 \, T\sb c$. The high polarization along the
rotation axis and the organized configuration of the twisted
vortex cluster is expected to reduce loop formation and
self-reconnection on individual vortex lines, but it also
suppresses reconnections between neighboring vortices from what
one would find in turbulent tangles. The amount of twist, in turn,
is reduced by the unwinding from the slip of the vortices along
the flat bottom end plate of the column.

The comparison of the two examples in
Fig.~\ref{TurbFront03Tc+04Tc} is continued further in
Table~\ref{t:1}. As a rule we expect that with decreasing
mutual-friction damping turbulent disturbances are expected to
cascade down to ever smaller length scales. In Table~\ref{t:1}
this is examined by defining the quantity $\delta$ which measures
on an average the distance over which the vortex front moves
during the time interval between two inter-vortex reconnections.
Since the axial motion of the front slows down and the
reconnection rate increases with decreasing temperature, $\delta$
decreases rapidly below $0.4\,T_{\rm c}$. Thus $\delta$ provides a
measure between mutual friction and reconnections. The bottom line
in Table~\ref{t:1} uses the dimensionless ratio $\delta / \ell$ to
characterize the relative importance of mutual friction
dissipation to reconnection losses. When $\delta/\ell > 1$, the
energy loss from reconnections, which presumably excite Kelvin
waves, is not important in the total energy balance of the
propagation. At $0.4\, T\sb c$, where $\delta/\ell \sim 2$, this
is what one expects. If $\delta/\ell<1$, as is the case below
$0.3\, T\sb c$, then it becomes possible that inter-vortex
reconnections might play a role in the total energy balance.
Table~\ref{t:1} thus hints that new phenomena are expected to
emerge on length scales approaching the inter-vortex distance
$\ell$.

The characteristics of the propagating vortex front and the
trailing twisted cluster are further illustrated in
Figs.~\ref{TotalVel+Fluct}  -- \ref{RadialVel+Fluct}. In
Fig.~\ref{TotalVel+Fluct} it is the modulus of the total velocity
of the superfluid component, $\mid \!\B {v}_{\rm s} (r, \phi, z,
t)\! \mid$, whose $[r,z]$ profiles are displayed. The top panel
shows the velocity $\langle v_{\rm s}\rangle_{t,\phi}$ in the
laboratory coordinate frame, averaged over time $t$ and azimuthal
angle $\phi$, while the bottom panel displays its mean fluctuation
amplitude $[\langle (v_{\rm s} - \langle v_{\rm
s}\rangle_{t,\phi})^2\rangle_{t,\phi}]^{1 \over 2}$.  In
Fig.~\ref{AxialVel+Fluct} it is the axial component $v_{{\rm s}
z}$ and in Fig.~\ref{RadialVel+Fluct} the radial component
$v_{{\rm s} r}$ which are analyzed in the same fashion. The
largest component $v_{{\rm s} \phi} $ is omitted, since its
profiles are so similar to those of $\mid \!\B {v}_{\rm s} \!
\mid$. In fact, at large radii $r \gtrsim R/5$ the profiles of the
two velocities as a function of $[r,z]$ are almost
indistinguishable, while at small radii $r \lesssim R/5$ the axial
velocity $v_{{\rm s} z}$ in Fig.~\ref{AxialVel+Fluct} is the main
contributor to $\mid \!\B {v}_{\rm s}\! \mid$.

The vortex front becomes clearly defined in
Figs.~\ref{TotalVel+Fluct} -- \ref{RadialVel+Fluct}. For instance,
in Fig.~\ref{TotalVel+Fluct} the steep almost linear rise in $\mid
\!\B {v}_{\rm s} \! \mid$ in the interval $-1 < z/R < 0$ signifies
the transition from the non-rotating state $v_{{\rm s}\phi} = 0$
at $z > 0$ to almost equilibrium rotation at $v_{{\rm s}\phi}
\approx \Omega r$ at $z < - R$. This is the vortex front with its
narrow thickness $\sim R$ and strong shear flow, created by the
vortices terminating within the front on the cylindrical side wall
perpendicular to the axis of rotation. Below the front the number
of vortices threading through any cross section of the cylinder
remains roughly constant. This is seen from the fact that at
$0.3\, T_{\rm c}$  the maximum value of $v_{\rm s}$, $\langle
v_{{\rm s} \phi} \rangle_{\rm max} \approx 0.6\, \Omega R$ remains
stable over the entire length of the twisted cluster, starting
from $z/R < -1 $ immediately behind the front. At higher
temperatures some vortices tend to fall behind the front and thus
at $0.4\,T_{\rm c}$ one finds that $v_{\rm s}$ slowly increases to
$\langle v_{{\rm s} \phi} \rangle_{\rm max} \approx 0.75\, \Omega
R$ at $z/R < -5$. Note also that radially the azimuthal flow
$v_{{\rm s} \phi}$ increases monotonously behind the front up to
the edge of the cluster at $\sim 0.9\,R$ and then slightly
decreases towards the cylindrical wall.

The fluctuations of $\mid \!\B {v}_{\rm s} \! \mid$ around its
mean value, as shown in the bottom panel of
Fig.~\ref{TotalVel+Fluct}, are substantial, of order 30\,\% in the
sharp peak created by the vortex front. The same applies to other
velocity components, their fluctuations also tend to be largest in
the front region $-1 < z/R < 0$ and typically only half as large
over the length of the twisted cluster. In
Figs.~\ref{TotalVel+Fluct} -- \ref{RadialVel+Fluct} the
fluctuations are sampled at a relatively slow rate of 2\,Hz. As
seen in Fig.~\ref{TurbFront03Tc+04Tc}, in the front the vortices
are not perfectly distributed: It is this disorder in the
structure of the front and the variations in its number of
vortices which gives rise to the fluctuation peak. In
Fig.~\ref{RadVelFluctuation} the radial distribution of the
fluctuations in $\mid \!\B {v}_{\rm s} \! \mid$ in the front
region $-1 < z/R < 0$ is analyzed. As expected, they grow rapidly
towards large radii. In Sec.~\ref{s:basicEq} we make use of this
radial distribution.

However, not only large-scale disorder contributes to velocity
fluctuations, but also Kelvin waves start to expand on individual
vortices at temperatures $\lesssim 0.3\,T_{\rm c}$, as seen in
Fig.~\ref{TurbFront03Tc+04Tc} \citep{Hanninen:2006}. In
Fig.~\ref{CurvatureDistribution} a momentary vortex configuration
(when the front has reached the height $z \approx 28\,$mm in
Fig.~\ref{TurbFront03Tc+04Tc}) is broken down in curvature radii
$\cal{R}$$_{\rm c}$ and the average curvature $\langle
\cal{R}$$_{\rm c}^{-1} \rangle$ is plotted as a function of $z$.
At $T \geq 0.4\,T_{\rm c}$ the sharp peak at the front is caused
by the vortices curving to the cylindrical side wall with
$\cal{R}_{\rm c}$ $\approx R$. However, at $0.3\,T_{\rm c}$ the
Kelvin wave contribution at shorter length scales becomes dominant
and the average radius of curvature drops to $\langle \cal{R}_{\rm
c} \rangle$ $\approx 0.4\,$mm. The characteristic Kelvin wave
frequency $\omega_{\rm k} \sim \kappa k^2$ corresponds at this
length scale to 0.4\,Hz (which is less than the sampling frequency
of 2\,Hz). Thus the dominant Kelvin waves are included in the
velocity fluctuations in Figs.~\ref{TotalVel+Fluct} --
\ref{RadialVel+Fluct}. Nevertheless, in these figures Kelvin waves
represent only a small part of the total fluctuations $\langle
(v_{\rm s} - \langle v_{\rm s}\rangle_{t,\phi})^2\rangle_{t,\phi}$
in the sharp peak created by the front. This is seen from the fact
that the peak is localized in the $z$ direction within the vortex
front at $-1 < z/R < 0$, while the Kelvin-wave excitations in
Fig.~\ref{CurvatureDistribution} extend half way down the twisted
cluster to $z/R \sim -10$, as can also be directly seen from
Fig.~\ref{TurbFront03Tc+04Tc}.

Fig.~\ref{Reconnections} shows that also reconnections between
neighboring vortices become rapidly more frequent at temperatures
$\lesssim 0.3\,T_{\rm c}$ (see also \cite{Hanninen:2006}). The
reconnections peak in the region around $z/R \lesssim -1$, where
the front ends and the twisted cluster starts. The increasing
reconnection rate with decreasing temperature is one factor which
helps to reduce the helical twist behind the front and might be
one reason why the measured twist appears to be reduced towards
low temperatures below $0.45\,T_{\rm c}$ (Fig.~\ref{overshoot}).

A further characteristic of front propagation in the rotating
column is that fluctuations in vortex length are small, since
these are mainly restricted to the transverse plane. The total
vortex length $L(t)$ in the moving front and in the twisted
cluster behind it increases linearly in time and thus the
fluctuations can be expressed with respect to a linearly
increasing fitted average $L_{\rm fit} (t)$. The average deviation
from the mean, $\langle \Delta L \rangle = [\langle (L - L_{\rm
fit})^2 \rangle]^{1 \over 2}$ turns out to be small: In the
conditions of Fig.~\ref{TurbFront03Tc+04Tc} at $0.3\,T_{\rm c}$
one obtains $\Delta L /L \sim 10^{-3}$. This is as expected since
fluctuations in length are energetically expensive in polarized
vortex motion.

Finally, we point out two features which are seen in the
calculations, which both grow in prominence with decreasing
temperature below $0.3\,T_{\rm c}$, but which have not yet been
searched for in measurements. The calculations are started with
the number of seed vortices equal to or larger than $N_{\rm eq}$,
but nevertheless, the number of vortex lines does not remain
stable at $N_{\rm eq}$ during steady-state
propagation.\footnote{In the equilibrium vortex state $N_{\rm eq}
\approx \pi (R-d_{\rm eq})^2 \,n_{\rm v}$, where the width of the
vortex-free annulus around the central cluster is $d_{\rm eq}
\approx n_{\rm v}^{-1/2}$ \citep{Ruutu:1998b}.}

The first feature concerns the number of vortices which in free
steady-state propagation thread through any cross section of the
column behind the front. This number is roughly constant, but less
than in equilibrium (for instance in Fig.~\ref{TurbFront03Tc+04Tc}
at $0.3\,T_{\rm c}$, there are about 130 vortices per cross
section while $N_{\rm eq} \approx 160$ vortices). In the simplest
model of the twist in Eq.~(\ref{uniftw}) the number of vortices in
the twisted state is smaller than in the equilibrium vortex state
by the factor $\approx (QR)^2 /\left([1 + (QR)^2] \;
\ln{[1+(QR)^2]}\right)$. This reduction is of correct order of
magnitude compared to that seen in the calculations at $\gtrsim
0.3\,T_{\rm c}$. In addition to improving the stability of the
twisted state, the reduction in the number of twisted vortices
decreases the axial flow velocity $v_{{\rm s} z}$
(Fig.~\ref{AxialVel+Fluct}), which has the effect of reducing the
longitudinal front velocity $v_{\rm f}$ and counteracts its
increase above the laminar extrapolation in Fig.~\ref{FrontVel}.
While the number of vortices per cross section of the twisted
cluster decreases with decreasing temperature below $0.3\,T_{\rm
c}$, the density of vortices, nevertheless, remains constant over
the cross section (but less than the equilibrium value: $n_{\rm v}
\lesssim n_{\rm v,eq} = 2 \Omega / \kappa$). When the front
finally reaches the upper end plate in
Fig.~\ref{TurbFront03Tc+04Tc} and the twist starts to relax,
simultaneously $N$ gradually recovers and approaches $N_{\rm eq}$
from below.

The second feature of the calculations is that in steady state
propagation the average length of the vortices is less than that
of the twisted cluster (which means that the total number of
individual vortices in the column may be well above $N_{\rm eq}$).
For instance at $0.3\,T_{\rm c}$ in Fig.~\ref{TurbFront03Tc+04Tc},
the average length is $\sim 15\,$mm and only $\sim 6\,$mm for
those vortices with both ends on the cylindrical side wall.
Nevertheless, the polarization of all vortices along
$\hat{\mathbf{z}}$ is high, $\sim 90\,$\%. After the front has
reached the upper end plate in Fig.~\ref{TurbFront03Tc+04Tc} and
the twist starts relaxing, the short vortices provide an ample
storage from which to add more vortices so that $N_{\rm eq}$
is reached.

To summarize the calculations, we return to our starting point,
namely the increasing deviation of the calculated front velocity
above the laminar extrapolation below $0.4\,T_{\rm c}$
(Fig.~\ref{FrontVel}). The analysis of the calculated vortex
configurations shows increasing turbulent disorder at large length
scale $\sim R$, growing Kelvin-wave amplitudes on individual
vortices on scales $\sim \ell$ and the presence of inter-vortex
reconnections. Combined these changes from the increasing
turbulent influence with decreasing temperature make up for the
difference, in the presence of a still finite mutual friction
dissipation. An additional sink of energy is the cascading of
Kelvin-wave excitations below the resolution limit of the
numerical calculations (usually $\sim 0.05\,$mm), which are lost
from the energy balance. The calculations are in reasonable
agreement with measurements down to $0.3\,T_{\rm c}$, but at lower
temperatures changes in the twisted cluster propagation appear to
occur, to maintain stability and polarization. These have not been
adequately studied and may lead to revisions of the current model
at the very lowest temperatures.

\subsection{\label{s:basicEq} Analytical model of turbulent front}

The measured front velocity in Fig.~\ref{FrontVel} displays below
$0.4\,T_{\rm c}$ two plateaus which are separated by a transition
centered at $0.25\,T_{\rm c}$. The numerical analysis in
Sec.~\ref{NumFront} places this transition in the temperature
region where the Kelvin-wave cascade acquires growing importance,
\textit{i.e.} in the regime where sub-intervortex scales start to
influence energy transfer. The peculiar shape of the measured
$v_{\rm f}(T)$ curve prompted \cite{LNR} to examine energy
transfer from the quasi-classical length scales (where bundles of
vortices form eddies of varying size with a Kolmogorov spectrum)
to the quantum regime (where Kelvin waves expand on individual
vortex lines). The result is a bottleneck model \citep{LNR2} which
matches energy transfer across the crossover region from
super-intervortex length scales to sub-intervortex scales. It can
be fitted with reasonable parameters to the measured $v_{\rm
f}(T)$ curve \citep{Eltsov:2007}.

\subsubsection{\label{ss:2fluids} Two-fluid coarse-grained equations}

To describe the superfluid motions on length scales exceeding the
inter-vortex distance $\ell$ we make use of  the coarse-grained
hydrodynamic equation for the superfluid component. This equation
is obtained from the Euler equation for the superfluid velocity
${\bf U}\equiv {\bf U}_{\rm s}$,  after averaging over the vortex
lines which are assumed to be locally approximately aligned,
forming vortex bundles which mimic the eddies in viscous flow (see
the review by \cite{Sonin}),
\begin{equation}
\frac{\p \B U}{\p t}+(\B U \cdot \B \nabla) \B U+ \B \nabla \mu=
\B f\sb{mf}\,. \label{SH1}
\end{equation}
Here   $\mu$ is the chemical potential and $ \B f\sb{mf}$ the
mutual friction force~\eq{MutFrictionForce}, which can be
rewritten as follows:
\begin{equation}
\B  f\sb{mf}= -\alpha'({\bf U} -{\bf U}_{\rm n})\times \B \omega+
\alpha~\hat{\B  \omega}\times[\B \omega \times({\bf U} -{\bf
U}_{\rm n}) ] ~. \label{SH2}
\end{equation}
We use $\B \omega=\nabla\times {\bf U}$ for the superfluid
vorticity; $\hat{\B \omega}\=\B \omega/\omega$ is the unit vector
in the direction of the mean vorticity; ${\bf U}_{\rm n}$ is the
velocity of the normal component (which is fixed). In flow with
locally roughly aligned vortices the mutual friction parameters
define the reactive ($\propto \a'$) and dissipative ($\propto \a$)
forces acting on a bunch of vortex lines, as it moves with respect
to the normal component in a field of slowly varying vortex
orientation.

We shall work in the rotating reference frame where ${\bm U}_{\rm
n}=0$. In this frame, the reactive  first term in Eq.~(\ref{SH2})
renormalizes the inertial term ${\bm U} \times \B \omega$ on the
left hand side (LHS)  of Eq.~(\ref{SH1}), introducing the factor
$1-\alpha'$ \citep{Finne:2003}. The relative magnitude of the two
non-linear terms, the ratio of the inertial term and of the
friction term [the latter is the second term on the RHS of
Eq.~(\ref{SH2})], is the Reynolds number in this hydrodynamics. It
proves to be the flow velocity independent ratio of the
dimensionless mutual friction parameters: $ \zeta
=(1-\alpha')/\alpha$ (which was introduced by \cite{Finne:2003} in
the form $q =1/\zeta$). To arrive at a qualitative description of
the front we follow \cite{Lvov:2004} and simplify the vectorial
structure of the dissipation term  and average \Eq{SH2} over the
directions of the vorticity $\B \omega$ (at fixed direction of the
applied counterflow velocity $\B v$). With the same level of
accuracy, omitting a factor $2/3$ in the result, we get
\begin{eqnarray}\label{V7} 
\B  f\sb{mf} \Rightarrow \< \B  f\sb{mf} \>  _{\B \omega /
|\omega|}&=&
  -   \frac23  \,\a\, \omega \sb{eff}\, \B v\Rightarrow
- \,\a\,  \omega\sb{eff}\,\B v \,, \br
 \omega\sb{eff}  &\equiv& \sqrt { \<
|\omega|^2\>}\,.
\end{eqnarray}
Here and henceforth $\< \dots \>$ denotes ``ensemble averaging" in
the proper theoretical meaning.  From the experimental point of
view $\< \dots \>$ can be considered as time averaging in a frame
which moves with the front (over a time interval, during which the
front propagates a distance equal to its width). The resulting
mean values, to be discussed in \Eqs{V8} and \eq{defs}, can be
considered as functions of (slow) time and space. We follow
\cite{Lvov:2004} and neglect the fluctuating turbulent part of $
\omega\sb{eff}$ in \Eq{V7}. We thus replace $|\omega|$ by its mean
value, the vorticity from rotation. In this approximation \Eq{V7}
takes the simple form:
\begin{equation}\label{V8} 
\B  f\sb{mf}  = - \Gamma\, \B U \,, \quad  \Gamma \, \equiv \alpha
\, \omega\sb{eff} \,.
\end{equation} 

\subsubsection{\label{ss:geometry}
Flow geometry, boundary conditions and Reynolds decomposition}
\vskip .3cm

\paragraph{\emph{a.} Axial flow geometry and flat idealization.}
As a simplified model of the rotating cylinder we consider first
the flat geometry with $\bf \hat x$ as the  stream-wise, $\bf \hat
y$ as the cross-stream, and  $ \bf \hat z$ as the front-normal
directions. A more realistic axial symmetry will be discussed
later. We denote the  position of the front with $\C Z(t)$ and
look for a stationary state of front propagation at constant
velocity $- V\sb f$: 
\begin{equation}\label{SteadyFrontMotion} \C Z(t)= -V\sb {\,f}\,  t\ .
\end{equation}

The normal velocity is zero everywhere, $\B U_n(\B r,t)=0$, which
corresponds to co-rotation with the rotating cylinder. The vortex
front propagates in the region $z< \C Z(t)$, where the superfluid
velocity is $-V_\infty$, while far away behind the front at $z >
\C Z(t)$ the superfluid velocity $\B U$ tends to zero (i.e.
towards the same value as the normal fluid velocity).

\paragraph{\emph{b.} Reynolds decomposition and definitions of the model.} 
Following the customary tradition [see, \textit{e.g.}
\cite{Pope}], we decompose the total velocity field $ \B U$ into
its mean part $\B V$ and the turbulent fluctuations $\B v$ with
zero mean:
\begin{subequations}\label{Rd}
\begin{equation}\label{RdA} 
\B U= \B V + \B v\,,\quad \B V =\<\B U \>\,,\quad \<\B v\>=0\ .
\end{equation} 
 In our model 
 \begin{equation}\label{RdB}
 \B V = {\bf \hat x }\, V \,, \quad V \approx \Omega\,  r\ .
 \end{equation} \ese
The following mean values are needed: the mean velocity shear
$S(\B r, t) $, the  turbulent kinetic energy density (per unite
mass) $K(\B r, t)$ and   the Reynolds stress $W(\B r, t)$. Their
definitions are as follows:
 \BSE{defs} \begin{equation}\label{defsC} S(\B r, t) \= \frac {\p V }{\p z}, \
  K(\B r, t) \= \frac12 \<|\B v|^2\>,
  \  W(\B r, t)\= -\< v_x v_z\>\ .
\end{equation} 
 Also we will be using the kinetic energy density of the mean
 flow,
 \begin{equation}\label{defsD}
 K\Sb V (\B r, t)\= \hf\,  \big[ V(\B r, t)\big ]^2\,,
 \end{equation}  
 and the total density of the kinetic energy
 \BE{defCK}
 \C K(\B r, t) \=  K\Sb V (\B r, t)+ K  (\B r, t)\ .
 \ee
  \ese
  Clearly, the total kinetic energy in the system, $\C E(t)$, is:
  \BE{E}
  \C E(t)= \int  \C K(\B r, t)\,  d\B r\ .
  \ee

\subsubsection{\label{ss:model}Balance of mechanical momentum
and kinetic energy }\vskip .3cm

\paragraph{\label{sss:mom}\emph{a.} Balance of mechanical momentum.}
Averaging \Eq{SH1}, with the dissipation term in the form of
\Eq{V8}, one gets for the planar geometry:

\begin{equation}\label{mom-bal} 
\frac{\p V }{\p t} - \frac{\p W}{\p z}+\alpha \,  \omega\sb{eff}
\, V=0\ . \end{equation}

\paragraph{\label{sss:V-en}\emph{b.} Mean-flow kinetic energy balance.}
Multiplying \Eq{mom-bal} with $V $ one gets the balance for the
kinetic energy of the mean flow, $K\Sb V$, defined in
\Eq{defsD}:
\begin{equation}\label{balKV} 
\frac{\p K\Sb V}{\p t} - \frac{\p }{\p z}\big[V W \big]+ S\, W
+\alpha \,  \omega\sb{eff} \, K\Sb V=0\ .\end{equation} 

The second term on the LHS of this equation describes the spatial
energy flux and does not contribute to the global energy balance
of the entire rotating cylinder.  The next term, $S\, W$, is
responsible for the energy transfer from the mean flow to the
turbulent subsystem; this energy finally dissipates into heat. The
last term, $ \alpha \, \omega\sb{eff} \, K\Sb V$, represents
direct dissipation of the mean flow kinetic energy into heat via
mutual friction.

\paragraph{\label{sss:en}\emph{c.} Turbulent kinetic energy balance. }
Let us take  \Eq{SH1} for the velocity fluctuations, with the
dissipation term in the form~\Eq{V8}, and we get an equation for
the energy balance which involves triple correlation functions. It
describes the energy flux in (physical) space.  In the theory of
wall bounded turbulence~\citep{Pope} these triple correlations are
traditionally approximated with second order correlation
functions.  For more details about the closure procedures  we
refer to \cite{Lvov:2006a,Lvov:2007b} and references cited
therein.

To be brief, notice that the total rate of kinetic energy
dissipation in the vortex front has two contributions in well
developed turbulence,
 \BSE{diss}\BE{dissA}
  \ve\sb{diss}=\ve\sb{diss,1}+\ve\sb{diss,2}\ .
 \ee

The first term, $\ve\sb{diss,1}$, arises from mutual friction
which acts on the global scale. It can be estimated from
 \Eq{SH1} as
 \BE{dissB}
 \ve\sb{diss,1}\simeq\a\o\sb{eff}K(z,r)\,,
 \ee
where $K(z,r) $ is the turbulent kinetic energy from
Eq.~\eq{defsC}. The second contribution, $\ve\sb{diss,2}$,
originates from the energy cascade toward small scales, where the
actual dissipation occurs.  In well developed turbulence of
viscous normal fluids this dissipation is caused by viscosity. It
dominates at the smallest length scales, known as the Kolmogorov
microscope. In superfluid turbulence the viscous contribution is
absent. Instead, at moderate temperatures it is replaced by mutual
friction. When mutual friction becomes negligibly small at the
lowest temperatures, the turbulent energy, while cascading down to
smaller length scales, is accumulated into Kelvin waves at some crossover scale. Ultimately the
energy then cascades further down where it can dissipate, e.g.  by
the emission of excitations (phonos in $^4$He II and
quasiparticles in $^3$He-B). In any case, the energy has to be
delivered to small length scale motions owing to the nonlinearity
of \Eq{SH1}. Clearly in steady state conditions, the rate of
energy dissipation at small scales, $\ve\sb{diss,2}$, is equal to
the energy flux, $\ve\sb{flux}$, from the largest scales, where
the energy is pumped into the system [from the mean flow via the
shear $S$, as follows from \Eq{balKV}].   The nonlinearity of
\Eq{SH1} is of standard hydrodynamic form and the associated
energy flux can be estimated by dimensional reasoning, as
suggested by Kolmogorov in 1941 [see e.g.~\cite{Pope}]:
 \BE{dissC} \ve\sb{diss,2}=\varepsilon\sb{flux}\simeq b\, K^{3/2}(\B  r)/\C L(\B
 r)\,.
 \ee
Here $b$ is a dimensionless parameter, which in the case of
classical (viscous) wall bounded turbulence can be estimated as
$b\sb{cl}\simeq 0.27$ [see e.g. \cite{Lvov:2006a}]. In \Eq{dissC}
$\C L(\B  r)$ is the outer scale of turbulence  (which defines the
length scale of the largest eddies containing the main part of the
turbulent kinetic energy). Clearly  near the centerline of the
cylinder, $\C L(\B  r)$ is determined by the thickness $\D(r)$ of
the turbulent front at given radius $r$: $\C L(\B r)\simeq \D(r)$.
Near the wall of the cylinder the size of the largest eddies is
limited by the distance to the wall, $R-r$. Therefore in this
region, $\C L\simeq R-r$. In the whole cylinder, $\C L$ should be
the smaller of these two scales, so that one can use an
interpolation formula
 \BE{dissD}
 \C L^{-1}(z)=\D(r)^{-1}+ (R-r)^{-1}\ .
 \ee
 \ese

The resulting energy balance equation, which accounts for the
energy dissipation in \Eqs{diss}, can be written (in cylindrical
coordinates $r$, $z$, and $\varphi$) as  follows: 
\begin{eqnarray}\label{en-bal}
&& \frac{\p K }{\p t}+\a \, \o\Sb {eff} \,K + \frac{
b\,K^{3/2}}{\C L }\br &-&g \,\Big [ \frac{\p }{\p z}\,  \C L \sqrt
{K}\frac{\p }{\p z}\,K + \frac 1 r \,\frac{\p }{\p r } \, r \,\C L
\sqrt {K}\, \frac{\p }{\p r } \, K \Big ]= S \, W\ .
 \end{eqnarray}
Here $K$, $W$, and $S$ are defined in \Eqs{defsC}. Generally
speaking, they are functions of $r$, $z$, and time $t$. The outer
scale of turbulence, $\C L $ depends on $r$ and $t$ in our
approximation~\eq{dissD}.

The right-hand-side (RHS) of \Eq{en-bal} represents the energy
flux (per unite mass) from the mean flow to the turbulent
subsystem. This expression rigorously  follows from \Eq{SH1} and
is exact. The two terms on the LHS  of \Eq{en-bal} (proportional
to $\a$ and $b$) describe the energy dissipation~\eq{diss}. The
last two terms on the RHS  of \Eq{en-bal} are proportional to the
phenomenological dimensionless parameter $g\approx 0.25$, which
was estimated by  \cite{Lvov:2007b}. These terms model turbulent
diffusion processes (in the differential approximation) with the
effective turbulent diffusion parameter $ \C L \, g \,\sqrt{K}$,
now expressed in cylindrical coordinates. The first term in
parenthesis describes turbulent diffusion in the direction normal
to the front, which is the main reason for the propagation of the
turbulent front. The second term, which vanishes in the flat
geometry, describes the turbulent energy flux in the radial
direction toward the centerline of the sample cylinder. This term
plays an important role in the propagation of the front as a
whole. The reason is that the central part of the front, where the
direct conversion of the mean flow energy to turbulence is small,
does not contain enough energy to exist without the radial  flux
of energy.

\paragraph{\label{sss:tot}\emph{d.} Total  kinetic energy balance. }
Adding $\p K\Sb V/\p t$ from \Eq{balKV} and $\p K / \p t$ from
\Eq{en-bal} one can see that the term $S\,W$, responsible for the
transfer of energy from the mean flow to the turbulent flow,
cancels and we arrive at a balance equation for $\C K\= K\Sb V+K
$: \BSE{totB}
\begin{eqnarray}\label{totBa}
&& \frac{\p \C K }{\p t}+\a \, \o\Sb {eff} \,\C K + \frac{
b\,K^{3/2}}{\C L }- \frac{\p}{\p z}\left[V W\right]\br &-&g \,\Big
[ \frac{\p }{\p z}\,  \C L  \sqrt {K}\frac{\p }{\p z}\,K + \frac 1
r \,\frac{\p }{\p r } \, r \,\C L  \sqrt {K}\, \frac{\p }{\p r }
\, K \Big ]= 0\ .
 \end{eqnarray}
 Integrating this relation over the cylinder, we see that the energy diffusion terms
 do not contribute to the total energy balance:
 \BEA{defK} \C E (t) &\=&   \int  \C K(\B r,t) d \B r \,, \\
 \label{totBc}
 \frac{d\C E}{d\, t}&=&- \int d \B r\Big[ \a \, \o\Sb {eff} \,\C K + \frac{  b\,K^{3/2}}{\C L }\Big]\ .
 \eea
 \ese

To summarize this section, we note that \Eqs{mom-bal} and
\eq{en-bal} allow us, at least in principle, (i) to describe the
propagating turbulent front in a rotating superfluid, (ii) to find
the front velocity $V\sb f$, and (iii) to describe the structure
of the front: its effective width $\D(r)$, the profiles for the
mean shear $S(r,z)$, the Reynolds stress $W(r,z)$, and the kinetic
energy $K(r,z)$. Here we present only some preliminary steps in
this direction, based on a  qualitative analysis in \Sec{ss:an} of
the global energy balance~\eq{totBa},  and discuss the  role of
the radial turbulent diffusion of  energy in \Sec{ss:dif}. In
\Sec{ss:X} we finally specify the model, accounting explicitly for
the bottleneck at the classical-quantum crossover, as described by
\cite{LNR}, and explain how it is influenced by the temperature
dependent mutual friction.

\subsubsection{\label{ss:an}Qualitative analysis of global energy
balance }\vskip .3 cm 
The total kinetic energy $\C E$ is dissipated by the propagation
of the turbulent front at constant velocity $V\sb f$, as expressed
in \Eq{KineticEnergy}. This means that during time $t$ the length
of the cylinder with unperturbed mean flow $V=\O\,r$ decreases by
$V\sb f\, t$. Using \Eq{KineticEnergy} we can write the overall
energy budget~\eq{totBc} as: \BSE{3} \BE{3A} V\sb f \, = \frac {4
} {\pi\, \O^2   R^4}\int   d \B r \Big [\a \, \o\sb{eff}\,  \C
K(z,r)+ \frac{ b\, K^{3/2}(z,r)}{\C L(r)} \Big]\ . 
\ee
This equation requires some corrections, because in its derivation
we  were a bit sloppy. First of all, approximation~\eq{V8} for the
mutual friction force~$\B f\sb{mf}$ cannot be applied for the
analysis of laminar flow, where the actual orientations of
vorticity and velocity are important. To fix this we present the
RHS of \Eq{3A} as a sum of two contributions,  \BE{3B} V\sb f=
V\sb{f,lam}+V\sb {f,turb}\,,
\ee and for $V\sb{f,lam}$ we use \Eq{Vlam}. To get the corrected
equation for $V\sb{f,turb}$  we replace in \Eq{3A} $\C K$ with its
turbulent part $K$ and replace $b$ with $\~b\=b (1-\a') $ in order
to account for the correction to the nonlinear term in \Eq{SH1}
from the reactive part  of the mutual friction~\eq{SH2}
(proportional to $\a'$).  Integrating the result over the
azimuthal angle $\varphi$ and  making use of the axial symmetry of
the problem, we get: 
\BE{3C} V\sb {f,turb} \, = \frac 8 {\O^2   R^4}\int_0^{R-\ell}\!
r \,  d r\, dz \Big [\a \, \o\sb{eff}\, K(z,r)+ \frac{ \~ b\,
K^{3/2}(z,r)}{\C L(r)}   \Big]\ . 
\ee
The hydrodynamic equation \eq{SH1} is not applicable near the
wall, where $R-r$ is less than the mean inter-vortex distance
$\ell$. Therefore this region is excluded from the
integration~\eq{3C}.

In the limit of a fully developed turbulent boundary layer (TBL),
the turbulent kinetic energy and Reynolds stress are independent
of the distance in the $z$ direction to the boundary of
turbulence. Here we consider the turbulence in the front to be
bounded in the $z$ direction, \textit{i.e.} we assume that the
front thickness $\Delta (r) < R$ and ignore the influence of the
cylindrical container wall. Therefore it is reasonable in the
qualitative analysis of \Eq{3C} to ignore the $z$ dependence of
$K(z,r)$ and $W(z,r)$, replacing these objects with their mean
values across the TBL:
  \begin{equation}\label{rep}
   K(r,z)\Rightarrow \overline{K}(r)\,,  \quad
   W(r,z)\Rightarrow \overline{W}(r)\,, \quad  \o\sb {eff}(r,z)\Rightarrow
   \overline{\o }\sb {eff}(r)\ .\end{equation} 
As seen from Eq.~(\ref{rep}), we use the same approximation also
for $\o\sb {eff}$. Now we can trivially integrate \Eq{3C} with
respect to $z$:
 \BE{3D} V\sb {f,turb} \, = \frac 8  {\O^2   R^4}\int_0^{R-\ell}\!
\Big [\a \, \overline{\o }\sb {eff}(r)\, \overline{K}(r)+ \frac{
\~ b\,  \overline{K}^{3/2}(r)}{\C L(r)}   \Big] \D(r) r \,  d r\,\
. 
\ee
\ese Here $\Delta (r)$ is the characteristic width of the TBL at
the distance $r$ from the central axial.   To perform the next
integration over $r$, one needs to know the $r$-dependence of
$\Delta$, $\overline{\o }\sb {eff}$, and  $K$. Notice first that $
\Delta (r)\overline{\o} \sb {eff} (r)$ has the dimension of
velocity. Therefore one expects that in the self-similar regime of
a fully developed TBL, $ \Delta (r)\overline{\o} \sb {eff} (r)$
has to be proportional to the characteristic velocity, which is
$\O\, r$. In other words, we expect that $\Delta (r)\overline{\o}
\sb{eff} (r)\propto r$. As we will see below in \Sec{ss:dif},
$\Delta (r)\propto r$, while $\overline{\o} \sb {eff}$ is
$r$-independent. With the same kind of reasoning we can conclude
that $K$ and $W$ have to be proportional to $r^2$, since they have
the dimensionality of velocity squared:
\begin{equation}\label{not1}
\Delta (r)\overline{\o} \sb{eff}= a\, \O r \,,  \quad
\overline{K}(r)= \frac{c}{2} (\O r)^2\,, \quad  \overline{W}(r)=
\~c \,  (\O r)^2\ . \end{equation} 
Now integrating~\eq{3D} one gets 
 \BSE{5} 
\BEA{5A}v\sb f &\=&  \frac{V\sb f }{  \O R}   \simeq (2\,
c)^{3/2}b\,(1-\a')A(R/\ell) +  \frac{4\,c \, \a}{5\, a} \,,\\
\label{5B} 
 A(R/\ell)&=& 0.2 +d
\big[\ln (R / \ell)  - 137/60 +5 \ell / R +\dots\big]\ ,\eea\ese
where we used $d = \Delta (r) /r$. At  $\O=1\,$rad/s, the ratio
$R/\ell\approx 17$. This gives $ A(R/\ell)\approx A(17)\approx
1.8$. We take $b=b\sb{cl}$ and choose the parameters $a=0.2$,
$c=0.25$, and $ d =2$ to fit the measurement in the region
(0.3\,--\,0.4)\,$T_{\rm c}$. With these parameters  \Eq{5} gives
$v\sb f\approx 0.16$ in the limit $T\to 0$ (when $\a=\a'=0$) and a
very weak temperature dependence up to $T\simeq 0.45\,T\sb c$.

\subsubsection{\label{ss:dif} Role of the  radial turbulent diffusion of
energy }
According to experimental observations, in steady state the front
propagates as a whole with the velocity $v\sb f$, independent of
the radial position $r$. To make this possible, the turbulent
energy has to flow from the near-wall region, where the azimuthal
mean velocity and consequently the energy influx into the
turbulent subsystem are large, toward the center, where the influx
goes to zero.

To clarify the role of the radial energy flux, consider
\Eq{en-bal}, averaged in the $z$ direction:
\begin{equation}\label{en-bal1}
 \a \, \overline{\o} \sb {eff} \,\overline{K}(r) + \frac{  \~b\,[\overline{K}(r)]^{3/2}}{\C L(r)}
-   \frac g r \,\frac{\p }{\p r } \, r \,\C L(r) \sqrt
{\overline{K}(r)}\, \frac{\p }{\p r } \, \overline{K}(r) =\frac{\O
\,r}{\D(r)}
 \, \overline{W}(r)\ .
 \end{equation}
On the RHS of this equation we replaced $\overline{S}(r)$ with its
natural estimate $ \O \,r/\D(r)$. According to \Eq{dissD}, close
to the cylinder axis $\C L(r)\approx \D(r)$. In this region
\Eq{en-bal1} has a self-similar scaling solution~\eq{not1}, in
which $\o\sb {eff}$ is indeed $r$-independent, $\D(r)\propto r $,
and for the Reynolds-stress constant  $\~c$ in Eq.~\eq{not1} one
finds the relationship:
 \BSE{sol1}
 \BE{tc}
 \frac{\~c}{c}=\a \,\frac{\o\sb{eff}}{\O}\, \frac{\D(R)}{R}+ \frac{\~   b }2 \sqrt{\frac {c }{2}}+ 2\, d\, \sqrt {2\, c}\Big[\frac{\D(R)}{R}\Big ]^2\ .
 \ee
In the vicinity of the wall, where $R-r\ll R$, the outer scale of
turbulence is $\C L\approx R-r$ and goes to zero when $r\to R$. To
account for the influence of the cylindrical wall in this region
we assume that, similar to a classical TBL, the two pair velocity
correlations $W(r)$ and $K(r)$ have the same dependence on
distance from the wall. Therefore their ratio is approximately
constant: $W(r)/K(r)\approx$ const. If so, the energy balance
\Eq{en-bal} dictates \BE{wall-est} W(r)\sim K(r)\sim   (R-r)^2 \,,
\ee\ese
with some relation between constants, similar to \Eq{tc}.
Actually, when $R-r$ becomes smaller than the inter-vortex
distance $\ell$, the whole hydrodynamic approach~\eq{SH1} together
with \Eq{en-bal1} fails.  Therefore we cannot expect that $W(r)$
and $K(r)$ really go to zero when $r\to R$. Rather they should
approach some constant values dominated by the vortex dynamics at
distances $R-r\simeq \ell$.

To summarize, we should say that \Eq{en-bal1} and its
solution~\eq{sol1} cannot be taken literally as a rigorous result.
They give a qualitative description of the kinetic energy
profiles. Indeed, the predicted profiles are in qualitative
agreement with numerical simulation calculations, as seen
\textit{eg.} in \Fig{RadVelFluctuation}:  Here $K(r)$ goes to zero
when $r\to 0$, increases at small $r$ approximately as $r^2$, as
concluded in \Eqs{not1} and \eq{tc}, reaches a maximum at
$r/R\lesssim 1$ and then decreases in qualitative agreement with
\Eq{wall-est}. Similar comparison with other numerical results in
Sec.~\ref{NumFront} makes it believable that the analysis of the
global energy balance in \Sec{ss:an} and the predicted plateau for
$v\sb f$ in the $T\to 0$ limit are reasonable.

\begin{figure}
\centerline{\includegraphics[width=0.8\linewidth]{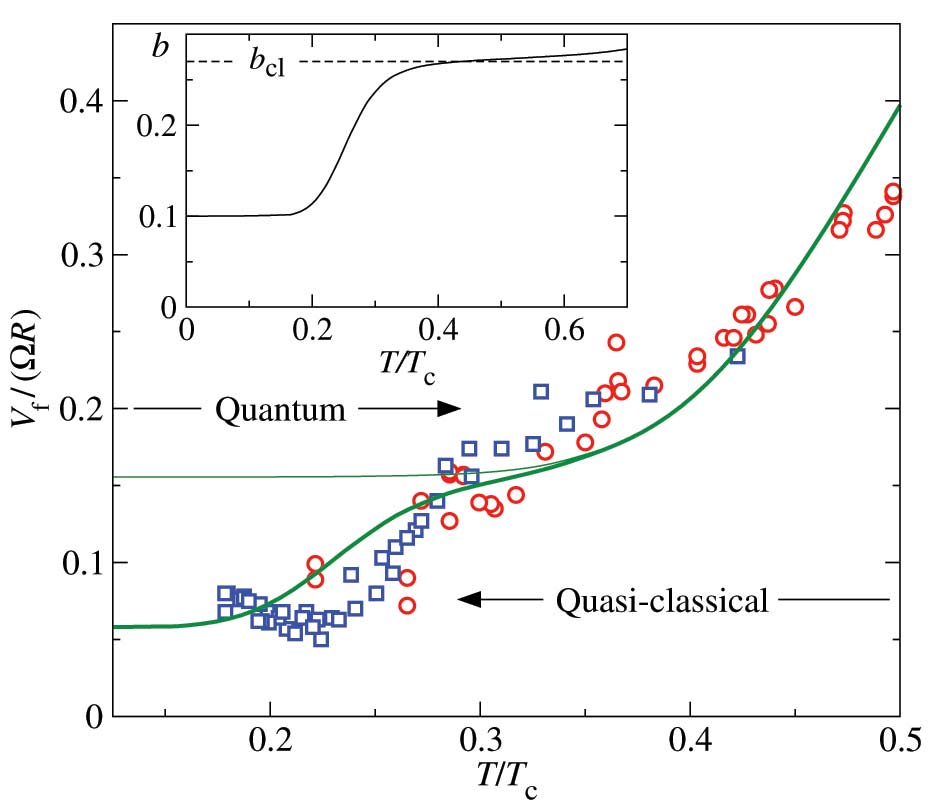}}
\caption{Scaled front velocity \textit{vs} temperature, similar to
Fig.~\ref{FrontVel}, but emphasizing the comparison of
measurements in Sec.~\ref{FrontVelocityMeasurement} to analytical
results in Sec.~\ref{s:basicEq} in the turbulent regime at low
temperatures. 
The thin and thick solid lines show consecutive model
approximations which sequentially account for dissipation in
turbulent energy transfer Eq.~(\ref{5B}) and bottleneck effect in
Sec.~\ref{ss:X}. \textit{(Insert)} Value of parameter $b(T)$ in
Eq.~(\ref{5}), which was used for the thick solid line in the main
panel.} \label{f:Vf}
\end{figure}

Nevertheless, both in the region of lower and higher temperatures
the experiment shows deviation from this ``plateau"
(Fig.~\ref{f:Vf}). The reason for this deviation at $T>0.35\, T\sb
c$ is that turbulence is not well developed near the cylinder axis
where the shear of the mean velocity, responsible for the
turbulent excitation, decreases. Therefore in the intermediate
temperature region only part of the front volume is turbulent.
When the temperature decreases, the turbulence expands toward the
axis. We should therefore account also for the laminar
contribution to the front velocity~\eq{Vlam} near the axis. Notice
furthermore that the simple sum of laminar and turbulent
contributions to $V\sb f$ in \Eq{3B} oversimplifies the situation,
by not accounting for the turbulent motions in \Eq{3C}. We will
not discuss this issue, but just suggest an interpolating formula
between the laminar and turbulent regimes, which has a shorter
intermediate region than \Eq{3B}: \BE{inter1} v\sb f=\sqrt{ v^2\sb
{f, lam}+ v^2\sb {f, turb}}\,,
\ee
  where $v\sb {f, lam}$  and $v\sb {f, turb}$  are given by \Eqs{Vlam}
and  \eq{5}.  This interpolation is shown in \Fig{f:Vf} as a thin
line for $T<0.3\, T\sb c$ and as a thick line for $T>0.3\, T\sb
c$. The agreement with measurements above $0.3 \, T\sb c$ is good,
but there is a clear deviation below $0.25\,T\sb c$, where $\a
\lesssim 10^{-2}$. Below an explanation is outlined which takes
into account the quantum character of turbulence, since as shown
by \citet{Volovik:2003}, individual vortex lines become important
below $0.3\,T_{\rm c}$ in the region where the transition to the
lower plateau occurs in Fig.~\ref{f:Vf}.

To appreciate the last aspect note that the mean free path of
$^3$He quasiparticles at the conditions of the measurements at
29\,bar pressure and $T\simeq 0.3\,T_{\rm c}$ is close to $\ell$,
while at $0.2\,T_{\rm c}$ it exceeds $R$. This change from the
hydrodynamic to the ballistic regime in the normal component may
influence the mutual friction force acting on individual vortices.
However, what is important for the interpretation of the
measurements is that the effect of the normal component on the
superfluid component becomes negligibly small. Therefore it
actually does not matter what is the physical mechanism of this
interaction: ballistic propagation of thermal excitation with
scattering on the wall or mutual friction that can only be
described in the continuous-media approximation, as given by
\Eq{SH2}. For simplicity we use the hydrodynamic
approximation~\eq{SH2} in \Sec{ss:X} to describe the turbulent
kinetic energy dissipation at low temperatures.

\subsubsection{\label{ss:X}Mutual friction and bottleneck crossover from classical to quantum cascade}
\vskip .3cm

\paragraph{\label{p:X}\emph{a.} Bottleneck at zero temperature.}
At low  temperatures, when mutual friction becomes sufficiently
small, the energy flux toward small length scales (or large wave
vectors $k$) $\ve_k$  propagates down to the quantum scale $\ell$
and vortex discreteness and quantization effects become most
important. Even though some part of the energy is lost in
intermittent vortex reconnections, the dominant part proceeds to
cascade below the scale $\ell$ by means of nonlinearly interacting
Kelvin waves [see
\cite{Vinen:2003,Svistunov:2008a,Svistunov:2008b} and references
there]. The Kelvin waves are generated by both slow vortex
filament motions and fast vortex reconnection events. As shown by
\cite{LNR}, the important point for the rate of energy dissipation
(and consequently for the turbulent front velocity) is that Kelvin
waves are much less efficient in the down-scale energy transfer
than classical hydrodynamic turbulence: in order to provide the
same energy flux as in the hydrodynamic regime, the energy density
of Kelvin waves at the crossover scale $\ell$ has to be
$\L^{10/3}$ times larger than that of hydrodynamic motions. For
$^3$He-B, $\L\=\ln(\ell / a_0) \simeq 10$, where $a_0$ is the
vortex-core radius. Assuming that the energy spectrum is
continuous at the crossover scale $\ell$ and that no other
mechanisms intervene between the classical and Kelvin-wave
cascades, then to maintain the same value of energy flux there
must be a bottleneck pile-up of the classical spectrum near this
scale by the factor $\L^{10/3}\!$. To account for this phenomenon
of energy pileup, we construct the ``warm cascade'' solution which
will be described in what follows.

For a qualitative description, we first define the hydrodynamic
kinetic energy density (in the one-dimensional $k$-space) $\C
E_k$, related to the total kinetic energy  $\C E$ as follows:
\BSE{K41}\BE{e-dens} \C E  \= \int \C E_k \, dk \,.
\ee 
The Kolmogorov-1941 cascade of hydrodynamic turbulence (with
$k$-independent energy flux, $\ve_k\Rightarrow \ve$=const) is
described with the spectrum: \BE{19}
\C E_k\Sp {K41}\simeq \ve^{2/3} |k|^{-5/3}\,. \ee 
\ese The energy flux carried by the classical hydrodynamic
turbulence with the K41 spectrum~\eq{19}  cannot adequately
propagate across the crossover region at $\ell$. Therefore,
hydrodynamic motions on larger length scales (smaller
wave-vectors) will have increased energy content up to the level
$\C E_{1/\ell}$, as required when the same energy flux has to be
maintained by means of Kelvin waves. As a result, for $k\le
1/\ell$ the spectrum of hydrodynamic turbulence $\C E\Sp{HD}_k\!$
will not have the K41 scale-invariant form $\C E\Sp{K41}_k\!$
given by \Eq{19}. To get a qualitative understanding of the
resulting bottleneck effect, we use the so called ``warm cascade''
solutions found by \cite{ConnaughtonNazarenko}. These solutions
follow from the \citet{Leith} differential model for the energy
flux of hydrodynamical turbulence,
\BE{22} \ve_k=- {1  \over  8 } \,
   \sqrt{|k|^{13}F_k}\  {d F_k  \over   d k}  \,,
\quad F_k\= \frac{\C E\Sp{HD}_k\!\!}{k^2}\,,  \ee 
where  $F_k$ is the 3-dimensional spectrum of turbulence. The
generic spectrum with a constant energy flux can be found as the
solution to the equation ~$\ve_k=\ve$: 
\BE{23}
 F_k =  \Big[ \frac{24\ve }{11 |k|^{11/2}}+
\Big(\frac{T}{\pi \rho}\Big)^{3/2}\Big]^{2/3}\ .
 \ee 
Here the range of large $k$ values belongs to the thermalized part
of the spectrum, with equipartition of energy characterized by an
effective temperature $T$, namely $T/2$ of energy per degree of
freedom, thus, $F_k = T \big/ \pi \r$ and $ \C E_k = T k^2 \big/
\pi \r$. At low $k$, \Eq{23} coincides with the K41
spectrum~\eq{19}.

 This ``warm cascade'' solution describes reflection of the K41
cascade and stagnation of the spectrum near the bottleneck scale
which, in our case, corresponds to the classical-quantum crossover
scale. To obtain the spectrum in the classical range of scales, it
remains to find $T$, by matching \Eq{23} with the value of the
Kelvin wave spectrum at the crossover scale $\C E_k \sim
\k^2/\ell$. This gives $T/\r \sim \k^2  \ell \sim (\k^{11} /\L^5
\ve)^{1/4}.$

Obviously, the transition between the classical and quantum
regimes is not sharp and in reality we should expect a gradual
increase of the role of the self-induced wave-like motions of
individual vortex lines with respect to the collective
classical-eddy type of motions of vortex bundles. Thus, the
high-wavenumber part of the thermalized range is likely to be wave
rather than eddy dominated. However, the energy spectrum for this
part should still be of the same $k^2$ form which corresponds to
equipartition of thermal energy. This picture, as explained below,
relies on the assumption that the self-induced wave motions have
small amplitudes and, therefore, do not lead to reconnections.

\begin{figure}
\begin{center} \vskip -.3cm
\includegraphics[width=8 cm]{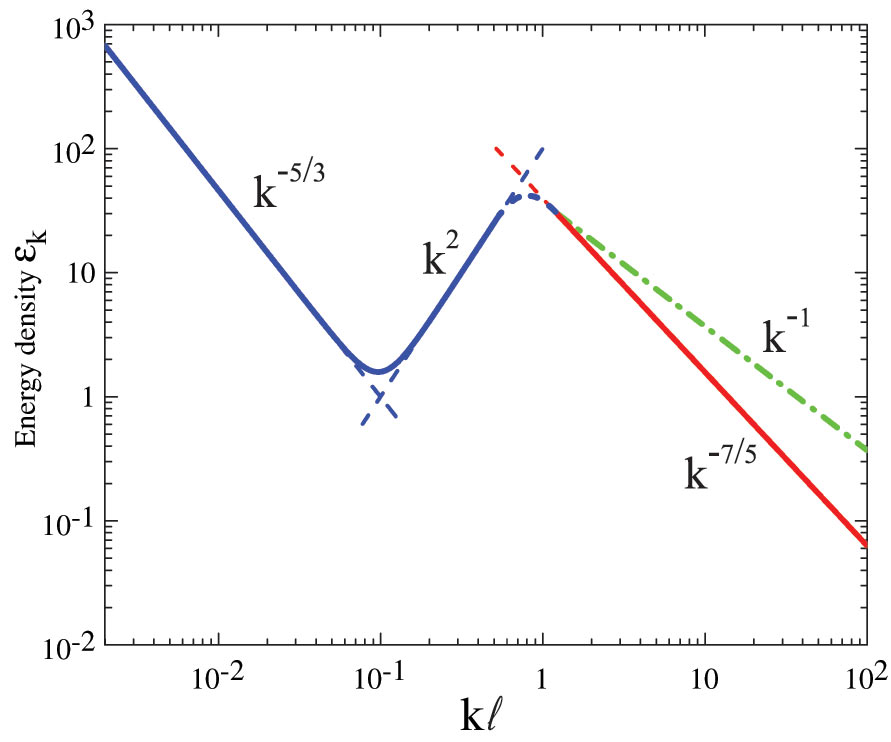}
\caption{
Energy spectra $\C E_k$ in the classical, $k < 1/\ell$, and
quantum, $k > 1/\ell$, ranges of scales. The two straight lines in
the classical range indicate the pure K41 scaling $\C E_k\Sp
{K41}\propto k^{-5/3}$ of \Eq{19}  and the pure thermodynamic
scaling $\C E_k \propto k^{2}$. In the quantum range, the  solid
line indicates the Kelvin wave cascade spectrum $\propto
k^{-7/5}$, whereas the dash-dotted line marks the spectrum
corresponding to the non-cascading part of the vortex tangle
energy $\propto k^{-1}$. }\label{f:24}
\end{center}
\end{figure}

The resulting spectrum, including its classical, quantum and
crossover parts,  is shown in \Fig{f:24} as a log-log plot. It is
important to note that in the quantum range $k > 1/\ell$, in
addition to the cascading energy associated with Kelvin waves,
there is also energy associated with the tangle of vortex
filaments (shown in \Fig{f:24} with the dash-dotted line). The
energy spectrum of this part is $\sim |k|^{-1}$, which is simply
the spectrum associated with a singular distribution of vorticity
along 1D curves in 3D space \citep{araki}. It does not support a
down-scale cascade of energy. The cascading and non-cascading
parts have similar energies at the crossover scale, that is the
wave period and the amplitude are of the order of the
characteristic time and size of evolving background filaments. In
other words, the scales of the waves and of the vortex
``skeleton'' are not separated enough to treat them as independent
components. This justifies matching the classical spectrum at the
crossover scale with the Kelvin wave part alone, ignoring the
vortex ``skeleton''. This is valid up to an order-one factor and
justifies the way of connecting the ``skeleton'' spacing $\ell$ to
the cascade rate $\ve$.

\paragraph{\label{p:mf}\emph{b.} Effect of mutual friction on
the bottleneck crossover.} Returning back to the propagating
turbulent vortex front, recall that in the measurements of
Sec.~\ref{FrontVelocityMeasurement} the inertial interval $R/\ell$
is about one decade. Therefore the distortion of the energy
spectrum owing to the bottleneck reaches the outer scale of
turbulence. This leads to an essential suppression of the energy
flux at any given turbulent energy or, in other words, to a
decrease in the effective parameter $b$, which relates $\ve$ and
$K$ in the estimate~\eq{dissC}. The effect is more pronounced at
low temperatures when mutual friction is small; thus $b(T)$ should
decrease with temperature. We analyze this effect with the help of
the stationary energy balance equation for the energy spectrum $\C
E_k$ in $ \B k$-space \BE{en-b2} \frac{d \ve_k} {d k} = - \G(T) \,
\C E_k\,,\ \ve_k \= -(1-\a')\sqrt{k^{11}\C E_k} \, \frac{d ( \C
E_k / k^2 ) } { 8\, d k} , \ee 
where $\G(T)$ is the  temperature dependent damping in \Eq{V8},
and the energy flux over scales $\ve(k)$ is taken in the
\citet{Leith} differential approximation. In addition to  \Eq{22},
we included in \Eq{en-b2} the mutual friction correction factor
$(1-\a')$ and substituted $F_k=\C E_k/k^2$.

\begin{figure}[t]
\begin{center}
\centerline{\includegraphics[width=0.49\linewidth]{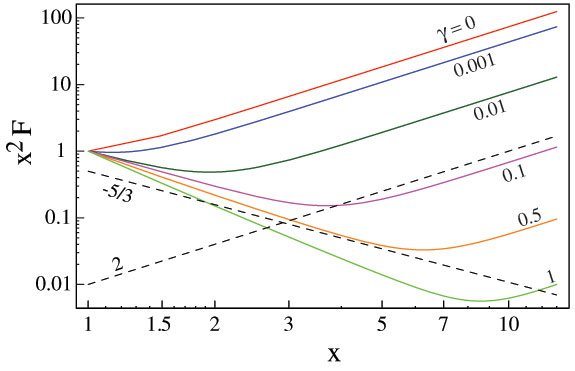}
\includegraphics[width=0.51\linewidth]{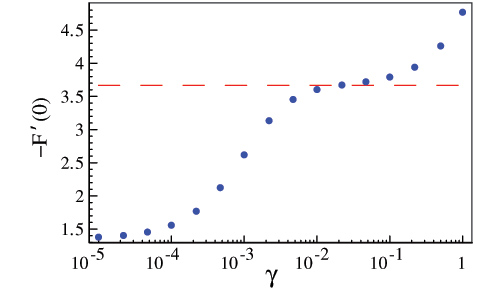}}
\caption{\label{f:25} \textit{On the left} the solutions of the
differential \Eq{en-b2} for different values of the dimensionless
mutual friction parameter $\g \equiv \G /\sqrt{k^3_+ \C E_{k_+}}$
have been plotted. The smallest wave vector $k_+$ corresponds to
the outer scale of turbulence: $\C L\approx 2\pi /k_+$. The wave
vectors are normalized with respect to $k_+$: $x\equiv k/k_+$ and
$F(x)\equiv F_{x\, k_+}/F_{k_+}$. The dashed lines denote slopes
of $\C E_k$ with K41 scaling $\C E_k\propto k^{-5/3}$ and
thermodynamic scaling $\C E_k \propto k^{2}$. \textit{On the
right} the slope $F'\equiv d F(x)/ d\, x$ is shown at $x=1$ for
different values of mutual friction $\gamma$. The horizontal
dashed line marks the K41 value of slope, $11/3$.}
\label{Solutions}
\end{center}
\end{figure}

Figure~\ref{f:25} displays the set of solutions for \Eq{en-b2}. We
use $\C L/\ell=12$ as the ratio of  the outer and  crossover
scales and characterize the bottleneck with the boundary condition
$\C E_k /[ k^3  d(\C E_k/k^2)/ d k ] = -4\cdot10^5$ at the
crossover scale. One goal of these calculations is to find the
slope of the function $F_k$ at the beginning of the inertial
interval $k=k_+\=2\pi/\C L$ which, according to \Eq{en-b2},
characterizes the rate of energy input into the system at fixed
value of the total energy $\C E$, \textit{i.e.} the
phenomenological parameter $b$ in the estimate~\eq{dissC}. The
dashed line in the right panel shows the Kolmogorov value of this
slope, $-11/3$, which is associated with the classical value of
$b=b\sb{cl}\approx 0.27$. Thus the ratio $3 b\sb{cl}|F'|/11$ can
be interpreted as the effective value of $b$ for a given value of
damping $\Gamma$. To relate $\Gamma$ with the temperature
dependent value $\a(T)$, we substitute the $r$ independent
$\omega_{\rm eff}$ from Eq.~(\ref{not1}) to Eq.~(\ref{V8}). This
gives $\Gamma = \alpha \, (a/d)\, \Omega$. The resulting function
$b(T)$, shown in the inset of Fig.~\ref{f:Vf} (for $\Omega =
1\,$rad/s), decreases from its classical value $b\sb {cl}\approx
0.27$ to $\simeq 0.1$ at $T<0.2 \, T\sb c$. Now, after accounting
for the temperature dependence of $b(T)$ in \Eq{5}, we get the
temperature dependence for the propagation velocity $v\sb f$ of
the quantum-turbulent front shown in Fig.~\ref{f:Vf} by the
bold-solid line below $0.3\,T_{\rm c}$. This fit is in good
agreement with the measured data. We thus have to conclude that in
this particular measurement the rapid drop in the dissipation rate
on entering the quantum regime can be explained as a consequence
of the relatively close proximity of the outer and quantum
crossover scales in this measuring setup.

\subsection{\label{ss:conclusionSec4}Summary: turbulent vortex front propagation}

In $^3$He-B strong turbulence is restricted to the lowest
temperatures below $0.4\,T_{\rm c}$. Depending on the type of
flow, turbulence varies in form and losses. The usual reference
point is an isotropic and homogeneous turbulent vortex tangle.
Here we have discussed an opposite extreme, the conversion of
metastable rotating vortex-free counterflow to the stable
equilibrium vortex state in a long circular column. The task is to
explain how increasing turbulence with decreasing temperature
influences this type of polarized vortex motion in steady state
propagation. The deterministic part of the motion takes place in
the form of a spiraling vortex front followed by a helically
twisted vortex cluster. When mutual friction decreases, turbulent
losses start to contribute to dissipation, concentrating in the
propagating front and immediately behind it.

Experimentally the conversion of a metastable  vortex-free state
to an equilibrium array of rectilinear vortex lines can be
arranged to occur at constant externally adjusted conditions. This
is done with different types of injection techniques, which
trigger the formation of a propagating vortex front and the
trailing twisted cluster behind it. In a long rotating column the
propagation can be studied in steady state conditions. At
temperatures above $0.4\,T_{\rm c}$ the propagation is laminar,
but below $0.4\,T_{\rm c}$ a growing influence from turbulence
appears. This is concluded from the measured propagation velocity
which provides a measure of the dissipation in vortex motion. In
addition to the increasing turbulent dissipation below
$0.4\,T_{\rm c}$, the measurement shows a peculiar transition at
$\sim 0.25\,T_{\rm c}$ between two plateaus, and a temperature
independent finite value of dissipation on approaching the
$T\rightarrow 0$ limit.

So far measurements on vortex propagation exist only for a column
with circular cross section. This is an exceptional case of high
stability (if the cylindrical symmetry axis and the rotation axis
are sufficiently well aligned). Here the crossover from the
laminar to the turbulent regime is smooth as a function of mutual
friction dissipation. The same smooth behavior is also seen in the
response to spin down, after a step-like stop of rotation. This is
in stark contrast to the spin down of a column with square cross
section, as we will see in Sec.~\ref{TurbulentDecay}. Similar
differences seem to apply in viscous hydrodynamics where flow in a
circular pipe is believed to be asymptotically linearly stable for
all Reynolds numbers, in contrast to a pipe with square cross
section \citep{Peixinho:2006}. In these measurements the formation
and decay of turbulent plugs is monitored. Vortex plugs and fronts
have also been observed in pipe flow of superfluid $^4$He through
long circular capillaries \citep{van Beelen:1987}.

Numerical vortex dynamics calculations have been used to examine
the turbulent contributions to dissipation in vortex front
propagation. This analysis demonstrates the increasing role of
turbulent excitations on sub-intervortex scales with decreasing
temperature. Analytical arguments have been developed which
explain that the transition to the lower plateau in the measured
overall dissipation is caused by the difficulty to bridge the
energy cascade from the quasi-classical to the quantum regime as a
function of decreasing mutual friction dissipation. This
bottleneck-scenario of \cite{LNR} in energy transfer was thus
directly inspired by the experimental result. It is expected to
apply foremost to polarized vortex motion in the rotating column
where vortex reconnections are suppressed. This situation differs
from the case of an ideal homogeneous vortex tangle, where
\cite{Svistunov:2008a,Svistunov:2008b} suggest that a bottleneck
is avoided owing to the high reconnection rate.

As for the second leveling off at the lowest temperatures, at
present time three different types of measurements
[\cite{Bradley:2006,Eltsov:2007,Walmsley:2007}] conclude that
dissipation in vortex motion remains finite at the lowest
temperatures. This applies for both the fermion superfluid
$^3$He-B and the boson case of $^4$He. Although the mechanisms for
dissipation in the $T\rightarrow 0$ limit in these two superfluids
are different and still under discussion, the results demonstrate
that coherent quantum systems can be inherently lossy even in the
$T=0$ state.




\section{Decay of homogeneous turbulence in superfluid $^4$He}
\label{TurbulentDecay}

\subsection{Introduction and experimental details}

\subsubsection{Quasiclassical and ultraquantum types of superfluid turbulence}

In this section we switch to nearly homogeneous and isotropic
superfluid turbulence with an emphasis on the recent experiments
in Manchester. A truly homogeneous turbulence is of course
impossible to achieve in real experiments because the tangle is
always confined by the container walls and can have other
inhomogeneities specific to the particular process of its
generation.

The modern paradigm of homogeneous isotropic turbulence at high
Reynolds numbers $Re$ in classical liquids is that there is a
broad range of wavenumbers $k$ within which the energy, pumped at
large scales, gets continuously redistributed without loss towards
smaller length scales thus arranging a steady-state distribution
in $k$-space of the Kolmogorov type
\citep{Kolmogorov:1941a,Kolmogorov:1941b,Batchelor:1953,Frisch:1995} (Eq.~\eq{19})
\begin{equation}
\C E_k\Sp {K41} = C \ve^{2/3} |k|^{-5/3},
 \label{Ek}
\end{equation}
where the Kolmogorov constant was found to be $C\approx 1.5$
\citep{Sreeni}. In the steady state, the energy flux $\epsilon$ is
equal to the dissipation of the kinetic energy  through the shear
strain in the flow at short length scales; its rate integrates to
\begin{equation}
\dot{\C E} = -\nu_{\rm cl} \omega_{\rm cl}^2, \label{NuCl}
\end{equation}
where $\nu_{\rm cl}$ is the kinematic viscosity and $\omega_{\rm
cl}$ is the r.m.s. vorticity. In realistic systems, the
distribution Eq.~\ref{Ek} is truncated at small $k=k_1$ by either the length
scale of forcing or the container size $h$, and at large $k=k_2$
by the Kolmogorov dissipation scale $\lambda_{\rm cl}(Re) \sim
k_2^{-1}$ which decreases with increasing $Re=(k_2/k_1)^{4/3}$. In
superfluids, the inertial cascade is expected to operate if the
mutual friction parameter $\alpha$ is sufficiently small
\citep{Vinen:2000}. At large length scales $>\ell$ the cascade is
{\it classical}. However, at short length scales $<\ell$ the
cascade becomes {\it quantum} as the discreteness of the vorticity
in superfluids adds new behaviour. At sufficiently low temperatures,
when the energy reaches very small scales $\ll \ell$ without
dissipation, it is expected that the quantum cascade takes the
form of a non-linear cascade of Kelvin waves
\citep{Svistunov:1995,Svistunov:2004}, eventually being truncated
at some quantum dissipation scale $\lambda_{\rm q}(\alpha)$. While
the theory of this regime at very high $k$ is now established, no
direct observations exist so far. The most complicated is, of
course, the transitional region between these two clear-cut
limits, the Kolmogorov and Kelvin-wave cascades. The question
being currently debated is whether the energy stagnates at $k <
\ell^{-1}$ due to the poor matching in the kinetic times of the
two cascades \citep{LNR}, or gets efficiently converted from 3-d
classical eddies to 1-d waves along quantized vortex lines with
the help of various reconnection processes
\citep{Svistunov:2008a}.

When comparing the two superfluid isotopes from the
experimental point of view, the $T=0$ limit in superfluid $^4$He
is achievable at some $T<0.5$~K \citep{Vinen:2000,Walmsley:2007}
while for $^3$He-B much lower temperatures $T<0.5$~mK are
required. Owing to the two or three orders of magnitude smaller
core size in $^4$He, $a_0 \sim 0.1$\,nm, the Kelvin-wave-cascade
is expected to extend over a broader range of length scales, down
to $\sim 3$\,nm \citep{Vinen:2001,Svistunov:2008b}. It has been
suggested \citep{Vinen:2002b,Vinen:2002} that in $^3$He-B Kelvin
waves should become overdamped at frequencies $\sim 10$~kHz
(corresponding to wavelengths $\sim 2$~$\mu$m) due to the resonant
scattering on core-bound quasiparticles
\citep{KopninSalomaa:1991}).

Flow on a scale greater than the inter-vortex distance $\ell$,
which is initially typically between $\sim 10$ and 100~$\mu$m in
experiments, can be obtained by mechanical stirring of the liquid.
So far the following methods have been used: counter-rotating
agitators with blades \citep{MT1998}, pipe flow
\citep{VanSciver:1999,Roche2007}, flow through an orifice
\citep{GueninHess}, towed grids
\citep{Smith:1993,Stalp1999,Niemela2005}, vibrating grids
\citep{McClintockGrid,Bradley:2006,FisherPLTP2008,VinenTsubotaHanninen2007,VinenSkrbekThisVolume},
as well as wires \citep{Bradley2000,SkrbekLanc}, and microspheres
\citep{SchoepeSphere}, plus most recently impulsive spin-down to
rest of a rotating container \citep{Walmsley:2007}. It is also
possible to initiate large-scale flow in superfluid $^4$He without
any moving parts or rotation of the cryostat: by either running
thermal counterflow in wide channels at $T>1$~K
\citep{BarenghiSkrbek,ChagovetzSkrbek} or a jet of injected
current \citep{Walmsley2008}.

Actually, the turbulence can take two very different forms
depending on whether the forcing is at scales above or below
$\ell$. So far we were discussing the flow on classical scales  $>
\ell$, where the energy cascades towards shorter length scales
like in the Richardson cascade in classical turbulence; the large
quasiclassical eddies being the result of correlations in
polarization of vortex lines. On the other hand, when forced on
quantum scales $< \ell$, the resulting uncorrelated tangle has no
classical analogs and should have completely different dynamics
first described by
\cite{Vinen1957a,Vinen1957b,Vinen1957c,Vinen1958} and later
investigated numerically by \cite{Schwarz:1988}. In both cases,
the dissipation of flow energy is through the motion of vortex
lines; its rate per unit mass can be assumed to be
\citep{Stalp1999,Vinen:2000,Stalp2002}
\begin{equation}
    \dot{\C E} = -\nu(\kappa L)^2,
    \label{EDot}
\end{equation}
where $\kappa^2 L^2$ is an effective total mean square vorticity
and the parameter $\nu$ is believed according to various models to
be approximately constant for a given temperature and type of
flow. This formula is the quantum analog of the classical
expression Eq.~(\ref{NuCl}). As we will show below, the efficiency
of the vortices in dissipating energy in the $T=0$ limit,
expressed through the ``effective kinematic viscosity'' $\nu$, is
different for these two regimes of turbulence.

To measure the values of $\nu$, one can monitor the free decay of
homogeneous tangles of both types. In any tangle, the {\it
quantum} energy associated with the quantized flow on length
scales $r < \ell$ is $\C E_{\rm q} = \gamma L/\rho_s$ (per unit
mass), where the energy of vortex line per unit length is $\gamma
= B\rho_s\kappa^2$ and $B  \approx  \ln(\ell/a_0)/4\pi$ is
approximately constant. This is the same energy that is shown in
the quantum range $k>1/\ell$ (including both non-cascading vortex
``skeleton'' and Kelvin-waves) in Fig.~\ref{f:24}. If the total
energy is mainly determined by $\C E_{\rm q}$, from
Eq.~(\ref{EDot}) we arrive at the late-time free decay
\begin{equation}
    L = B \nu^{-1}t^{-1}.
    \label{L-t-V}
\end{equation}
This type of turbulence, without any motion on classical scales,
will be called \textit{ultraquantum} (in the literature one can
find \textit{e.g.} ``Vinen turbulence''
\citep{Volovik:2003,Volovik:2004} and ``random'' or
``unstructured'' vortex tangle). For $^4$He, $\kappa_4 =
2\pi\hbar/m_4$, $a_0 \sim 0.1$\,nm and $B \approx 1.2$, while for
$^3$He-B, $\kappa_3 = 2\pi \hbar/(2m_3)$, $a_0 \sim 13-65$\,nm and
$B \approx 0.7$.

Now suppose that the tangle is not random but structured due to
the presence of flow on classical length scales $r>\ell$, and the
additional energy of this {\it classical} flow $\C E_{\rm c}$ is
much greater than $\C E_{\rm q}$. This type of turbulence will be
called {\it quasiclassical} (e.g. ``Kolmogorov turbulence'' and
``structured'' or ``polarized'' vortex tangle). For the Kolmogorov
spectrum between wavenumbers $k_1$ and $k_2$ ($k_1 \ll k_2$),
while the size of the energy-containing eddy stays equal to the
size of container $h$ (\textit{i.e.} $k_1 \approx 2\pi/h$), the
late-time free decay follows \citep{Stalp1999}
\begin{equation}
    L = (3C)^{3/2}\kappa^{-1}k_1^{-1}\nu^{-1/2} t^{-3/2}.
    \label{L-t-K}
\end{equation}
Eventually, when the energy flux from the decay of classical
energy $\C E_{\rm c}$ will become smaller than that from the
quantum energy $\C E_{\rm q}$ (which should typically happen when
only a couple of vortex lines are left in the container,
\textit{i.e.} $L\sim h^{-2}$), the quasiclassical regime will
cross over to the ultraquantum one, so $L\propto t^{-3/2}$ decay
will be replaced by $L \propto t^{-1}$. Deviations from the
described scenario are also possible if some fraction of the
classical energy is stored in non-cascading ``thermal spectrum''
\citep{LNR}.

In this Section we first review the relevant experimental
techniques of detecting and generating turbulence. Next we
describe the more recent new developments at the very lowest
temperatures. The results of these measurements will be discussed
in detail in the last part of this section, where we conclude with
a general discussion of the significance of the observed
temperature dependence of $\nu$.

\subsubsection{Techniques of measuring the vortex line density $L$}

The attenuation of second sound in superfluid $^4$He, which is
proportional to the total length of vortex lines, has been the
prefered technique of measuring $L$ in the temperature range 1 --
2~K \citep{Vinen1957a,Stalp1999,Niemela2005,BarenghiSkrbek} since
\cite{Vinen:1956}, however it cannot be extended below 1~K. Very
recently, scattering of negative ions off vortex lines has been
succesfully utilized at temperatures 80~mK -- 1.6~K
\citep{Walmsley:2007}, showing good agreement with previous
measurements of $\nu$ at overlapping temperatures
\citep{Stalp2002}. Actually, the first use of ions for the
observations of a vortex tangle was reported almost fifty years
ago \citep{CareriTangle}, and numerous further experiments
revealed their potential for investigating turbulence in
superfluid $^4$He (see the reviews by \cite{ToughReview} and
\cite{Donnelly:1991} for more references on early studies).

It is worth mentioning possible alternative techniques of
detecting turbulence in $^4$He in the $T=0$ limit that are
currently being developed, such as calorimetric
\citep{IhasQFS2007} and others \citep{Vinen:2006}. The design of
towed grid for turbulence generation, suitable for experiments at
$T<0.5$~K in $^4$He is presently being attempted by McClintock at
Lancaster and Ihas at Gainsville. Optical vizualization of
individual vortex lines in the turbulent tangle
\citep{Lathrop,Bewley:2008,VanSciver,excimers,Barenghi} (see the
contribution by \cite{VanSciverBarenghiThisVolume} in this volume)
is another promising development, although the prospects of
extending this technique down to at least $T=0.5$~K seem distant.
As we mentioned in Sec.~\ref{VortexInstability} and
\ref{s:FrontMotion}, computer simulations have contributed a great
deal to our understanding of certain processes in the dynamics of
tangles and of the Kelvin-wave cascade at $T=0$ in particular
\citep{Schwarz:1988,BarenghiVassilicos,Tsubota2000,Vinen:2003,VinenTsubotaHanninen2007}
(see also the contribution by Tsubota and Kobayashi in this
volume); however the proper modelling of the inertial cascade
pumped at large (quasiclassical) length scales and dissipated at
short (quantum) scales requires a considerable range of length
scales and hence remains a challenge.

\subsubsection{General properties of injected ions in liquid helium}

Injected ions (see reviews
\citep{SchwarzReview,FetterReview,Donnelly:1991,Borghesani:2007})
are convenient tools to study elementary excitations and quantized
vortices in superfluid helium. They were used to detect vortex
arrays in rotating helium
\citep{CareriArray,PackardVizualization}, quantized vortex rings
\citep{RR1964}, and vortex tangles \citep{CareriTangle} in $^4$He.
To create a {\it negative ion}, an excess electron is injected
into liquid helium, where it self-localizes in a spherical cavity
(``electron bubble'') of radius $\sim 2\,$nm at zero pressure. To
create a {\it positive ion}, an electron is removed from one atom,
which results in a positively-charged cluster ion (``snowball'')
of radius $\sim 0.7\,$nm. These objects are attracted to the core
of a quantized vortex, resulting in trapping with negligible
escape rate provided the temperature is low enough ($T<1.7$~K for
negatives and $T<0.4$~K for positives in $^4$He at zero pressure).
Because of this interaction, both species have been used a great
deal to study quantized vortices although negative ions are often
more convenient, especially if one wants to cover a wider range of
temperatures. The binding energy of a negative ion trapped by a
vortex is about 50~K \citep{Donnelly:1991}. In superfluid $^3$He,
thanks to much larger diameters of vortex cores this binding is
much weaker, hence there were no observations of ion trapping on
vortex cores yet. Still, the anisotropy of ions motion in $^3$He-A
made it possible to use negative ions to detect textural changes
around vortex cores \citep{HUTIons}.

In this section, we focus on negative ions in superfluid $^4$He.
Their mobility is limited by scattering thermal excitations and is
hence rapidly increasing with cooling. Below some 0.8~K it is so
high that they quickly reach the critical velocity for creation of
(depending on pressure and concentration of $^3$He impurities)
either rotons (and then move with the terminal Landau velocity
$\sim 60$~m/s, shedding off rotons continuously) or quantized
vortex rings (and then get trapped by such a ring and move with it
as a new  entity -- a charged vortex ring)
\citep{McClintockCritVelocity,McClintockVortexNucleation,Winiecki2000,Berloff2000a,Berloff2000b}.
Above 1~K, the trapping diameter $\sigma$ of an ion by a vortex
line is inversely proportional to the electric field, $\sigma
\propto E^{-1}$. In a field $E=33$~V/cm it is decreasing with
temperature from $\sigma = 100$\,nm at $T=1.6$~K to $\sigma =
4$\,nm at $T=0.8$~K \citep{OstermeierGlaberson:1974}.

\cite{RR1964} produced singly-charged singly-quantized vortex
rings in $^4$He at $T=0.4$~K and, by measuring the dependence of
their self-induced velocity $v \propto \kappa^3 K^{-1}$,
Eqs.~(\ref{RingEnergy}) -- (\ref{RingVelocity}), on their energy
$K$, confirmed the value of the circulation quantum $\kappa =
1.00\times10^{-3}$~cm$^2$s$^{-1}$. \cite{SchwarzDonnelly1966}
investigated the trapping of charged vortex rings by rectilinear
vortices in a rotating cryostat and found the trapping diameter to
be of order of the ring radius, meaning that the interaction is
basically geometrical. They wrote: ``quantized vortex rings are
very sensitive ``vortex-line detectors,'' making them suitable
probes for a number of problems in quantum hydrodynamics''.
\cite{GueninHess} used charged vortex rings to detect turbulent
vortex tangles in $^4$He at $T=0.4$~K, created by forcing a jet of
superfluid through an orifice at supercritical velocities.

\begin{figure}[t]
\begin{center}
\includegraphics[width=0.8\linewidth,keepaspectratio]{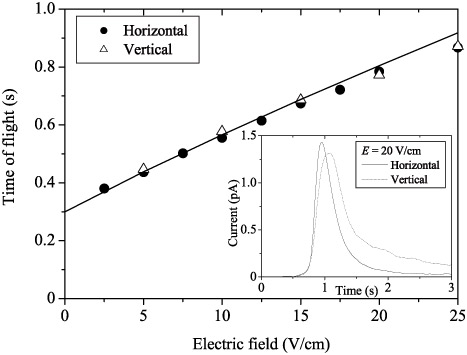}
\caption{Time of flight (corresponding to the leading edge of the
pulse of collector current shown in inset) of charged vortex rings
as a function of electric field at $T=0.15$~K in both the
horizontal and vertical direction. The solid line show the times
of flight for rings with initial (as injected into the drift
volume) radii and energies of 0.53~$\mu$m and 21~eV, calculated
using Eqs.~(\ref{RingEnergy}), (\ref{RingVelocity}), and
(\ref{RingImpulse}).}
\label{MancCVRTime}
\end{center}
\end{figure}

While charged vortex rings are more convenient than free ions
because of their much greater trapping diameter, their dynamics is
more peculiar (Fig.~\ref{MancCVRTime}; all figures in Sections
4.1--4.3 represent the Manchester experiments
\citep{Walmsley:2007,Walmsley2008}). The energy $K$, velocity $v$
and impulse $P$ of a quantized vortex ring of radius $\cal{R}$
with a hollow core of radius $a_0$ and no potential energy in the
core are \citep{GlabersonDonnelly1986,Donnelly:1991}
\begin{equation}
K=\frac{1}{2}\rho_\mathrm{s}\kappa^2 \mathcal{R}
\left(\ln\frac{8\mathcal{R}}{a_0} - 2\right), \label{RingEnergy}
\end{equation}
\begin{equation}
v=\frac{\kappa}{4\pi \mathcal{R}}\left(\ln\frac{8\mathcal{R}}{a_0}
- 1\right), \label{RingVelocity}
\end{equation}
\begin{equation}
P=\pi\rho_\mathrm{s}\kappa \mathcal{R}^2. \label{RingImpulse}
\end{equation}
By integrating these equations, one can calculate the trajectories
of the charged vortex rings subject to a particular electric field
\citep{WalmsleyCVRTrajectories}.

\subsubsection{Measuring the vortex line density $L$ by ion scattering}

In the absence of vortex lines the ions would propagate either
freely (these dominate at $T>0.8$~K) or riding on small vortex
rings (dominate at $T<0.6$~K). Without other vortex lines in their
way, the time of flight of both species over a particular distance
is a well-defined  function of temperature, electric field, and
initial radius of the attached vortex ring (if any). Hence, after
injecting a short pulse of such ions, a sharp pulse of current
arrives at the collector as shown in Fig.~\ref{MancPulseExamples}.
The interaction with other existing vortices, that happened to be
in the way of the propagating ions, is characterized by a
``trapping diameter'' $\sigma$ \citep{SchwarzDonnelly1966,
OstermeierGlaberson:1974} and leads to the depletion of the
pulses of the collector current. This is used to measure the
average density of vortex lines $L$ between the injector and
collector through
\begin{equation}
L(t) = (\sigma h)^{-1}\ln(I(\infty)/I(t)).
\end{equation}
To calibrate the value of $\sigma$ for either free negative ions
or ions trapped on a vortex ring at different temperature and
electric field, one can measure the attenuation of the pulses of
the collector current when the cryostat is at continuous rotation
at angular velocity $\Omega$ (thus having an equilibrium density
of rectilinear vortex lines $L=2\Omega/\kappa$) as in figure
\ref{MancCrossSections}. This is best done in the direction
perpendicular to the rotation axis.

\begin{figure}[t]
\begin{center}
\includegraphics[width=0.8\linewidth,keepaspectratio]{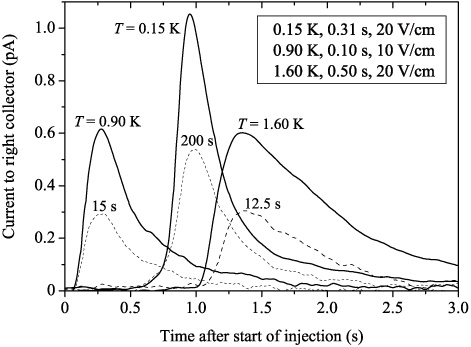}
\caption{Examples of current transients produced by short pulses
for three different temperatures (the temperatures, pulse
durations and mean driving fields are indicated in the legend).
The solid lines show the transients without a vortex tangle in the
ions' path, while the dashed ones represent the transients
suppressed by the vortex  tangle which has been decaying a
specified time (indicated near curves) after stopping generation.
The charge carriers are either free ions ($T=1.60$~K and
$T=0.90$~K) or charged vortex rings ($T=0.15$~K). The electronics
time constant is 0.15~s, hence the time of arrival of the fastest
peak ($T=0.90$~K) cannot be resolved.}
 \label{MancPulseExamples}
\end{center}
\end{figure}

\begin{figure}[t]
\begin{center}
\includegraphics[width=0.8\linewidth,keepaspectratio]{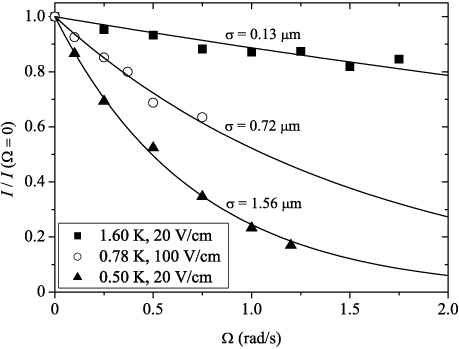}
\caption{Dependence of pulse amplitude on the angular velocity of
rotation, $\Omega$, in examples of calibration measurements. The
temperatures, driving fields and trapping diameters $\sigma$ are
indicated. The charge carriers are either free ions ($T=1.60$~K)
or charged vortex rings ($T=0.78$~K and $T=0.50$~K). }
 \label{MancCrossSections}
\end{center}
\end{figure}

At low temperatures in quantum cascade Kelvin waves of a
broad range of wavelengths are excited, however the main
contribution to the total length $L$ converges quickly at scales
just below $\ell$ \citep{Svistunov:2008b}. Hence, a probe ion with
the trapping diameter $\sigma \ll \ell$, moving at a speed ($v
\sim 10$~cm/s) much greater than the characteristic velocities of
vortex segments ($\sim \kappa/\ell < 10^{-1}$~cm/s), should sample
the full length $L$. Experimental measurements with different
$\sigma = 0.4$--1.7~$\mu$m (using charged vortex rings in a range
of driving electric fields) indeed produce consistent values of
$L$.

There are evidences that small concentrations of trapped space
charge (when the ratio of trapped space charge density $n$ to the
vortex line length $L$, does not exceed $n/L \sim 10^5$~cm$^{-1}$)
do not affect the tangle dynamics \citep{WalmsleyChargedTangle}.
For example, experiments generating quasiclassical tangles at
$T\geq 0.7$~K by a jet of ions (and hence resulting in a
substantial trapped space charge) revealed that the late-time
decay is identical to that for the tangles generated by an
impulsive spin-down to rest without injecting any ions. Still, one
might worry that the very presence of trapped charge, as it
interacts with the beam of probe ions, might affect the result of
measuring $L$. Hence, in order not to introduce extra turbulence
and not to contaminate the tangle with any ions, the measurement
was performed by probing each realization of turbulence only once
-- after a particular waiting time $t$ during its free decay. Then
the contaminated tangle was discarded and a fresh tangle
generated.

\subsubsection{Design of ion experiment}

\begin{figure}[t]
\begin{center}
\includegraphics[width=0.8\linewidth,keepaspectratio]{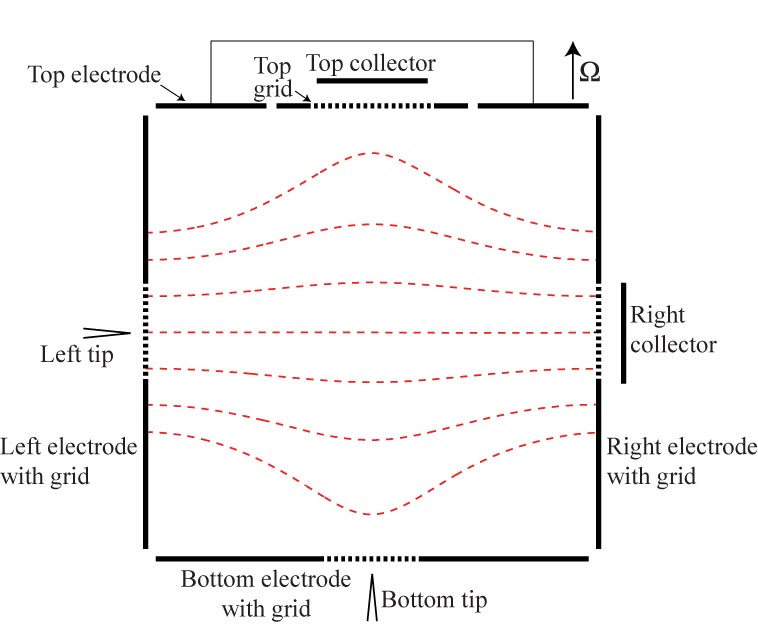}
\caption{A side cross-section of the Manchester cell. The distance
between opposite electrode plates is 4.5~cm. An example of the
driving electric field is shown by dashed lines. Such a
configuration was used in all the Manchester experiments described
in this section. It was calculated for the following potentials
relative to the right electrode: left electrode at -90~V, side,
top and bottom electrodes at -45~V, and the right electrode and
collector at 0 producing a 20 V/cm average driving field in the
horizontal direction. To inject ions, the left tip was usually
kept at  between -500 to -350 V relative to the left grid. When
ions traveling vertically across the cell (injected from bottom
tip and detected at the top collector) were required, the
potentials on the electrodes were rearranged as appropriate. }
\label{MancCellSchematic}
\end{center}
\end{figure}

The cell used in the Manchester experiments had cubic geometry
with sides of length $h=4.5$~cm. A schematic drawing of the cell
is shown in Fig.~\ref{MancCellSchematic}. The relatively large
size of the cell was important to enhance the efficiency of ion
trapping and the time resolution of vortex dynamics, and was
really instrumental to ensure that the continuum limit $\ell \ll
h$ holds even when the vortex line density drops to as low as just
$L \sim 10$~cm$^{-2}$. This also helped ensure that the presence
of the walls, which might accelerate the decay of turbulence
within some distance $\sim \ell$, does not affect the dynamics of
the turbulent tangle in the bulk of the cell. In order to probe
the vortex densities along the axial and transverse directions,
there were two independent pairs of injectors and collectors of
electrons. Both injectors and collectors were protected by
electrostatic grids, enabling injecting and detecting pulses of
electrons. The injectors were field emission tips made of 0.1~mm
diameter tungsten wire \citep{GolovIshimoto}.  The threshold for
ion emission was initially $\simeq$ -100~V and -210 V for the
bottom and left injectors respectively; however, after over two
years of almost daily operation they changed to some -270~V and
-520~V. The fact that the two injectors had very different
threshold voltages helped investigate the dependence (or rather
lack of it) of the radius of initial charged vortex rings on the
injector voltage.
 The six side plates
(electrodes) that make up the cube can be labeled ``top'',
``bottom'', ``left'', ``right'', and two ``side'' electrodes. The
top, bottom, left and right electrodes had circular grids in their
centers. All grids were made of square tungsten mesh with period
0.5~mm and wire diameter 0.020~mm, giving a geometrical
transparency of 92\%. The grids in the bottom, left and right
plates had diameter 10~mm and were electrically connected to those
plates. The grid in the top plate had diameter of 13~mm and was
isolated from the top plate. The injector tips were positioned
about 1.5--2~mm behind their grids. The two collector plates were
placed 2.5~mm behind their grids and were typically biased at +10
-- +25~V relative to the grids. Further details can be found in
\cite{WalmsleyCVRTrajectories}.

\begin{figure}[t]
\begin{center}
\includegraphics[width=0.98\linewidth,keepaspectratio]{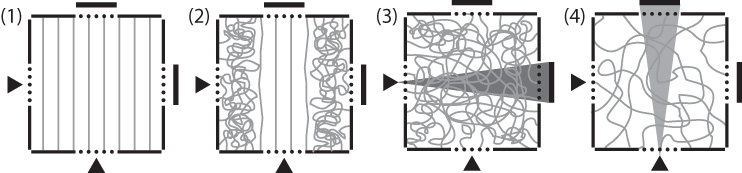}
\caption{Cartoon of the vortex configurations produced by
spin-down in the experimental cell (side view) at different
stages. (1) Regular array of vortex lines during rotation at
constant $\Omega$ before deceleration. (2) Immediately after
stopping rotation ($0 < \Omega t < 10$), turbulence appears at the
outer edges but not on the axis of rotation. (3) After about
30~rad of initial rotation ($\Omega t\sim 30$), 3D homogeneous
turbulence is everywhere. (4) Then ($\Omega t \sim 10^3$) the 3D
turbulence decays with time. Shaded areas indicate the paths of
probe ions when sampling the vortex density in the transverse
($L_{\rm t}$, 3) and axial ($L_{\rm a}$, 4) directions.}
\label{MancCartoonSpinDown}
\end{center}
\end{figure}

\subsubsection{Tangles generated by an impulsive spin-down}

This novel technique of generating quasiclassical turbulence,
suitable for any temperatures down to at least 80~mK, relied on
rapidly bringing a rotating cubic-shaped container of superfluid
$^4$He to rest \citep{Walmsley:2007}. The range of angular
velocities of initial rotation $\Omega$ was 0.05--1.5~rad/s. In
classical liquids at high $Re$, spin-down to rest is always
unstable, especially at high deceleration and in axially
asymmetric geometries. It is known that, within a few radians of
initial rotation upon an impulsive spin-down to rest, a nearly
homogeneous turbulence develops with the energy-containing eddies
of the size of the container. In a cylindrical container,
centrifugal and Taylor-G\"ortler instabilities usually break up at
the perimeter, thus facilitating slowing down of the outer region
of the initially rotating liquid. Simultaneously, because of the
Ekman pumping of non-rotating liquid into the central axis via the
top and bottom walls, the central core of the initial giant vortex
slows down too \citep{Donnelly:1991}. In a cubic container, the
turbulence becomes homogeneous much faster. Some residual rotation
of the central region in the original direction might still
survive for a while but, as we show below, the generated
turbulence is in general pretty homogeneous.

This behaviour is well-documented for classical liquids
\citep{RectContainer}, and one expects similar processes to occur
in a superfluid liquid providing the process of initial
multiplication of vortices does not affect the dynamics. As
spin-down experiments always begin from already existing dense
rectilinear vortex arrays of equilibrium density
$L=2\Omega/\kappa$, and rapid randomization and multiplication of
these vortices is expected due to the lack of axial symmetry of
the container, as well as surface pinning (friction) in the
boundary region, these seem to be sufficient for the superfluid to
mimic the large classical turbulent eddies. In
Fig.~\ref{MancCartoonSpinDown} we show the four different stages
of the evolution from a vortex array to a decaying homogeneous
tangle upon an impulsive spin-down to rest, which are in agreement
with the observations outlined below.

\begin{figure}[t]
\begin{center}
\includegraphics[width=0.98\linewidth,keepaspectratio]{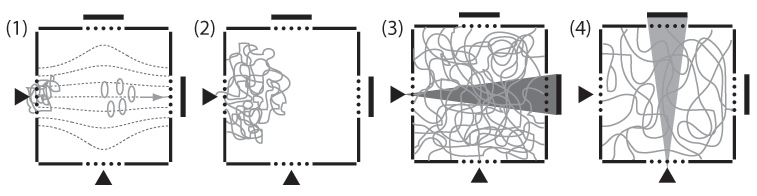}
\caption{Cartoon of the vortex configurations produced by a pulse
of injected ions at $T<0.5$~K in the experimental cell (side view)
at different stages. (1, $<1$~s) A pulse of charged vortex rings
is injected from the left injector. While most make it to the
collector as a sharp pulse, some got entangled near the injector.
(2, $\sim 5$~s) The tangle spreads into the middle of the cell.
(3, $\sim 20$~s) The tangle has occupied all volume; from now on
it is nearly homogeneous (as probed in two directions). (4, up to
1000~s) The homogeneous tangle is decaying further. The shaded
areas indicate the trajectories of ions used to probe the tangle
along two orthogonal directions.} \label{MancCVRCartoon}
\end{center}
\end{figure}

Before making each measurement, the cryostat was kept at steady
rotation at the required $\Omega$ for at least 300~s  before
decelerating to full stop, then waiting a time interval $t$ and
taking the data point. Then the probed tangle was discarded and a
new one generated. Hence, different data points represent
different realizations of the turbulence. The deceleration was
linear in time taking 2.5~s for $\Omega = 1.5$~rad/s and 0.1~s for
$\Omega = 0.05$~rad/s. The origin $t=0$ was chosen at the start of
deceleration.

\subsubsection{Current-generated tangles}

An alternative technique of generating turbulence, by a jet of
injected ions that does not require any moving parts in the
cryostat, has been developed \citep{Walmsley2008}
(Fig.~\ref{MancCVRCartoon}). In early experiments with injected
ions at low temperatures \citep{Bruschi:1966,McClintockTangle}, it
was observed that a pulse of negative ions through superfluid
$^4$He leaves behind a tangle of vortices. \cite{Walmsley:2007}
found that the properties of these tangles can be quite different.
As we explain below, the tangles generated after long injection at
high temperatures possess the properties of developed
quasiclassical turbulence while those produced after short
injection at low temperatures have signatures of random tangles
with little large-scale flow.

\begin{figure}[t]
\begin{center}
\includegraphics[width=0.8\linewidth,keepaspectratio]{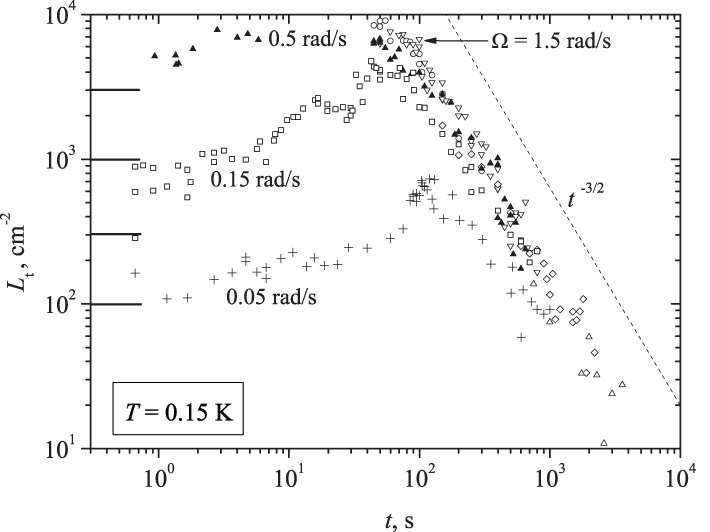}
\caption{$L_{\rm t}(t)$ at $T=0.15$~K for four values of $\Omega$.
The average driving fields used for $\Omega = 1.5$~rad/s: 5~V/cm
($\circ$), 10~V/cm ($\bigtriangledown$), 20~V/cm ($\diamond$),
25~V/cm ($\bigtriangleup$). At other values of $\Omega$ the
electric fields used were either 20~V/cm (0.05~rad/s) or 10~V/cm
(0.5 and 0.15~rad/s). The dashed line shows the dependence
$t^{-3/2}$. The horizontal bars indicate the initial vortex
densities at steady rotation, $L=2\Omega/\kappa$, at
$\Omega=1.5$~rad/s, 0.5~rad/s, 0.15~rad/s and 0.05~rad/s (from top
to bottom).}
\label{MancRawSpinDownT}
\end{center}
\end{figure}

At high temperatures $T>0.7$~K, while moving relative to the
normal component, the ions experience viscous drag through
scattering of excitations -- this means that the ions entrain the
normal component along. They can also get trapped by an existing
vortex line (provided $T < 1.8$~K), after which the vortex line
will be pulled along with the current of such ions. The existing
vortices and hence the superfluid component, through the action of
mutual friction, will be pulled by the already entrained normal
component too, and vice versa. Then one can expect the injected
current to produce a large-scale jet-like flow of liquid helium
which is an efficient way of driving quasiclassical turbulence at
large scales. This is similar to jet flow through an orifice --
one of the traditional means of generating turbulence in classical
liquids \citep{Grenoble}. On the other hand at low temperatures
the ions always bring small vortex rings along, hence, upon
formating a tangle, can directly pump up the vortex length $L$
without introducing substantial large-scale flow -- at least for
not too-long injections.

\subsection{Experimental results}

\subsubsection{Quasiclassical turbulence generated by spin-down to rest}

Above 1\,K, the $L\propto t^{-3/2}$ free decay was monitored by
second sound for towed grid turbulence in Oregon
\citep{Stalp1999}. The recent spin-down experiments at Manchester
are in good agreement with them in the overlapping temperature
regime. Below 1\,K, only scattering of ions off vortex lines has
been used to measure $L$ so far.

\begin{figure}[t]
\begin{center}
\includegraphics[width=0.8\linewidth,keepaspectratio]{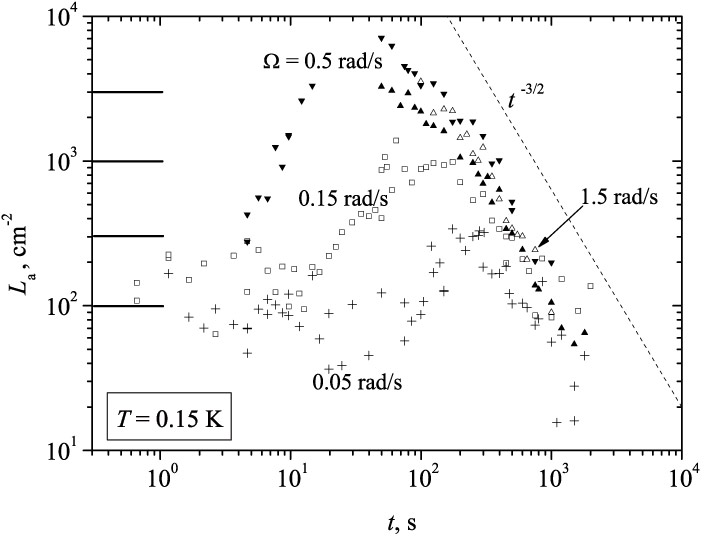}
\caption{$L_{\rm a}(t)$ at $T=0.15$~K for four values of $\Omega$.
The average driving field used was 20~V/cm in all cases except at
0.5~rad/s where both 10~V/cm ($\bigtriangledown$) and 20~V/cm
($\bigtriangleup$) were used. The dashed line shows the dependence
$t^{-3/2}$. The horizontal bars indicate the initial vortex
densities at steady rotation, $L=2\Omega/\kappa$, at
$\Omega=1.5$~rad/s, 0.5~rad/s, 0.15~rad/s and 0.05~rad/s (from top
to bottom).}
\label{MancRawSpinDownA}
\end{center}
\end{figure}

In Fig.~\ref{MancRawSpinDownT}, the measured densities of vortex
lines along the horizontal axis (transverse, $L_{\rm t}$) are
shown for four different initial angular velocities $\Omega$.
During the transient, which lasts some $\sim 100\Omega^{-1}$,
$L_{\rm t} (t)$ goes through the maximum after which it decays
eventually reaching the universal late-time form of $L \propto
t^{-3/2}$. For $\Omega \geq 0.5$~rad/s, the values of $L$ at
maximum were too high to be detected. The initial vortex densities
at steady rotation, $L=2\Omega/\kappa$, are shown by horizontal
lines. Similarly, the densities of vortex lines measured along the
vertical axis (axial, $L_{\rm a}$) are shown in
Fig.~\ref{MancRawSpinDownA}.

\begin{figure}[t]
\begin{center}
\includegraphics[width=0.8\linewidth,keepaspectratio]{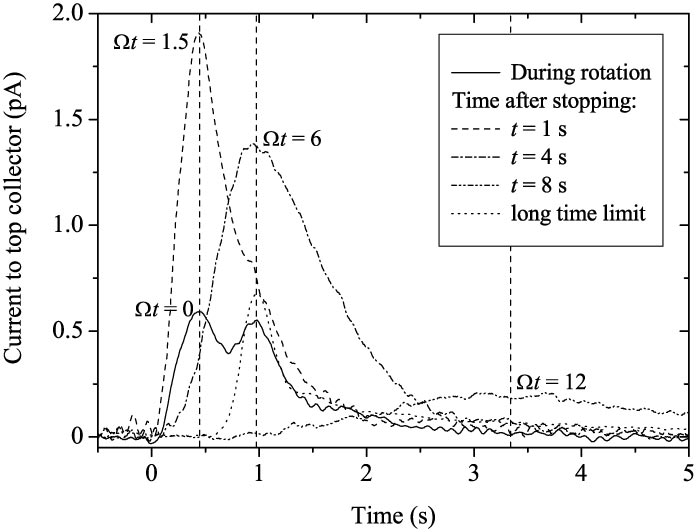}
\caption{Records of the current to the top collector, injected
from the bottom injector as a 0.1~s-long pulse at the time $t$
after an impulsive spin-down from $\Omega = 1.5$~rad/s to rest. $T
=0.15$~K, $E= 20$~V/cm.}
\label{MancSpinDownTransients}
\end{center}
\end{figure}

To illustrate what is happening near the vertical axis of the
container at different stages of the transient after a spin-down,
in Fig.~\ref{MancSpinDownTransients} we show five records of the
current to the top collector arriving after a short (0.1~s) pulse
of probed ions was fired from the bottom injector. Each is
characteristic of a particular configuration of vortex lines near
the rotational vertical axis of the container during the
transformation from an array of parallel lines to a homogeneous
decaying vortex tangle. One can see three different characteristic
times (vertical dashed lines in Fig.~\ref{MancSpinDownTransients})
of arrival of ions via different means. The first peak at $\approx
0.4$~s (determined by the time constant of the current
preamplifier, 0.15~s) corresponds to the ions trapped on the
rectilinear vortex lines which can slide along those lines very
quickly, provided those lines are continuous from the injector to
collector as during steady rotation. The second peak at $\approx
1$~s corresponds to the coherent arrival of the ballistic charged
vortex rings from the bottom to the top. The third broad peak at
times $\sim 3$~s (but with a long tail detectable until $\sim
40$~s) corresponds to the charge trapped on the vortex tangle and
drifting with the tangle. Hence, we have the following regimes
(and curves in Fig.~\ref{MancSpinDownTransients}) labeled by the
time $t$ after the spin-down:
\begin{enumerate} \item{$\Omega t\leq 0$, steady
rotation. The nearly equal first and second peaks tell that there
exist a vortex array without disorder (otherwise the second peak
would get suppressed and the third would appear).}
\item{$\Omega t = 1.5$. The first peak gets enhanced three-fold
while the second one is still there (and no third peak) -- meaning
more trapped ions can now reach the collector along the
rectilinear vortex lines while there is not much turbulence in
this region yet.} \item{$\Omega t = 6$. The first peak has
disappeared in favour of the second one which has got broadened --
at this stage the rectilinear vortices should have become
scrambled in the Ekman layers  near the top and bottom walls/grids
while the ballistic charged vortex rings are still the dominant
transport of charge.} \item{$\Omega t = 12$. Now both fast peaks
have disappeared completely while the third broad peak has emerged
-- this means that a turbulent tangle has finally reached the
axial region.} \item{$\Omega t \rightarrow \infty$. The sharp
second peak has recovered but all others vanished -- after the
turbulence has decayed only ballistic charged vortex rings carry
the charge, neither the rectilinear array nor the turbulent tangle
contribute to the transport any more.} \end{enumerate}

\begin{figure}[t]
\begin{center}
\includegraphics[width=0.8\linewidth,keepaspectratio]{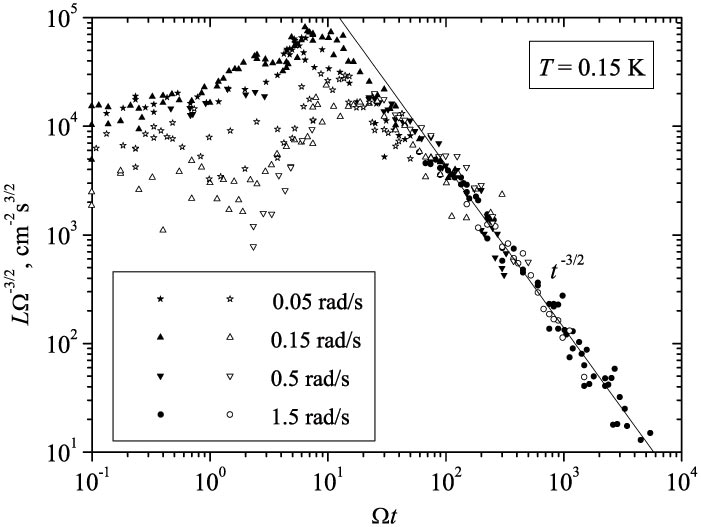}
\caption{$\Omega^{-3/2}L_{\rm t}(t)$ (filled symbols) and
$\Omega^{-3/2}L_{\rm a}(t)$ (open symbols) vs. $\Omega t$ for four
values of $\Omega$ at $T=0.15$~K. The straight line $\propto
t^{-3/2}$ guides the eye through the late-time decay.}
\label{MancSpinDownAT}
\end{center}
\end{figure}

In Fig.~\ref{MancSpinDownAT} the measured densities of vortex
lines along the horizontal, $L_{\rm t}$, and vertical, $L_{\rm
a}$, axes are shown, by solid and open symbols respectively. To
stress the scaling of the characteristic times with the initial
turn-over time $\Omega^{-1}$ and the universal late-time decay
$\propto t^{-3/2}$, the data for different $\Omega$ are rescaled
accordingly. We can see that at all $\Omega$ the transients are
basically universal. Immediately after deceleration, $L_{\rm t}$
jumps to $\approx 10^4\Omega^{3/2}$~cm$^{-2}$s$^{3/2}$, indicating
the appearance of the turbulent boundary layer at the perimeter,
while $L_{\rm a}$ is stable at $L_{\rm i} \approx 2\Omega/\kappa$.
Only at $\Omega t \approx 3$, the latter starts to grow,
signalling the destruction of the rotating core with vertical
rectilinear vortices. After passing through a maximum at $\Omega t
= 8$ and $\Omega t = 15$ respectively, $L_{\rm t}$ and $L_{\rm a}$
merge at $\Omega t \sim 30$ and then become indistinguishable.
This implies that from now on the tangle density is distributed
nearly homogeneously. The scaling of the transient times with the
turn-over time $\Omega^{-1}$ indicates that transient flows are
similar at different initial velocities $\Omega$, which is
expected for flow instabilities in classical inviscid liquids.
Eventually, after $\Omega t \sim 100$, the decay takes its
late-time form $L \propto t^{-3/2}$ expected for quasiclassical
isotropic turbulence, whose energy is mainly concentrated in the
largest eddies bound by the container size $h$, but homogeneous on
smaller length scales. We hence assume that the turbulence in
$^4$He at this stage is nearly homogeneous and isotropic, and can
apply Eq.~(\ref{L-t-K}) to extract the effective kinematic
viscosity $\nu$.
The cross-over to the $L\propto t^{-1}$ regime at late time has
never been observed, probably because the measured values of $L$
never dropped below 10~cm$^{-2} \gg h^{-2} \sim 0.05$~cm$^{-2}$.

\begin{figure}[t]
\begin{center}
\includegraphics[width=0.8\linewidth,keepaspectratio]{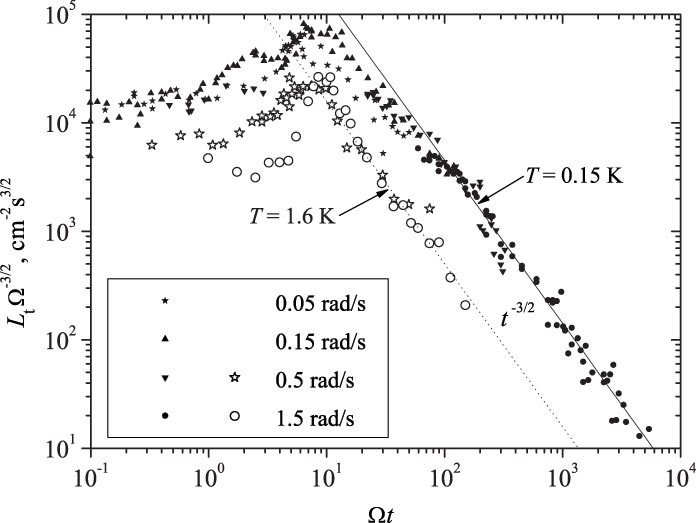}
\caption{$\Omega^{-3/2}L_t(t)$ vs. $\Omega t$ for $T=0.15$~K
(filled symbols) and $T=1.6$~K (open symbols). Dashed and solid
lines $\propto t^{-3/2}$ guide the eye through the late-time decay
at $T=1.6$~K and 0.15~K, respectively.}
\label{MancSpinDownHighLowTemp}
\end{center}
\end{figure}

At $0.08<T<0.5$~K the measured $L(t)$ were independent of
temperature. In Fig.~\ref{MancSpinDownHighLowTemp}, we compare the
transients $L_{\rm t}(t)$ at low ($T=0.15~K$) and high ($T=1.6~K$)
temperatures. The prefactor in the late-time dependence $L \propto
t^{-3/2}$ at $T=0.15$~K is almost an order of magnitude larger
that that for $T=1.6$~K. This implies that at low temperatures,
the steady-state inertial cascade with a saturated
energy-containing length and constant energy flux down the range
of length scales  requires a much greater total vortex line
density. This means that in the $T=0$ limit the effective
kinematic viscosity $\nu$ is approximately 70 times smaller than
at $T=1.6$~K (provided $C \approx 1.5$ and $k_1 \approx 2\pi/h$
are independent of temperature).

\begin{figure}[t]
\begin{center}
\includegraphics[width=0.8\linewidth,keepaspectratio]{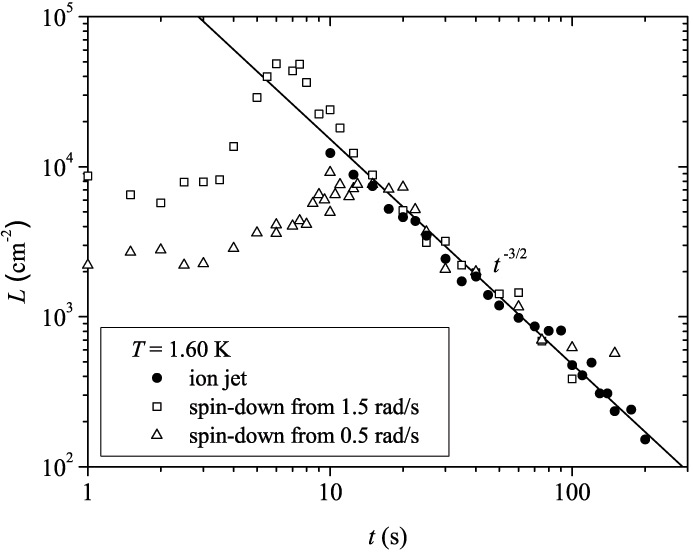}
\caption{Free decay of a tangle produced by a jet of free ions
from the bottom injector ($\bullet$) \citep{Walmsley2008}, as well
as by an impulsive spin-down to rest ($\Box$)
\citep{Walmsley:2007} from 1.5~rad/s and 0.5~rad/s, at $T=1.60$~K.
All tangles were probed by pulses of free ions in the horizontal
direction. The line $L\propto t^{-3/2}$ corresponds to
Eq.~(\ref{L-t-K}) with $\nu = 0.2\kappa$.}
\label{MancIonJetHT}
\end{center}
\end{figure}

\begin{figure}[t]
\begin{center}
\includegraphics[width=0.8\linewidth,keepaspectratio]{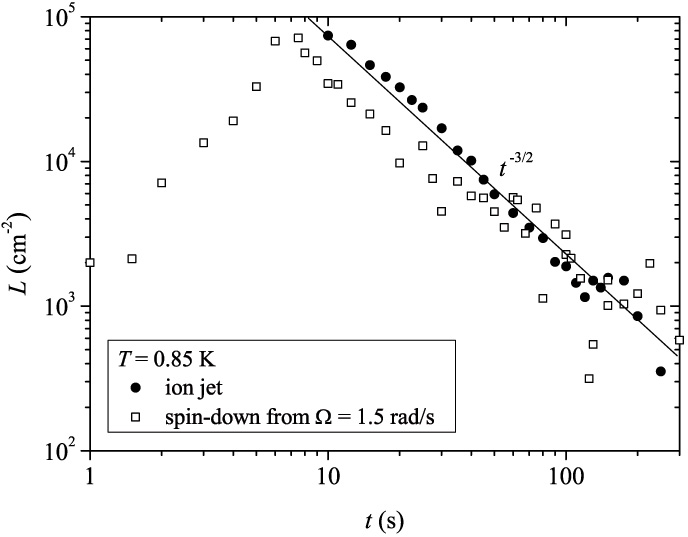}
\caption{Free decay of a tangle produced by a jet of free ions
from the bottom injector ($\bullet$) \citep{Walmsley2008}, as well
as by an impulsive spin-down to rest \citep{Walmsley:2007} from
1.5~rad/s, at $T=0.85$~K. All tangles were probed by pulses of
free ions in the horizontal direction. The ion jet data are the
average of nine measurements at each particular time but the
spin-down data show individual measurements.}
\label{MancIonJetLT}
\end{center}
\end{figure}

At all temperatures the transients $L_{\rm t}(t)$ after a
spin-down are, in first approximation, universal, \textit{i.e.}
the timing of the maximum is the same $\Omega  t\approx 7$ and the
amplitudes of the maximum are comparable. This supports our
approach to turbulent superfluid helium at large length scales as
to an inviscid classical liquid, agitated at large scale and
carrying an inertial cascade down the lengthscales, independent of
temperature. On the other hand, as the temperature increases and
interactions with the viscous normal component become stronger,
changes in the shape of transients might be expected. Indeed, one
can see that the slope of $L(t)$ after the maximum but before
reaching the ultimate late-time decay $L\propto t^{-3/2}$
(\textit{i.e.} for $10 < \Omega t < 100$) is changing gradually
with increasing temperature from being less steep than $t^{-3/2}$
at $T=0.15$~K to more steep than $t^{-3/2}$ at $T=1.6$~K
(Fig.~\ref{MancSpinDownHighLowTemp}). At temperatures around $T
=0.85$~K, it nearly matches $t^{-3/2}$, thus making an erroneous
determination of $\nu$ possible by taking this part of the
transient for the late-time decay $L \propto t^{-3/2}$. Indeed, in
the first publication \citep{Walmsley:2007}, parts of some
transient at $T=0.8-1.0$~K for as early as $\Omega t
> 15$ were occasionally used to be fitted by $L \propto t^{-3/2}$
that often resulted in overestimation of the value for $\nu$. To
rectify this, we have re-fitted the datasets used in the original
publication as well as subsequent measurements with $L\propto
t^{-3/2}$ for spin-downs using the following rules: for
$T\leq0.5$~K only points for $\Omega t > 300$ were used, between
0.5~K and 1.0~K only points for $\Omega t>150$ used and at
$T>1.0$~K only $\Omega t > 75$ were used. This resulted in slight
systematic reduction of the extracted values of $\nu(T)$ at
temperatures 0.8--1.2~K from those published in
\cite{Walmsley:2007}; what looked as a rather steep drop in
$\nu(T)$ at 0.7--0.8~K, now occurs at 0.85 -- 0.90\,K and is
somewhat reduced in magnitude.

\subsubsection{Quasiclassical turbulence generated by an ion jet}

As mentioned above, ion jets can produce quasiclassical tangles
decaying as $L\propto t^{-3/2}$ \citep{Walmsley2008}.
Examples of the decay of such tangles alongside with those
obtained by a spin-down at $T=1.6$~K and $T=0.8$~K are shown in
Figs.~\ref{MancIonJetHT} and \ref{MancIonJetLT}, demonstrating
good quantitative agreement between the late-time decays. At temperatures
0.7 -- 1.6~K the late-time decays of quasiclassical
turbulences generated by these two different techniques were
identical within the experimental errors.

\subsubsection{Ultraquantum turbulence generated by a jet of charged vortex rings}

\begin{figure}[t]
\begin{center}
\includegraphics[width=0.8\linewidth,keepaspectratio]{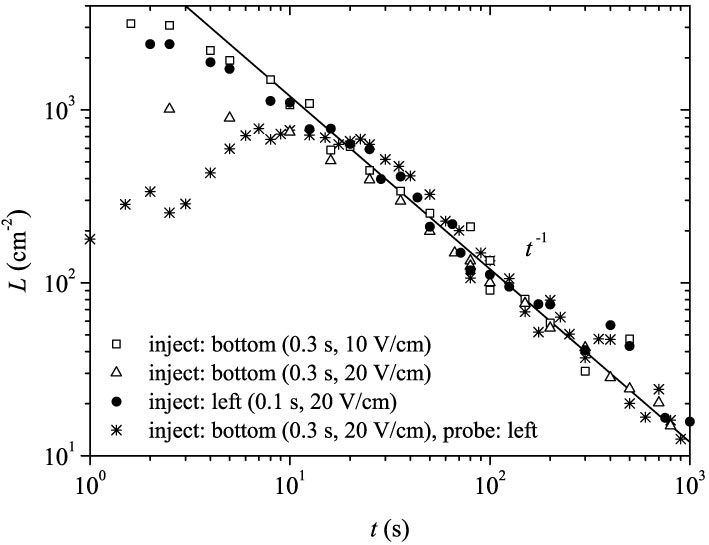}
\caption{Free decay of a tangle produced by beams of charged
vortex rings of different durations and densities, $T=0.15$~K. The
injection direction and duration, and driving field are indicated.
Probing with pulses of charged vortex rings of duration 0.1--0.3~s
were done in the same direction as the initial injection, except
in one case ($*$).  The line $L\propto t^{-1}$ corresponds to
Eq.~(\ref{L-t-V}) with $\nu = 0.1\kappa$.}
\label{MancCVRTransient}
\end{center}
\end{figure}

In the temperature range $0<T<0.5$~K, all tangles produced by a
pulse of injected current of duration 0.1--1~s, intensity
$10^{-12}$--$10^{-10}$~A and in the driving field $0-20$~V/cm
revealed the late-time decay $L\propto t^{-1}$, all with the same
prefactor (Fig.~\ref{MancCVRTransient}) \citep{Walmsley2008}. This
universality of the prefactor for all initial vortex densities is
a strong argument in favour of the random character of the tangles
whose decay is described by Eq.~(\ref{L-t-V}) (ultraquantum
turbulence). During the injection, the tangle originates near the
injector, presumably as the result of colliding many charged
vortex rings all of the same radius $R\approx 0.5$~$\mu$m, and
then spreads into the centre of the cell in 3--5~s, eventually
filling all container and becoming nearly homogeneous after $\sim
20$~s. This can be seen at the transient in
Fig.~\ref{MancCVRTransient} ($*$) where the tangle was initiated
at the bottom injector but then sampled along the horizontal axis
of the container. The dynamics of the tangle spreading was found
to be independent of the driving field 0--20~V/cm.

The tangles produced by the bottom and left injectors had very
nearly identical late-time decays $L\propto t^{-1}$
(Fig.~\ref{MancCVRTransient}). The corresponding values of
$\nu/\kappa$, inferred using Eq.~(\ref{L-t-V}), are $0.120 \pm
0.013$ and $0.083 \pm 0.004$. As the scattering diameters $\sigma$
of charged vortex rings produced only by the left injectors could
be calibrated directly {\it in situ} on the arrays of rectilinear
vortex lines at steady rotation, the absolute value of $\nu=
0.08\kappa$ for these tangles probed along the horizontal axis is
treated as more reliable.

\subsection{Discussion: dissipation in different types of turbulence}

\subsubsection{Dissipation in $^4$He at different temperatures}

\begin{figure}[t] \begin{center}
\includegraphics[width=0.8\linewidth,keepaspectratio]{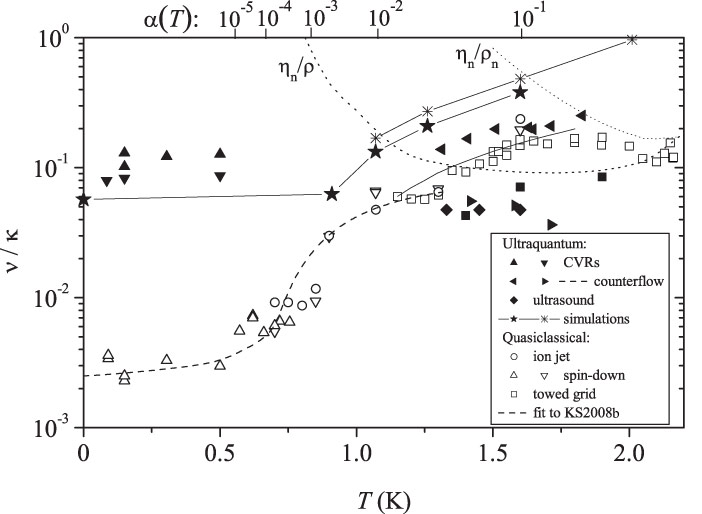}
\caption{The effective kinematic viscosity $\nu(T)$ for various
types of turbulence in $^4$He. \textbf{(i)} Quasiclassical
turbulence: $\nu(T)$ inferred from the $L\propto t^{-3/2}$
dependent free decay of tangles using Eq.~(\ref{L-t-K}) and
produced by impulsive spin-down (open triangles
\cite{Walmsley:2007}); an ion jet (open circles
\cite{Walmsley2008}); towed grids of two different designs (open
squares \cite{Stalp2002,Niemela2005}). The solid theoretical curve
represents Eq.~(\ref{EqVinenNiemela}) \citep{Vinen:2002} and the
dashed curve is a fit to the \cite{Walmsley:2007} data (open
triangles and circles) by \cite{Svistunov:2008b}.\textbf{(ii)}
Ultraquantum turbulence: $\nu(T)$ inferred using Eq.~(\ref{L-t-V})
from $L\propto t^{-1}$ dependent free decay of tangles produced by
colliding charged vortex rings (filled up- and down- triangles
\cite{Walmsley2008}); counterflow (filled right-pointing triangles
\cite{Vinen1957b} and filled squares \cite{RozenSchwarz});
ultrasound (filled diamonds \cite{Milliken}); computer simulations
(line-connected stars \cite{Tsubota2000}); as well as from the
analysis of the measured tangle density in applied counterflow
(filled left-pointing triangles \cite{Vinen1957b}); and computer
simulations (line-connected asterisks \cite{Schwarz:1988}). Note
that in simulations ($\star$ and $*$) the normal component was
artificially clamped to laminar flow, while in most experiments at
$T>1$~K it is involved in turbulent motions to different extent.
The values of the kinematic viscosities, $\eta_{\rm n}/\rho_{\rm
n}$ of the normal component
 and $\eta_{\rm n}/(\rho_{\rm n}+\rho_{\rm s})$ of
the ``coupled normal and superfluid components'' are shown by
dotted lines \citep{Donnelly:1998}. }

\label{MancNuPrime} \end{center} \end{figure}

Let us now summarize what is known about the dissipation of
various types of turbulence in superfluid $^4$He at different
temperatures. As the relevant parameter, we plot in
Fig.~\ref{MancNuPrime} the effective kinematic viscosity $\nu$ (as
defined in Eq.~(\ref{EDot})) which is a function of the mutual
friction parameter $\alpha(T)$ but can be different for different
types of flow. One can see that at high temperatures, $T>1$~K, all
experimental points group around $\sim 0.1\,\kappa$, give or take
a factor of two. The fact that different types of experiments seem
to produce slightly different absolute values and temperature
dependences might have various reasons. Firstly, experimental
techniques rely on the means of calibrating the sensitivity to the
absolute value of $L$ and on the knowledge of other parameters in
the model. The calibration is normally performed on an array of
rectilinear vortex lines in the direction perpendicular to the
lines but not on tangles of vortex lines. For example, recent
refined calculations \citep{ChagovetzSkrbek} suggested to correct
all previous second sound measurements of $L$ on isotropic tangles
by a factor of $3\pi/8 \approx 1.2$, which we apply here.
Secondly, the exact values and dependence on temperature of the
effective parameters relating $L(t)$ to $\nu$ (such as $B\approx
1.2$ in Eq.~(\ref{L-t-V}), and $C\approx 1.5$ and $k_1/h \approx
2\pi$ in Eq.~(\ref{L-t-K})) are not known, thus complicating
precise comparison of $\nu$ for different temperatures and types
of flow. Thirdly, at $T>1$~K the normal
component can also become turbulent
\citep{RozenSchwarz,Barenghi:2002,Vinen:2000,Lvov:2006b}, hence
its velocity field might vary for different means of generating
turbulence.

Theoretically, one can consider two limiting cases: either
completely locked (${\bf v}_{\rm n} = {\bf v}_{\rm s}$)  turbulent
flow of both components or an absolutely laminar normal component
(${\bf v}_{\rm n}=0$ in a local reference frame) that slows down
the segments of quantized vortices moving past it. The former is
favourable at high mutual friction $\alpha(T)$ and low viscosity
of the normal component $\eta_{\rm n}(T)$ (\textit{i.e.} at the
high-temperature end) and is expected to be described by the
kinematic viscosity $\eta_{\rm n}/\rho$ (dotted line in
Fig.~\ref{MancNuPrime}), where $\rho = \rho_{\rm n} + \rho_{\rm
s}$, while the latter is favourable at low $\alpha$ and high
$\eta_{\rm n}$ (\textit{i.e.} at the low-temperature end) and is
expected to follow the predictions of mutual-friction-controlled
models which assume a laminar normal component ({\it e.g.}
line-connected stars and asterisks for random tangles in
Fig.~\ref{MancNuPrime}). The experimental situation is obviously
somewhere in between (and the measured $\nu$ should probably tend
to the smaller value of those predicted by the two limits),
although the particular means of driving might tip the balance
towards either of the limits. This might be one reason why
different techniques show systematic disagreement in the
temperature range above 1\,K.

One can speculate that quasiclassical turbulent flow generated by
mechanical stirring is less prone to such uncertainties, since
both the superfluid and normal large-scale flows are generated
simultaneously and are probably locked. On the other hand, the
generation of quasi-random tangles was attempted by a variety of
techniques: those continuously pumped by counterflow in narrow
channels might have the normal component more laminar, while those
with freely decaying turbulence in wide channels were shown to
evolve towards coupled superfluid and normal turbulence at high
temperatures \citep{BarenghiSkrbek}, especially so if created with
ultrasound which equally pumps turbulence in both components. In
any case, on approaching $T=1$~K from above, when the normal
component progressively becomes more viscous and less dense, all
experimental and numerical values of $\nu(T)$ for quasi-random
tangles seem to converge, which supports these reasonings.

At temperatures below 1~K, the values of $\nu(T)$ appear to split
rapidly: those for ultraquantum turbulence apparently stay at
nearly the same level, $\sim 0.1 \, \kappa$, as at high
temperatures, while those for quasiclassical turbulence keep
decreasing until the zero-temperature limit of $\sim 0.003 \,
\kappa$. The mean free path of excitations in $^4$He rapidly
increases with decreasing temperature and becomes comparable with
$\ell\sim 100$~$\mu$m at $0.7$~K and with the container size
$h=4.5$~cm at $0.5$~K. At these temperatures, the normal component
is too viscous to follow the turbulent superfluid component and
hence the convenient model of a laminar normal component becomes
adequate here (in this respect this regime of $^4$He is similar to
that in superfluid $^3$He-B below $ 0.4\,T_{\rm c}$ discussed in
previous sections). It is indeed comforting that numerical
simulations of tangles initiated at short scales (without
large-scale turbulence \citep{Tsubota2000}) are in good agreement
with experiments for random tangles at these temperatures
\citep{Walmsley2008}. As yet, there are no satisfactory computer
simulations of homogeneous quasiclassical turbulence in the limit
of zero temperature.

Between 1.1 -- 1.6~K, the values of $\nu(T)$ for quasiclassical
turbulence generated with towed grids of different designs
\citep{Stalp1999,Niemela2005}, impulsive spin-down to rest, and an
ion jet are reasonably consistent between each other and show an
increase from $\nu = 0.05 \, \kappa$ to $0.2 \, \kappa$. There was
an early attempt \citep{Stalp1999} to relate these values with the
kinematic viscosity  $\eta_{\rm n}/\rho = 10^{-4}$~cm$^2$/s $
\approx 0.1 \,\kappa$, assuming locked turbulence of the
superfluid and normal components. However, it was soon realized
\citep{Stalp2002} that the contribution of quantized vortex lines
to dissipation controlled by mutual friction and reconnections is
also important. This certainly dominates below 1.2\,K, where
$\eta_{\rm n}/\rho$ begins to increase with cooling (hence, making
the normal component effectively decoupled from vortices on short
length scales) while the measured $\nu(T)$ decreases with cooling.

In the mutual-friction-dominated regime and provided the normal
component is only locked to the superfluid velocity at scales $>
\ell$, the actual dissipation is through the self-induced motion
of vortex lines relative to the normal component -- whatever the
kinetics of the quantum cascade. In this regime, a segment of a
vortex line bent at radius $\cal{R}$ and hence moving at
self-induced velocity $v \sim B\kappa \cal{R}$$^{-1}$ dissipates
energy at the rate $\sim \rho_{\rm s}\alpha B^2 \kappa^3
\cal{R}$$^{-2}$ per unit length, where $B\approx 1.2$ is
introduced in Eq.~(\ref{L-t-V}). This approach leads to the
formula \citep{Vinen:2000,Vinen:2002}
\begin{equation}
    \nu = s B^2 c_2^2 \alpha \kappa \, .
    \label{EqVinenNiemela}
\end{equation}
The multiplier $s$ is meant to account for the degree of
correlation between the motion of individual segments of vortex
lines and the surrounding superfluid in a structured
quasiclassical tangle ($s \approx 0.6$), compared to  a random
tangle of the same density $L$ ($s = 1$). In the simulations of
counterflow-maintained tangles in the local induction
approximation by \cite{Schwarz:1988}, the measure of the relative
abundance of small scale kinks on vortex lines, $c_2^2 = L^{-1}\,
\langle \cal{R}$$^{-2} \rangle$, was found to increase from 2 to
12 as $\alpha$ decreased from 0.3 to $10^{-2}$ with decreasing
temperature from 2 to 1\,K. This formula nicely agrees with the
simulations by \cite{Tsubota2000}. It also captures the trend and
magnitude of $\nu(T)$ for all experimental quasiclassical tangles
as well as for the random tangle maintained by counterflow
\citep{Vinen1957a} (where the normal component is expected to be
nearly laminar); although to fit it to the data of quasiclassical
tangles in Fig.~\ref{MancNuPrime} (solid line) we used $s = 0.2$,
apparently owing to partial locking between the flows in the
normal and superfluid components.

As vortex segments with smaller radii of curvature $\cal{R}$ lose
their energy faster, for a given energy flux down the length
scales there is exists a  dissipative scale $\lambda_{\rm q}$
below which the cascade essentially cuts off
\citep{Svistunov:1995,Vinen:2000,Lvov:2006b,Svistunov:2008b}. For
a developed non-linear Kelvin-wave cascade ($ \lambda_{\rm q} \ll
\ell$), with the amplitude spectrum $b_{\rm k} \propto k^{-6/5}$
\citep{Svistunov:2004}, this implies that $\langle \cal{R}$$^{-2}
\rangle \sim \ell^{-2/5} \, \lambda_{\rm q}^{-8/5}$. In this sense
the Schwarz's parameter $c_2^2 \sim (\ell/\lambda_{\rm q})^{8/5}$
quantifies the range of wavelengths involved in the quantum
cascade and becomes another quantum analog of the classical
Reynolds number: $Re = (h/\lambda_{\rm cl})^{4/3}$. With
decreasing $\alpha$, the value of $\lambda_{\rm q}$ progressively
decreases until, at $T<0.5$~K
\citep{Vinen:2001,Lvov:2006b,Svistunov:2008b}, it reaches the
wavelength $\lambda_{\rm ph}\sim 3\,$nm at which phonons can be
effectively emitted. Below this temperature, as the dissipation of
Kelvin waves due to mutual friction becomes negligible compared to
the emission of phonons, $\nu(T)$ is independent of temperature.
However, in the presence of the quantum cascade at larger
wavelengths, it is the cascade's kinetics that controls the energy
flux, and it might happen that the vortex length $L$ already
saturates at higher values of $\alpha$, thus signalling the onset
of the regime with $\nu(T)= $ const as $T \rightarrow 0$,
Eq.~(\ref{EDot}).

Let us now discuss the specific models for the low-temperature
behaviour of $\nu(T)$. For the dissipative mechanisms that involve
individual quantized vortices, such as those maintained by vortex
reconnections, simple considerations \citep{Vinen:2002} usually
predict $\nu$ to be of order $\kappa$. However, more involved
models, such as that of a ``bottleneck'' for quasiclassical
turbulence, might suggest smaller values \citep{LNR}. If certain
configurations of vortices are inefficient in transferring energy
down to short wavelengths (\textit{e.g.} either there exists a
non-cascading part of the spectrum or reconnections are less
efficient in polarized tangles), the resulting accumulation of an
extra contribution to $L$ leads to a reduction of the parameter
$\nu$. This is what is observed for quasiclassical tangles in
$^4$He on approaching the $T=0$ limit.

An important question remains: how is the energy of classical
eddies passed over to the shorter quantum length scales? Two main
mechanisms are currently discussed: (i) the excitation of Kelvin
waves through purely non-linear interactions in classical eddies
or (ii) vortex reconnections (each leaving a sharp kink on both
vortex lines, hence effectively redistributing the energy to
smaller length scales). Both predict an increase in $L(T)$ at
constant energy flux down the cascade (\textit{i.e.} the decrease
of $\nu(T)$) with decreasing temperature. The former considers the
accumulation of extra ``non-cascading'' quasiclassical vorticity
at length scales above $\ell$ due to the difficulty in
transferring energy through wavenumbers around $\ell^{-1}$  if
reconnections do not ease the process (referred to as the
``bottleneck'' model in Sec.~\ref{s:FrontMotion}) \citep{LNR}. The
latter mainly associates the excess vortex length $L$ with the
contributions from the new self-similar structures produced by
vortex reconnections on length scales shorter than $\ell$, when
vortex motion becomes progressively less damped at low
temperatures \citep{Svistunov:2008b}. This model relies on four
types of processes: reconnections of vortex bundles, reconnections
of neighbouring vortices, self-reconnections of a vortex, and
non-linear interactions of Kelvin waves, which bridge the energy
cascade from the Kolmogorov to the Kelvin-wave regimes. It can be
successfully fitted to the experimental $\nu(T)$, as seen in
Fig.~\ref{MancNuPrime}. In short, both models predict an
enhancement of vortex densities but on different sides of the
cross-over scale $\ell$.

Finally we may ask why is $\nu(T \! = \! 0)$ larger for
ultraquantum (random) tangles than for quasiclassical tangles in
the $T=0$ limit? Whatever the  model, this means that, for the
same total density of vortex lines $L$, the rate of energy
dissipation in random tangles is larger. Within the bottleneck
model \citep{LNR}, this seems straightforward  as these tangles
have no energy on classical scales at all. Within the
reconnections/fractalization scenario \citep{Svistunov:2008b},
this is explained by the fact that reconnections in partially
polarized tangles are less frequent and less efficient. In the
framework of this latter model, to explain why there is apparently
no temperature dependence of $\nu(T)$ for ultraquantum tangles
below $T=1$~K (although no experimental data exist for the range
0.5 -- 1.3~K), one should assume that for this type of tangle all
significant increase in the total line length $L$ caused by the
fractalization with decreasing temperature has already occurred
above 1\,K.

\subsubsection{Comparing turbulent dynamics in $^3$He-B and $^4$He}

\begin{figure}[t]
\begin{center}
\includegraphics[width=0.8\linewidth,keepaspectratio]{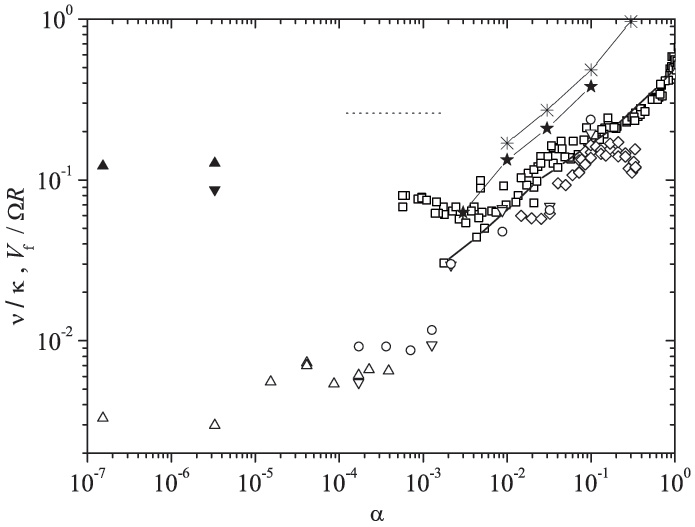}
\caption{Normalized measures of dissipation rate, the effective
kinematic viscosity $\nu$ and the normalized velocity of the
propagating vortex front $v_{\rm f} = V_{\rm f}/\Omega R$, as a
function of mutual friction dissipation $\alpha$ in $^4$He and
$^3$He-B: \textbf{(i)} $\nu(T)$ for quasiclassical turbulence in
$^4$He: spin-down, open triangles \citep{Walmsley:2007}; ion jet,
open circles \citep{Walmsley2008}; towed grid, open diamonds
\citep{Stalp2002,Niemela2005}; \textbf{(ii)} $\nu(T)$ for
ultraquantum turbulence in $^4$He: colliding charged vortex rings,
filled triangles \citep{Walmsley2008}; numerical simulations,
line-connected stars \citep{Tsubota2000}, and asterisks
\citep{Schwarz:1988}. \textbf{(iii)} $\nu(T)$ estimate for
ultraquantum turbulence in $^3$He-B: dashed line
\citep{Bradley:2006}. \textbf{(iv)} $v_{\rm f}$ in $^3$He-B:
experiment, open squares \citep{Eltsov:2007}; and numerical
simulations, line-connected squares (Sec.~\ref{s:FrontMotion}). }
\label{MancCompar3He4He}
\end{center}
\end{figure}

It would be instructive to provide a comparison of turbulent
dissipation in superfluid $^4$He and $^3$He-B. Unfortunately at
present, with only few measurements on varying types of flow, this
remains a task for the future. The Lancaster measurements on the
decay of an inhomogeneous tangle generated with a vibrating grid
by \cite{Bradley:2006} revealed a $L\propto t^{-3/2}$ late-time
decay \citep{FisherPLTP2008}. Assuming that the origin of the
$t^{-3/2}$ dependence is the same as in quasiclassical turbulence
with a saturated energy-containing length, and that this length is
equal to the spread of the turbulent tangle over a distance of
$1.5~$mm, using Eq.~(\ref{L-t-K}) they extract the
temperature-independent value of $\nu \approx 0.2 \, \kappa$ at
$0.156<T/T_{\rm c}< 0.2$. This is different from the $^4$He value
$\nu \approx 0.003 \, \kappa$. However, it is questionable whether
for this turbulence the condition for the applicability of
Eq.~(\ref{L-t-K}) is met, namely that the classical energy
$\C{E}_{\rm c}$ dominates over the quantum energy $\C{E}_{\rm q}$.
From their recent measurements \citep{Bradley:2008} of the initial
velocity of mean large-scale flow, $u \sim 0.5$\,mm/s, and vortex
density, $L \sim 10^4$\,cm$^{-2}$, we estimate ${\C{E}_{\rm c}}
\sim u^2/2 \sim 10^{-3}$\,cm$^2$/s while ${\C{E}_{\rm q}} \sim 0.7
\kappa^2 L \sim 10^{-2}$\,cm$^2$/s; \textit{i.e.} ${\C{E}_{\rm c}}
\ll {\C{E}_{\rm q}}$ even in the early stages of the decay. Hence,
it is possible that the dynamics of the decay of their localized
tangles is that of an ultraquantum turbulence but accelerated due
to diffusion or emission of vortex rings into space; thus the
decay $L(t)$ is steeper than $\propto t^{-1}$. More experiments
with grid turbulence in $^3$He-B are thus desirable, which could
shed light on the temperature dependence of turbulent dynamics,
\textit{e.g.} if there is a similar decrease in $\nu(T)$ below
$\alpha \sim 10^{-3}$ as measured for $^4$He, or to provide an
independent characterization of the energy-containing length
scale. It would also be important to reproduce the results for
$^4$He in the $T=0$ limit with another method of generating
turbulence (\textit{e.g.} by means of towed or vibrating grids).

The Helsinki experiments on the propagating vortex front
(Sec.~\ref{s:FrontMotion}) provide a measure of dissipation via
the front velocity $V_{\rm f}$. The observed relative enhancement
of the rate of decay below $0.4\,T_{\rm c}$ ($\alpha < 1$) proves
the efficiency of the turbulent cascade in dissipating the energy
of large-scale flow. The leveling off of this rate below
$0.28\,T_{\rm c}$ ($\alpha < 10^{-2}$) is qualitatively similar to
that observed in $^4$He at $T<0.5$~K ($\alpha < 10^{-5}$), and
might hint at the $T = 0$ limit in the inertial cascade. However,
the respective values of $\alpha$, at which the leveling-off
occurs, differ substantially. Various reasons can be given to
explain this difference, \textit{eg.} that different types of flow
have their own specific $\nu(\alpha)$ dependence (evidence for
this is the difference in the $\nu(T)$ values of quasiclassical
and random tangles). For instance, to test the  interpretation of
the temperature dependence of $V_{\rm f} (T)$ on approaching the
$T \rightarrow 0$ limit, given in Sec.~\ref{ss:X} in terms of the
proximity of the outer and quantum crossover scales, it would be
instructive to conduct similar measurements in containers of
different radii.

In front propagation the energy dissipation rate is proportional
to $V_{\rm f}(T)$, while in experiments described in this section
the rate is proportional to $\nu(T)$. Thus it is interesting to
compare the temperature dependences of $V_{\rm f}(T)$ and $\nu(T)$
in the two superfluids. We plot in Fig.~\ref{MancCompar3He4He} the
effective kinematic viscosity $\nu/\kappa$ and the normalized
velocity of the propagating front $v_{\rm f} = V_{\rm f}/\Omega R$
as a function of mutual friction dissipation $\alpha$. Note that
the role of turbulence comes into play in a very different manner
in these two types of flow: as nearly homogeneous and isotropic
turbulence in $^4$He in the Manchester experiments and, roughly
speaking, via the deviation from laminar vortex front propagation,
owing to turbulent excitations, in $^3$He-B in the Helsinki
experiments.

It is striking that for $\alpha < 0.1$ (\textit{i.e.} the regime
where developed turbulence becomes possible) and before leveling
off at $T \rightarrow 0$, the data for all experiments and
numerical simulations discussed here approach a similar slope of
$\sim \alpha^{0.5}$. For $\nu(\alpha)$, in the spirit of
Eq.~(\ref{EqVinenNiemela}), this implies that the parameter
$c_2^2$ is roughly proportional to $\alpha^{-0.5}$ -- even well
below the values of $\alpha$ of the calculations by
\cite{Schwarz:1988}. On a finer scale, particular models of
matching the classical and quantum cascades, as discussed in the
previous subsection, might produce specific dependences of
$c_2(\alpha)$, which perhaps can be tested in future experiments
in more detail, provided that the quality of data is improved.

We do not attempt to compare the absolute values of $\nu/\kappa$
and $v_{\rm f}$ in Fig.~\ref{MancCompar3He4He}. Instead, we
concentrate on the $\alpha-$dependences, especially on the values
of $\alpha$ at which the dissipation rates level off as $T
\rightarrow 0$. The fact that the measured $^3$He-B data level off
at higher values of $\alpha$ might suggest that the quantum
cascade in $^3$He-B is not as developed as in $^4$He and that the
ultimate dissipative mechanism, independent of $\alpha$ in the
$\alpha\rightarrow 0$ limit, is stronger  and takes over at larger
length scales in $^3$He-B. Indeed, the numerically calculated
velocity of front propagation $v_{\rm f}(\alpha)$, which does not
explicitly incorporate any excess dissipation beyond mutual
friction dissipation and hence is equally suitable for $^3$He-B
and $^4$He, is steadily decreasing with decreasing $\alpha$,
following the same trend as the experimental and numerical
$\nu(T)$ for $^4$He. In contrast, the measured $v_{\rm f}(\alpha)$
levels off below $\alpha \sim 1 \cdot 10^{-2}$, which seems to
imply that the $T=0$ regime prevails at $\alpha < 1 \cdot 10^{-2}$
in $^3$He-B, while this happens only below $\alpha < 1 \cdot
10^{-5}$ in $^4$He. The fermionic nature, the large vortex core
diameter, and the much lower absolute temperatures lead to
different inherent dissipation mechanisms in $^3$He-B. For
example, the energy loss during each reconnection event can be
substantial in $^3$He-B but not in $^4$He. This is supported by
the fact that the values of $\nu$ measured in Lancaster for
$^3$He-B with $a_0\sim 30$\,nm (at $P=12$\,bar) level off
(\textit{i.e.} branch away from the common trend of $\nu(\alpha)$
marked by the numerically calculated dependences) at larger values
of $\alpha \sim 3 \cdot 10^{-2}$, than the measured front velocity
$v_{\rm f}$ for $a_0\sim 16$\,nm (at $P=29$~bar) at $\alpha \sim 1
\cdot 10^{-2}$. We conclude by noting that further work on
turbulence in different types of flow in both $^4$He and $^3$He-B
is highly desired.

\subsection {Summary: decay of quasiclassical and ultraquantum turbulence}

The newly developed technique of measuring the density $L$ of a
vortex tangle by the scattering of charged vortex rings of
convenient radius $\sim 1$~$\mu$m has made it possible to monitor
the evolution of tangles in superfluid $^4$He down to below 0.5~K,
\textit{i.e.} deep into the zero-temperature limit. The dynamics
of two very different types of tangles can be studied.
Quasiclassical tangles, mimicking the flow of classical liquids on
large length scales were generated by an impulsive spin-down from
angular velocity $\Omega$ to rest of a rotating cubical container
with helium. After a transient of duration $\sim 100/\Omega$ the
turbulence becomes nearly isotropic and homogeneous and decays as
$L\propto t^{-3/2}$, as expected for turbulence posessing a
Kolmogorov cascade of energy from the energy-containing eddies of
constant size, set by the container size $h$. This was studied in
isotopically pure $^4$He in a broad range of temperatures of 80~mK
-- 1.6~K, corresponding to the range of mutual friction $\alpha$
from $\sim 10^{-10}$ to $10^{-1}$. Identical results were obtained
for the free decay of quasiclassical tangles generated by a
central jet of ions, although only at temperatures above 0.7~K so
far.

Alternatively, non-structured (ultraquantum) tangles of quantized
vortices, that have little flow at scales above the inter-vortex
distance and hence no classical analog, can be obtained by
colliding many small quantized vortex rings of radius $R\ll \ell$
at temperatures below 0.5~K, \textit{i.e.} in the zero-temperature
limit. These tangles took about 10~s to spread from an injector
into all experimental volume of size $h=4.5$~cm, followed by free
decay with the universal dynamics $L\propto t^{-1}$, independent
of the initial conditions. The relatively fast rate of spreading
is indeed surprising and might be due to the small polarization
(mean velocity) of the tangle.

Quantitative measurements of the free decay for both types of
tangles allow the extraction of the ``effective kinematic
viscosities'' $\nu$ -- the flow-specific parameter linking the
rate of energy dissipation with the total density of vortices $L$.
It turned out that at high temperatures $T>1$~K, where the quantum
cascade is not well-developed, the values of $\nu$ for both types
of turbulence are comparable. However, in the zero-temperature
limit, where the dissipation can only take place at very short
length scales $\ll \ell$, to which the energy can only be
delivered by a cascade of non-linear Kelvin waves on individual
vortex lines, the saturated value of $\nu$ for quasiclassical
turbulence seems to be smaller than that for ultraquantum
turbulence. On the microscopic level this is most probably related
to the fact that in the presence of quasiclassical eddies the
partial mutual alignment of vortex lines in ``bundles'' slows down
the process of exciting Kelvin waves for a given vortex density
$L$. For instance, this mutual alignment is expected to reduce the
frequency and efficiency of vortex reconnections, which are
believed to be the defining process of the dynamics of
non-structured (ultraquantum) tangles as they fuel the Kelvin-wave
cascade. At present, two microscopic models are developed to
describe the energy transfer from classical to quantum scales
(cascades) in quasiclassical tangles: one relying on reconnections
and the other assuming that additional ``non-cascading '' vortex
density helps maintain the continuity of the energy flux down to
the dissipative length scale.

We have also compared the rates of dissipation as a function of
mutual friction in superfluid $^4$He  and $^3$He-B. Both
similarities and differences in the behaviour are spotted and
discussed. However, a final quantitative verdict on the role of
specific mechanisms cannot be delivered at this stage, pending
further development of theoretical models and of experiments on
different types of turbulent flow in both superfluids.

\section{Summary}

In the last five years we have witnessed important advances in the
understanding of the appearance, growth, and decay of different
types of superfluid turbulence, especially in the fundamentally
important limit of zero temperature where the intrinsic processes
within the superfluid component set the dynamics of the tangle of
quantized vortices. As these are absolutely undamped on a broad
range of length scales down to wavelengths of order 10\,nm --
$1\,\mu$m, depending on the type of superfluid, a principally new
microscopic dynamics emerges -- set by the instabilities of
individual vortices, their reconnections and one-dimensional
cascades of energy from non-linear Kelvin waves (wave turbulence)
along individual vortex lines. Amazingly, even in this limit,
various turbulent flows behave classically on large length scales,
and often the observed rate of decay is no different from that at
high temperatures -- owing to the fact that the energy-containing
and dissipative lengths and times are well-separated.

However, also the discrete nature of quantized superfluid
vorticity becomes important in such processes as the growth of an
initial seed vortex tangle out of a single-vortex instability, the
efficiency of dissipating the energy of large-scale eddies through
short-wavelength Kelvin waves on individual vortex lines, or the
dynamics of non-structured tangles possessing no large-scale flow.
In the helium superfluids these underlying processes are
separately observable and are ultimately expected to become the
corner stones of a detailed theoretical framework. This should
make it possible to develop a consistent picture of turbulence in
helium superfluids, which describes the nonlinear turbulent
dynamics of discrete line vortices in a macroscopically coherent
quantum liquid of zero viscosity.

\textbf{Acknowledgments:} This work is supported by the Academy of
Finland (grants 213496,  124616, and 114887), by ULTI research
visits (EU Transnational Access Programme FP6, contract
RITA-CT-2003-505313), and by EPSRC (UK) (grants GR/R94855 and
EP/E001009).





\end{document}